\documentstyle[draft,epsf,psfig]{mn}

\def\hi{{\rm H\,{\sc i} }}
\def\hii{{\rm H\,{\sc ii} }}
\def\kd{$K_{dark}$ }
\def\gsim{\mathrel{\raise0.35ex\hbox{$\scriptstyle >$}\kern-0.6em 
\lower0.40ex\hbox{{$\scriptstyle \sim$}}}}
\def\lsim{\mathrel{\raise0.35ex\hbox{$\scriptstyle <$}\kern-0.6em 
\lower0.40ex\hbox{{$\scriptstyle \sim$}}}}

\title{The star formation histories of 
low surface brightness galaxies}

\author[Bell et al.]
{
Eric F. Bell$^1$, 
David Barnaby$^2$,
Richard G. Bower$^1$, 
Roelof S. de Jong$^{1,3}$\thanks{Hubble Fellow}, \cr
Doyal A. Harper, Jr.$^2$, Mark Hereld$^4$, 
Robert F. Loewenstein$^2$ and \cr
Bernard J. Rauscher$^{1,5}$\\
$^1$ Department of Physics, University of Durham, Science Labs, South Road, 
Durham DH1 3LE, UK\\
$^2$ Yerkes Observatory, University of Chicago, 373 W Geneva St., Williams
Bay, WI 53191, USA\\
$^3$ Steward Observatory, University of Arizona, 
949 N. Cherry Ave., Tucson, AZ 85719, USA\\
$^4$ Department of Astronomy and Astrophysics, 
University of Chicago, 5640 S. Ellis
Ave., Chicago, IL 60637, USA\\
$^5$ Space Telescope Science Institute, 3700 San Martin Drive, Baltimore,
MD 21218, USA
}

\begin{document}

\date{\fbox{{\sc Submitted to MNRAS}: \today}}

\maketitle

\begin{abstract}
We have performed deep imaging of a diverse sample of 26 low
surface brightness galaxies (LSBGs) in the optical and
the near-infrared.  
Using stellar population synthesis models, we find that it is possible
to place constraints on the ratio of young to old stars (which
we parameterise in terms of the average age of the galaxy), as well
as the metallicity of the galaxy, using optical and near-infrared colours.
LSBGs have a wide range
of morphologies and stellar populations, ranging from 
older, high metallicity earlier types to much younger and lower
metallicity late type galaxies.  Despite this wide range of
star formation histories, we find that colour gradients 
are common in LSBGs.  These are most naturally interpreted as 
gradients in mean stellar age, with the outer regions of LSBGs 
having younger ages than their inner regions.
In an attempt to understand what drives the differences in LSBG stellar 
populations, we compare LSBG average ages and metallicities
with their physical parameters.  Strong correlations
are seen between a LSBG's star formation history and its
$K$ band surface brightness, $K$ band absolute magnitude and gas fraction.
These correlations are consistent
with a scenario in which the star formation history 
of a LSBG primarily correlates
with its surface density and its metallicity correlates
both with its mass and surface density.  
\end{abstract}


\section{Introduction}

There has been much recent debate on the star formation histories (SFHs) of 
low surface brightness disc galaxies (LSBGs; galaxies with $B$ band
central surface brightnesses fainter than 22.5 mag\,arcsec$^{-2}$).
The best studied LSBGs are blue in the optical and the near-infrared
(near-IR) 
\cite{mcgaugh1994a,deblok1995,bergvall1999}, 
indicating a young mean stellar age
and/or low metallicity.  
Their measured \hii region metallicities are
low, at around or below $1/3$ solar abundance
\cite{mcgaugh1994b,ronnback1995,deblok1998spec}.
Morphologically, the best studied LSBGs have discs, but
little spiral structure \cite{mcgaugh1995mor}.  The current
massive star formation rates (SFRs) in LSBGs are 
an order of magnitude lower than those of high surface
brightness (HSB) galaxies \cite{vdh93,vanzeesfr}. 
\hi observations show that
LSBGs have high gas mass fractions, sometimes even approaching unity
\cite{deblok1996,mcgaugh1997}.
As yet, there have been no CO
detections of LSBGs, only upper limits on the CO abundances which
indicate that LSBGs have CO/\hi ratios significantly lower than those
of HSB galaxies \cite{schombert1990,deblok1998co}.
All of these observations are consistent with a scenario in which LSBGs are 
relatively unevolved, low mass surface
density, low metallicity systems, with roughly constant or even
increasing SFRs \cite{deblok1996,gerritsen1999}.

However, this scenario has difficulty accommodating giant
LSBGs (as, indeed, this scenario was not designed with giant
LSBGs in mind): 
with scale lengths typically in excess of 5--10 $h_{65}^{-1}$ kpc, 
these galaxies are similar to, but less extreme than, Malin 1.
Quillen \& Pickering \shortcite{quillen1997}, in an as yet unpublished 
work, obtained near-IR $H$ band imaging of two giant
LSBGs.  They concluded that the central optical--near-IR colours
of their galaxies were compatible with 
those seen in old stellar populations (such as E/S0
galaxies), and that the (more uncertain) outer colours were consistent
with somewhat younger stellar populations.  

Another difficulty for this scenario is posed by the 
recent discovery of a substantial population of red LSBGs
\cite{oneil1997a,oneil1997b}.  The optical colours of these
galaxies are similar to those of old stellar populations, 
but the red colours could be caused by age or metallicity 
effects (note that dust is not expected to be an 
important effect in most face-on LSBGs; section \ref{subsec:dust}).  
Either way, the existence of old or metal-rich 
LSBGs is difficult to understand if all LSBGs are 
unevolved and gas-rich.
This same age-metallicity
degeneracy plagues the analysis of the colours of blue LSBGs.  Padoan,
Jimenez \& Antonuccio-Delogu \shortcite{padoan1997} 
question the apparent youth of
blue LSBG stellar populations: they find that LSBG optical colours
are consistent with those of old, very low metallicity
stellar populations.  
\footnote{Note that Padoan et al.'s unconventional choice of 
IMF does not significantly affect their conclusions: blue 
colours for old, low metallicity stellar populations
are a general property of most stellar population synthesis models.}

This uncertainty caused by the age/metallicity degeneracy is partially
avoidable; for stellar populations with ongoing star formation, it
is possible to learn something of their SFH using
a combination of optical and near-IR colours.  Essentially, 
it is possible to 
compare the SFHs of galaxies in a relative sense, using a 
kind of `birthrate parameter' relating
the amount of recent star formation to the cumulative amount of 
previous star formation (in this work, we parameterise the 
SFHs using an exponential SFH with a timescale $\tau$).  
As a guide to interpreting these optical and near-IR colours, 
we use the latest multi-metallicity stellar population synthesis models of
e.g.\ Bruzual \& Charlot (in preparation) and Kodama \& Arimoto
\shortcite{ka97}.  The limiting factor in applying this technique
is typically the availability of near-IR imaging;  especially so for 
LSBGs, where the near-IR central surface brightness can be 
up to a factor of 500 fainter than the sky surface brightness 
at these wavelengths.

In this paper, we present optical and near-IR imaging for a diverse 
sample of 26 LSBGs in order to explore their SFHs.  
In our study we include examples of i) the well-studied
blue-selected LSBGs taken from 
de Blok et al.\ \shortcite{deblok1995,deblok1996}, 
de Jong \& van der Kruit \shortcite{dejong1994} and the ESO-LV 
catalogue, ii) the more poorly studied, intriguing red-selected LSBGs 
from O'Neil et al.\ \shortcite{oneil1997a,oneil1997b}, and iii)
the LSBG giants, taken from Sprayberry et al.\ \shortcite{spray95}.
Earlier results of this programme were presented in Bell et al.\
\shortcite{b99}, where the stellar populations in a subset
of five red and blue LSBGs were explored; red LSBGs were found to
be much older and metal-rich than blue LSBGs, indicating that 
the two classes of galaxy do not share a common origin.
Here, we explore the differences in SFH
between red, blue and giant LSBGs.  
We also attempt to 
understand which, if any, galaxy parameters (e.g.\ mass, or density) affect
the SFHs of LSBGs, driving the differences between
e.g.\ red and blue LSBGs. 

The plan of the paper is as follows.
The observations and data reduction are described in section 
\ref{sec:obs}.  In section \ref{sec:phot} we present the photometry
for our sample, compare with existing published photometry, and discuss
the morphologies of LSBGs in the optical and the near-IR.  In section
\ref{sec:sfh} we present the optical--near-IR colours for our sample
of galaxies, discussing them in terms of differences in SFH.  In section
\ref{sec:disc}, we elaborate on this colour-based analysis and 
explore trends in SFH with physical parameters.
Finally, we present our conclusions in section \ref{sec:conc}.

\section{Observations and data reduction} \label{sec:obs}

\subsection{Sample selection}

Our sample of 26 LSBGs was selected with a number of criteria in mind.  
They must be detectable using reasonable exposure times on 
4-m class telescopes in the near-IR,
and must span a wide range of observed LSBG
properties, such as physical size, surface brightness and colour.
In addition, we selected galaxies with as much existing optical
and \hi data as possible.  
Our sample is by no means complete, but is designed instead
to span as wide a range as possible of observed LSBG parameters.

Our northern sample was selected from 
a number of sources 
\cite{deblok1995,deblok1996,oneil1997a,oneil1997b,spray95,dejong1994}
to have moderately low published $B$ band inclination corrected
central surface brightness
$22.5\,\lsim\,\mu_{B,0}\,\lsim\,23.5$ 
mag\,arcsec$^{-2}$. 
Furthermore, the northern sample was selected to have major axis radii to the
25 $B$ mag\,arcsec$^{-2}$ isophote larger than $\sim$ 16 arcsec.
Galaxies with only moderately low
surface brightness were chosen as the near-IR sky background at temperate
sites is very high due to the large thermal flux from the 
telescope and sky.

In order to explore the properties of galaxies with even 
lower surface brightnesses, we imaged a small
sample of galaxies using the South Pole 0.6-m telescope.  The $K$ band
sky background at the South Pole 
is suppressed by a factor of $\sim\,20$ compared to
temperate sites, allowing unprecedented sensitivity for faint
extended $K$ band surface photometry (see e.g. Nguyen et al.\ 1996, 
Rauscher et al.\ 1998). 
Our southern hemisphere sample 
is selected from the 
ESO-Uppsula Catalogue \cite{esolv} to have larger
sizes and lower surface brightnesses compared to the northern sample: 
$\mu_{B,0}\,\ge\,23.0$ mag\,arcsec$^{-2}$, 
65 arcsec $\le$ R$_{{\it eff}}\,\le\,150$ arcsec, inclination less than
67$^{\circ}$ and galactic latitude $|b|\,\ge\,20^{\circ}$ to avoid
excessive foreground galactic extinction 
(where R$_{{\it eff}}$ denotes the galaxy half light radius).

Our sample is described in further detail in Table \ref{tab:samplepar}.
Galaxy types and heliocentric velocities were typically taken from
the NASA/IPAC Extragalactic Database (NED).  
Galaxy distances $D$ were 
determined from velocities centred on the Local Group 
\cite{richter} assuming
a a Hubble constant $H_0$ of 65 kms$^{-1}$Mpc$^{-1}$.
For galaxies within $\sim$ 150 Mpc, we further take into account the 
local bulk peculiar motions \cite{enzo}.  Typical distance uncertainties,
corresponding to the uncertainties in the peculiar motions, are
typically $\sim$ 7 Mpc.
\hi gas masses
were calculated using $M_{\rm{H{\sc i}}} = 2.36 \times 10^5 D^2(\rm{Mpc})
\int S(\rm{Jy}) dv(\rm{kms^{-1}})$ \cite{deblok1996}, 
where $S$ is the \hi line flux in Jy and  
$dv$ is the line width in km\,s$^{-1}$.
\hi fluxes were taken from NED, except for the \hi fluxes from 
de Blok et al.\ \shortcite{deblok1996}, Sprayberry et al.\ 
\shortcite{spray95} and O'Neil, Bothun \& Schombert \shortcite{oneil1999}.
Note that in Table \ref{tab:samplepar} all the galaxies without
\hi masses have not yet been observed in \hi {\it except} C1-4 and
C3-2, which were not detected with Arecibo in 5 $\times$ 5 minutes
and 12 $\times$ 5 minutes respectively (these limits roughly
correspond to 3$\sigma$ upper limits on the gas fraction of 
$\sim$ 15 per cent).  
In Table \ref{tab:samplepar} we have also presented the foreground 
galactic extinction in $B$ band, as estimated by Schlegel, Finkbeiner
and Davis \shortcite{sfd}. 
We have checked the Infra
Red Astronomical Satellite (IRAS) point source and small scale
structure catalogues \cite{iras} for our sample: only the Seyfert 
1 LSBG giant 2327-0244 was detected (at both 60$\mu$m and 100$\mu$m).

\begin{table*}
\caption{Sample parameters}
\label{tab:samplepar}
\begin{minipage}{175mm}
    \begin{tabular}{lcccccccc}
      \hline
      Object &  RA(2000)  &  Dec(2000)  & 
      Type & $D$(Mpc) & 
	  $\log_{10} M_{\sc Hi}$  & $A_B$ & Source   \\
      \hline
      UGC 128   & 00 13 51.3 & 35 59 41 & Sdm & 69 & 10.12$\pm$0.06$^d$ 
		& 0.28 & de Blok et al.\ (1995) \\
      ESO-LV 280140$^a$ & 00 19 58.5 & -77 05 27 & 
		        SB(s)d & 35 & 9.82$\pm$0.15$^e$ & 0.23 & ESO-LV \\
      UGC 334   & 00 33 54.9 & 31 27 04 & SAB(s)cd & 90 
	& 10.01$\pm$0.08$^f$ & 0.24 & de Jong \& van der Kruit (1994) \\
      0052-0119 & 00 55 08.9 & -1 02 47 & Sd & 213 & --- 
		& 0.15 & Sprayberry et al.\ (1995)\\
      UGC 628   & 01 00 52.0 & 19 28 37 & Sm: & 85 & 9.85$\pm$0.13$^f$ &
		0.19 & de Blok et al.\ (1995) \\
      0221+0001 & 02 24 01.3 & 00 15 07 & Sc & 575 & --- 
		& 0.19 & Sprayberry et al.\ (1995)\\
      0237-0159 & 02 40 11.0 & -1 46 27 & Sc & 197 & --- 
		& 0.13 & Sprayberry et al.\ (1995)\\
      ESO-LV 2490360$^a$ & 03 59 15.2 & -45 52 15 & 
			 IB(s)m & 24 & 9.56$\pm$0.21$^e$ & 0.04 & ESO-LV \\
      F561-1    & 08 09 41.3 & 22 33 33 & Sm & 82 & 9.39$\pm$0.05$^g$ &
		0.20 & de Blok et al.\ (1996) \\
      C1-4      & 08 19 24.4 & 21 00 12 & S0$^b$ & 56 & --- 
		& 0.21 & O'Neil et al.\ (1997a) \\
      C3-2      & 08 22 35.9 & 20 59 47 & SB0a$^b$ & --- & --- 
		& 0.16 & O'Neil et al.\ (1997a) \\
      F563-V2   & 08 53 03.5 & 18 26 05 & Irr & 74 & 9.52$\pm$0.06$^g$ &
		0.07 & de Blok et al.\ (1996) \\
      F568-3    & 10 27 20.5 & 22 14 22 & Sd & 99 & 9.66$\pm$0.04$^g$ &
		0.09 & de Blok et al.\ (1996) \\
      1034+0220$^a$ & 10 37 27.6 & 02 05 21 & Sc & 326 & --- &
		0.14 & Sprayberry et al.\ (1995)\\
      N10-2     & 11 58 42.4 & 20 34 41 & Sb$^b$ & --- & ---  &
		0.10 & O'Neil et al.\ (1997a) \\
      1226+0105 & 12 29 12.8 & 00 49 03 & Sc & 362
		& 10.63$\pm$0.01$^h$ & 0.10 & Sprayberry et al.\ (1995)\\
      F574-1    & 12 38 07.3 & 22 18 45 & Sd & 112 & 9.67$\pm$0.04$^g$ &
		0.10 & de Blok et al.\ (1996) \\
      F579-V1   & 14 32 50.0 & 22 45 46 & Sd & 103 & 9.47$\pm$0.04$^g$ &
		0.12 & de Blok et al.\ (1996) \\
      I1-2      & 15 40 06.8 & 28 16 26 & Sd$^b$ & 147$^i$ 
		& 9.73$\pm$0.05$^i$ & 0.11 & O'Neil et al.\ (1997a) \\
      F583-1    & 15 57 27.6 & 20 40 07 & Sm/Irr & 41 & 9.45$\pm$0.11$^g$ &
		0.20 & de Blok et al.\ (1996) \\
      ESO-LV 1040220 & 18 55 41.3 & -64 48 39 & 
		     IB(s)m & 23 & 9.54$\pm$0.21$^e$ & 0.36 & ESO-LV \\
      ESO-LV 1040440 & 19 11 23.6 & -64 13 21 & 
		     SABm & 23 & 9.56$\pm$0.22$^e$ & 0.16 & ESO-LV \\
      ESO-LV 1870510 & 21 07 32.7 & -54 57 12 & SB(s)m & 29 & --- &
		     0.15 & ESO-LV \\
      ESO-LV 1450250 & 21 54 05.7 & -57 36 49 & 
		     SAB(s)dm & 34 & 10.08$\pm$0.16$^e$ & 0.13 & ESO-LV \\
      P1-7      & 23 20 16.2 & 08 00 20 & Sm: & 43 
		& 9.32$\pm$0.11$^i$ & 0.45 & O'Neil et al.\ (1997a) \\
      2327-0244$^c$ & 23 30 32.3 & -2 27 45 & SB(r)b pec & 157
		    & 10.18$\pm$0.07$^j$ & 0.22 & Sprayberry et al.\ (1995) \\
      \hline
    \end{tabular}
\hspace{0.2cm} \\
We use H$_0 = 65$ km\,s$^{-1}$\,Mpc$^{-1}$ to compute the 
distances and \hi mass: distance uncertainties are $\sim$ 7 Mpc. \\
$^a$ Has a confirmed companion at a similar redshift \hspace{7.5cm}
$^b$ Our own classification\\
$^c$ Has been detected at 60 and 100 microns using IRAS \hspace{3.4cm}
$^d$ \hi Flux from Wegner, Haynes \& Giovanelli \protect\shortcite{wegner1993}\\
$^e$ \hi Flux from Huchtmeier \& Richter \protect\shortcite{huchtmeier1989} 
   \hspace{0.8cm}
$^f$ \hi Flux from Schneider et al.\ \protect\shortcite{schneider1992}
   \hspace{0.8cm}
$^g$ \hi Flux from de Blok et al.\ \protect\shortcite{deblok1996} \\
$^h$ \hi Flux from Sprayberry et al.\ \protect\shortcite{spray95} 
 	\hspace{1.07cm}
$^i$ \hi Flux from O'Neil et al.\ \protect\shortcite{oneil1999} \hspace{1.07cm}
$^j$ \hi Flux from Theureau et al.\ \protect\shortcite{theureau1998} 
\end{minipage}
\end{table*}

\subsection{Near-infrared data}

In this section, we describe the near-IR observations and data 
reduction.  As pointed out in the previous section, we used 
both the APO 3.5-m telescope and the South Pole 0.6-m telescope for
these observations.  Due to the widely different characteristics
of these two telescopes, the observation strategy and 
data reduction differed considerably between these two sets of data.

Before we discuss the observations and reduction of each data set
separately, it is useful to provide an overview of the observation
techniques and reduction steps common to our two sets of near-IR imaging data.
The near-IR sky background is much higher than the optical 
sky background.
Therefore, in order to be able to accurately 
compensate for temporal and positional variation in the high
sky background, offset sky frames were taken in addition to the
target frames.  

Our data were dark subtracted and were then flat fielded using 
a clipped median combination of all of a given
night's offset sky frames.  After this dark 
subtraction and flat fielding, large-scale structure in the images 
(on scales of $\ga$ 1/8 of the chip size) was readily visible 
at levels comparable to the galaxy emission.  This 
structure was minimised by subtracting an edited, averaged combination 
of nearby offset sky frames from each target frame.  This step is the most
involved one, and is the greatest difference between the 
reduction of the two datasets: detailed discussion of our
editing of the sky frames
is presented in sections \ref{subsubsec:apo} and \ref{subsubsec:sp}.  
The data were then aligned by centroiding 
bright stars in the individual frames, or if there are no bright
stars to align on, using the telescope offsets.  These aligned images were
then median combined together (with suitable scalings applied for
non-photometric data).  Finally, the data were photometrically calibrated
using standard stars.  Readers not interested in the 
details of the near-IR data reduction can skip to section \ref{subsec:opt}.

\subsubsection{Apache Point Observatory data} \label{subsubsec:apo}

\begin{table}
  \caption{Near-infrared observing log}
  \label{tab:nirobs}
  \begin{tabular}{lccc}
  \hline
   Galaxy & Date & Telescope$^a$ & Exposure \\ 
  \hline
  UGC 128 & 05/09/98 &  APO 3.5-m   & 360$\times$4.8s \\
  ESO-LV 280140 & 25/08/97 &  SP 0.6-m   & 11$\times$600s \\
                & 27/08/97 &  SP 0.6-m   & 12$\times$600s \\
  UGC 334 & 06/09/98 &  APO 3.5-m   & 288$\times$4.8s \\
  0052-0119 & 05/09/98 &  APO 3.5-m   & 312$\times$4.8s \\
  UGC 628 & 09/11/98 &  APO 3.5-m   & 138$\times$9.8s \\
  0221+0001 & 09/11/98 &  APO 3.5-m   & 144$\times$9.8s \\
  0237-0159 & 05/09/98 &  APO 3.5-m   & 264$\times$4.8s \\
  ESO-LV 2490360 & 28/08/97 &  SP 0.6-m   & 11$\times$600s \\
                 & 29/08/97 &  SP 0.6-m   & 10$\times$600s \\
  F561-1 & 21/03/97 &  APO 3.5-m   & 66$\times$9.8s \\
         & 08/03/98 &  APO 3.5-m   & 60$\times$9.8s \\
  C1-4 & 21/03/97 &  APO 3.5-m   & 96$\times$9.8s \\
  C3-2 & 22/03/97 &  APO 3.5-m   & 143$\times$9.8s \\
  F563-V2 & 21/03/97 &  APO 3.5-m   & 96$\times$9.8s \\
  F568-3 & 21/03/97 &  APO 3.5-m   & 96$\times$9.8s \\
  1034+0220 & 08/03/98 &  APO 3.5-m   & 110$\times$9.8s \\
  N10-2 & 21/03/97 &  APO 3.5-m   & 88$\times$9.8s \\
  1226+0105 & 08/03/98 &  APO 3.5-m   & 114$\times$9.8s \\
  F574-1 & 22/03/97 &  APO 3.5-m   & 130$\times$9.8s \\
         & 08/03/98 &  APO 3.5-m   & 84$\times$9.8s \\
  F579-V1 & 22/03/97 &  APO 3.5-m   & 126$\times$9.8s \\
          & 05/05/98 &  APO 3.5-m   & 90$\times$9.8s \\
  I1-2 & 06/09/98 &  APO 3.5-m   & 240$\times$4.8s \\
  F583-1 & 22/03/97 &  APO 3.5-m   & 204$\times$9.8s \\
  ESO-LV 1040220 & 12/08/97 &  SP 0.6-m   & 18$\times$600s \\
                 & 13/08/97 &  SP 0.6-m   & 6$\times$600s \\
                 & 14/08/97 &  SP 0.6-m   & 18$\times$600s \\
  ESO-LV 1040440 & 28/08/97 &  SP 0.6-m   & 600s \\
                 & 29/08/97 &  SP 0.6-m   & 2$\times$600s \\
                 & 30/08/97 &  SP 0.6-m   & 14$\times$600s \\
  ESO-LV 1870510 & 23/08/97 &  SP 0.6-m   & 17$\times$600s \\
                 & 24/08/97 &  SP 0.6-m  & 6$\times$600s \\
  ESO-LV 1450250 & 17/08/97 &  SP 0.6-m   & 6$\times$600s \\
                 & 19/08/97 &  SP 0.6-m   & 8$\times$600s \\
                 & 20/08/97 &  SP 0.6-m   & 14$\times$600s \\
                 & 24/08/97 &  SP 0.6-m  & 9$\times$600s \\
  P1-7 & 05/09/98 &  APO 3.5-m   & 252$\times$4.8s \\
  2327-0244 & 09/11/98 &  APO 3.5-m  & 168$\times$9.8s \\
  \hline
  \end{tabular}
\hspace{0.3cm} \\
$^a$ A $K'$ filter was used at the APO 3.5-m telescope, and a \kd 
filter was used with the South Pole 0.6-m telescope.
\end{table}

\begin{table}
  \caption{Photometric calibration for the Apache Point Observatory
    $K'$ data }
  \label{tab:apocalib}
  \begin{tabular}{lccc}
  \hline
   Date & $Z_{0,K'}$ & RMS & Sky Level \\ 
  \hline
   21/03/97 & $22.53\pm 0.04$ & 0.04 & 13.0$\pm0.1$ \\ 
   22/03/97$^{a}$ & --- & --- & $\sim$13.1 \\ 
   05/03/98$^{a}$ & --- & --- & $\sim$13.5 \\ 
   08/03/98 & $22.52\pm 0.05$ & 0.05 & 13.6$\pm0.2$ \\ 
   05/05/98$^{a}$ & --- & --- & $\sim$12.8 \\ 
   05/09/98 & $22.39\pm 0.04$ & 0.04 & 12.6$\pm0.2$ \\ 
   06/09/98 & $22.39\pm 0.04$ & 0.05 & 12.51$\pm0.08$ \\ 
   08/11/98 & $22.39\pm 0.04$ & 0.04 & 13.00$\pm0.04$ \\ 
  \hline
  \end{tabular}
	\\
  The $K'$ magnitude of a source giving 1 count\,sec$^{-1}$ is 
  $m_{K'} = Z_{0,K'} - 0.09 \sec z$\\
  Note that an airmass term of $-$0.09 mag\,airmass$^{-1}$ was assumed
  for these fits (if an airmass term was fit to each night's 
  data a value of $-$0.09 
  mag\,airmass$^{-1}$ was consistent with the data; 
  this value of airmass extinction
  coefficient is the average value for the nights with the most standard 
  stars). \\
  $^{a}$ Calibrated using United Kingdom Infrared Telescope
  service observations on 6$^{th}$ Dec.\ 1997 and 19$^{th}$ Feb.\ 1998.
\end{table}

\paragraph{Observations and preliminary reductions}

Our northern sample of 20 galaxies was imaged in the near-IR $K'$
passband (1.94--2.29\,$\mu$m) using the GRIM {\sc ii} instrument (with
a $256\,\times\,256$ NICMOS 3 detector and
0.473 arcsec\,pixel$^{-1}$) on the
Apache Point Observatory (APO) 3.5-m telescope.  
$K'$ band was used in preference to the more commonly used $K$ band to
cut down the contribution to the sky background from thermal emission.
The near-IR observing log is presented in Table \ref{tab:nirobs}.

The data were taken in blocks of six 9.8 second exposures (or if the sky
brightness was high, twelve 4.8 second exposures), yielding $\sim\,1$
minute of exposure time per pointing.  For these observations,
we took two pointings on the object (offset from each other 
by $\ga$ 20 arcsec, to allow better flat fielding accuracy and to 
facilitate cosmic ray and bad pixel removal), 
bracketed on either side by
pointings on offset sky fields (offset by $\ga$ 2 arcmin).

The six or twelve separate exposures at each pointing were
co-added to give reasonable signal in each image.
A 3$\sigma$ clipped average dark frame 
(formed from at least 20 dark frames taken before and
after the data frames) was subtracted from these co-added images.
The accuracy of this dark subtraction is $\sim$ 0.01 per cent 
of the sky level.  These
dark subtracted data were then
flat fielded with a scaled 2.5$\sigma$ clipped median combination of
the night's offset sky frames.  
We determined that the flat fielding is accurate to better than 
$\sim$ 0.5 per cent.

\paragraph{Sky subtraction}

Sky subtraction was then carried out on all of the target galaxy
frames using an automatically edited weighted average of the two nearest
offset sky frames.  Stars were automatically edited out of this 
averaged sky frame by comparing each pixel in a single sky frame with 
the same pixel in a 10 $\times$ 10 median filtered version of that frame;
pixels with ratios deviating by more than $+$0.5 per cent from unity  
were disregarded in formation of the weighted average.  

After this sky subtraction large scale gradients in the sky level
across the image, and on some occasions, residual large scale
variations in the dark frame, were visible in our sky subtracted images.
In order to take off these structures we used a preliminary,
mosaicked galaxy image to subtract off the galaxy emission in 
our sky subtracted frame, leaving only an image of the structure in 
the sky level.  We then fit a gradient to this image using the 
{\sc iraf} package {\sc imsurfit}.  The residual variations
in the dark frame
took the form of coherent drifts in two well-defined strips 
and/or quadrants of the detector.  For both types of  
variation in the dark frame, 
images were constructed to mimic these basic structures. 
These images were added or subtracted (with a number of different
trial amplitudes) from each image of the 
sky structure, and the overall image variance was determined.
The amplitude of the dark level pattern that minimised each image's
variance was then chosen as best representation of the
structure.
The best-fit gradient and dark level structure was then subtracted
from the object image, yielding a much improved
galaxy image.
The addition or subtraction of a gradient will not affect the 
photometry in a given individual object frame.
The subtraction of the structure in the dark frame 
could in principle change the galaxy photometry in a given
galaxy frame, however it is justified given the amplitude of the dark frame
variation, and the linear geometry of the structure
in the image.  It is impossible to introduce artifacts from either the
gradient or dark subtraction that will mimic a
centrally-concentrated galaxy light profile.  This suggests, and tests
carried out using corrected and uncorrected galaxy images show, that
these corrections do not significantly affect the galaxy photometry,
but reduce the size of the errors in the important outer
regions of the galaxy profile.

\paragraph{Mosaicking and calibration}

After sky subtraction, it is necessary to register and mosaic the
individual dithered object exposures together.  In most cases,
the centroids of bright stars were used to align the images,
which was typically accurate to $\sim\,0.25$ arcsec.  In a few cases,
there were no stars in the frame bright enough to centroid with:  in
these cases the telescope offsets were used to align the images.  This
procedure was typically accurate to $\sim\,0.5$ arcsec.  These images
were then median combined with a 2.5$\sigma$ clipping algorithm applied.

Calibration was achieved using standard stars from Hunt et al.\
\shortcite{Hunt1998}.  Their standard star list is an extension of the UKIRT
faint standard list of Casali \& Hawarden \shortcite{casali1992}, and
consistent results were obtained using both sets of standard star
magnitudes.  An airmass term of $-$0.09 mag\,airmass$^{-1}$ 
was assumed for the calibration (representing the typical airmass term
for most sites in $K$ band, and the average airmass term determined
from our two full nights with a sufficient number of 
calibration stars); 
because our observing time was often in half nights, the small number
of standard stars frequently did not allow accurate determination of both a
zero point and airmass term.  For those nights without photometric
calibration, the data were calibrated
using United Kingdom Infrared Telescope (UKIRT)
data (reduced in a similar way to the galaxy data) 
in $H$ and $K$ to determine the $K'$ magnitude of
bright stars in the field of our targets.  This calibration was
typically accurate to $\sim\,0.05$ mag.  
This calibration was cross-checked with galaxies with known
zero-points; the UKIRT and Apache Point Observatory calibration agree
to within their combined photometric uncertainties.
$K'$ calibration for the Apache Point Observatory data is given in
Table \ref{tab:apocalib}.  Note that our $K'$ magnitudes can readily
be translated into the more standard $K$ band using Wainscoat \& 
Cowie's \shortcite{wainscoat1992} relation 
$K' - K = (0.22 \pm 0.03) (H - K)$, assuming a typical
$H - K$ colour of $\sim$ 0.3 from de Jong \shortcite{dejong1996iv}.

\subsubsection{South Pole data} \label{subsubsec:sp}

\begin{table}
  \caption{Photometric calibration for the South Pole
    \kd data }
  \label{tab:spcalib}
  \begin{tabular}{lcccc}
  \hline
   Date & $Z_0$ & $A$ & $Z_0(\sec z = 1.2)$ & RMS  \\
  \hline
   12/08/97 & 15.60 & 0.07 & 15.52 & 0.08 \\
   13/08/97 & 15.56 & 0.07 & 15.48 & 0.10 \\ 
   14/08/97 & 15.80 & 0.18 & 15.58 & 0.06 \\ 
   17/08/97$^{a}$ & 16.10 & 0.32 & 15.72 & 0.07 \\ 
   19/08/97$^{b}$ & 16.26 & 0.86 & 15.23 & 0.07 \\ 
   20/08/97 & 16.29 & 0.68 & 15.47 & 0.10 \\ 
   23/08/97 & 16.67 & 0.69 & 15.84 & 0.07 \\ 
   24/08/97 & 16.23 & 0.35 & 15.81 & 0.06 \\ 
   25/08/97 & 15.88 & 0.15 & 15.70 & 0.08 \\ 
   27/08/97 & 15.97 & 0.26 & 15.65 & 0.17 \\ 
   28/08/97 & --- & --- & --- & --- \\
   29/08/97 & 15.77 & 0.04 & 15.72 & 0.12 \\ 
   30/08/97 & 16.25 & 0.32 & 15.87 & 0.10 \\ 
  \hline
  \end{tabular}
  \\
   The $K$ magnitude for an object giving 
   1 count\,sec$^{-1}$ is $m_K = Z_0 - A \sec z$ \\
  Most nights were non-photometric.  Despite the large
   variations seen in zero point calibration, repetition of objects on
   different nights indicated that the calibrations are repeatable to
   better than 0.1 mag. \\
   $^{a}$ First part only \\ 
   $^{b}$ Thick ice fog 
\end{table}

\paragraph{Observations and preliminary reductions}

The six galaxies in our
southern sample were imaged in the \kd passband (2.27--2.45\,$\mu$m)
using the GRIM {\sc i} instrument on the South Pole 0.6-m 
telescope.  GRIM {\sc i} uses a NICMOS 1 128 $\times$ 128 array
with a pixel scale of 4.2 arcsec\,pixel$^{-1}$.
The \kd filter was used in preference to the more
common $K$ filter to reduce the background: the \kd filter
selects a portion of $K$ band that is largely free from the intense and 
highly variable forest of OH lines  that dominate the sky background
between $\sim$ 1.9$\mu$m and 2.27$\mu$m.  
From 2.27--2.45$\mu$m, there are few OH lines, and the
background is almost entirely due to thermal emission from the
telescope and atmosphere. 
By observing at the South Pole, where mean
winter temperatures are $\sim$ $-$65$^\circ$C, this thermal emission is
greatly reduced, yielding a net background nearly 3~mag lower than at a
mid-latitude site, and over 1 mag lower than those achievable at the 
South Pole using a 
standard $K$ filter \cite{nguyen1996}.  
The observation dates and on-source exposure times
are presented in Table \ref{tab:nirobs}.

The data were taken using a 10 minute
exposure time, with every object frame bracketed by two offset sky
frames (with offsets $\sim$ 10 arcmin).  
Each object and sky pointing was offset by $\sim$ 1 arcmin 
to ensure accurate cosmic ray, bad pixel and flat fielding
artifact removal.
Dark subtraction was performed each night using 3$\sigma$
clip average dark frames.  
Due to the long exposure times (10 minutes), the bias
level varied significantly during the exposure, leading to
dark frame variations with amplitudes $\la$ 5 per cent. 
The data were flat fielded using a 2.5$\sigma$ clipped median
combination of a given night's offset sky frames.  The uncertainty in
the flat field was determined from the night-to-night variation in the
flat field: 
over the central region of the chip used for galaxy photometry,
the night-to-night 
RMS variations in the flat frame are $\sim$ 3 per cent.
Both the dark and flat field variations are much larger than
those typical of more modern near-IR chips at temperate
sites (where the exposure time is much shorter).  However, these 
variations in the dark frame and in the flat field are
largely compensated for during the sky subtraction step, and
so will not greatly affect the final accuracy of our galaxy 
photometry and sky level determination.
The flat field uncertainty will, however, affect the galaxy 
photometry slightly, 
potentially introducing spurious structure over reasonably
large spatial scales at the $\sim$ 0.03 mag level.

\paragraph{Sky subtraction}

Because of
the large field of view of the device, and the large pixel scale
($\sim\,4.2$ arcsec), the offset sky frames were heavily contaminated
with stars.  For this reason, the techniques used for sky subtraction
for the APO data are not suitable, as weighted averaging of the nearest
two star-subtracted sky frames gives a poor result, with many stellar
artifacts visible across the frame.  
We have used a modified version of the sky subtraction method of 
Rauscher et al.\ \shortcite{rauscher1998} as follows. 
\begin{enumerate}
\item Stars were automatically edited out from each sky frame.  
We fit a 5$^{th}$ order
polynomial in $x$ and $y$ (with cross terms enabled) to each sky frame,
rejecting pixels (and their neighbours within a 2 pixel radius) 
with larger than 2.5$\sigma$ deviations from 
the local ($5\,\times\,5$ pixel) median.  These surface
fits to the frame were used to replace the above rejected pixel regions.
These edited frames were then median filtered
using a $8\,\times\,8$ box, to further reject residual structure in
the wings of any bright star.  This process was carried out for each
sky frame in turn.  
\item For each pixel, its value in each of these edited
frames is determined, and a cubic spline is fit to the variation
of this pixel value as a function of time
(note that cubic spline fits are simply an alternative way of
interpolating between two data points that also takes into account longer
scale trends in the values).  
\item The time of each object frame is then used to `read off' 
the pixel values given by the 128$^2$ cubic 
spline fits.  This 
manufactured cubic spline fit image is used as the sky image.
\end{enumerate}
By using this method for background estimation, it is possible to use
long timescale trends in the data to better estimate the sky structure at
the time that the object frame was observed.
In this way, it is possible to reduce the limiting 1$\sigma$ noise over
large scales in the final mosaic by $\sim\,0.3$ mag (to between 
23.4 and 24.1 mag\, arcsec$^{-1}$, depending on exposure time), 
compared to simple
linear interpolation of the sky frames.  Further description of the
background subtraction
method can be found 
in Rauscher et al.\  \shortcite{rauscher1998,rauscher1999}.


\paragraph{Mosaicking and calibration}

These dark subtracted, flat fielded and 
sky subtracted images are then mapped 
onto a more finely-sampled grid, with 16
pixels mapping onto every input pixel.
Object image alignment is performed 
using centroiding on several bright stars
in the field near the galaxy; typical uncertainties in alignment are
$\lsim 0.5$ arcsec.  Conditions during August 1997 were only rarely photometric
at the South Pole, thus images were scaled to have common intensities
before combination.  This was achieved using the {\sc iraf} task {\sc
  linmatch}, using several high contrast areas of the image to define
the scaling of the images.  In this way, images were scaled to an
accuracy of $\sim\,2$ per cent before combination into a final mosaic.
These images are then median-combined with a 2.5$\sigma$ clip applied
to form the final object image.   

Photometric calibration was achieved using stars taken from the NICMOS
standard star list of Persson et al.\ \shortcite{nicmos} 
and Elias et al.\ \shortcite{elias}
and is presented in Table \ref{tab:spcalib}.
Due to the non-photometric conditions during our observing run 
at the South Pole and the 
the $\sim$ 3 per cent flat fielding uncertainty,
the photometry is expected to be relatively inaccurate.
However, each galaxy was observed on multiple nights, yielding
different estimates of the zero point,
allowing us to estimate the accuracy of the calibration of the 
final galaxy image.
Calibration typically repeated to better then
0.1 mag in absolute terms 
(see the zero point at $\sec z = 1.2$ in Table \ref{tab:spcalib}):  
the variations in the zero point
calibration are quite small when considered over the range of airmasses
of these observations ($1.1\,\la\,\sec z\,\la\,1.5$). 
The final zero points are accurate to better than 0.1 mag, except
for the zero point for ESO-LV 2490360, which is uncertain to $\sim$ 0.2 mag, as
it was observed on only one relatively clear night.
For further details of the calibration of 1997 observing season
South Pole data, see Barnaby et al.\ \shortcite{barn}.  

\subsection{Optical data} \label{subsec:opt}

\begin{table}
  \caption{Optical observing log}
  \label{tab:optobs}
  \begin{tabular}{lccc}
  \hline
   Galaxy & Date & Telescope/Filter & Expos. \\ 
  \hline
  ESO-LV  280140 & 4-5/10/97 & CTIO 0.9-m / $B$ & 6$\times$600s \\
                 & 4-5/10/97 & CTIO 0.9-m / $R$ & 6$\times$300s \\
  ESO-LV 2490360 & 4-5/10/97 & CTIO 0.9-m / $B$ & 6$\times$600s \\
                 & 4-5/10/97 & CTIO 0.9-m / $R$ & 6$\times$300s \\
  C1-4           & 25/11/97  & INT 2.5-m / $R$  & 480s \\
  C3-2           & 25/11/97  & INT 2.5-m / $R$  & 480s \\
  F563-V2        & 25/11/97  & INT 2.5-m / $V$  & 480s \\
                 & 25/11/97  & INT 2.5-m / $R$  & 480s \\
  N10-2          & 13/07/98  & JKT 1.0-m / $B$  & 600s \\
                 & 13/07/98  & JKT 1.0-m / $V$  & 300s \\
                 & 25/11/97  & INT 2.5-m / $R$  & 2$\times$480s \\
  F574-1         & 19/12/97  & INT 2.5-m / $B$  & 600s \\
                 & 19/12/97  & INT 2.5-m / $V$  & 480s \\
  F579-V1        & 13/07/98  & JKT 1.0-m / $B$  & 2$\times$900s \\
                 & 13/07/98  & JKT 1.0-m / $V$  & 900s \\
  I1-2           & 13/07/98  & JKT 1.0-m / $B$  & 900s \\
                 & 13/07/98  & JKT 1.0-m / $R$  & 600s \\
  F583-1         & 13/07/98  & JKT 1.0-m / $B$  & 2$\times$900s \\
                 & 13/07/98  & JKT 1.0-m / $V$  & 900s \\
  ESO-LV 1040220 & 4-5/10/97 & CTIO 0.9-m / $B$ & 6$\times$600s \\
                 & 4-5/10/97 & CTIO 0.9-m / $R$ & 6$\times$300s \\
  ESO-LV 1040440 & 4-5/10/97 & CTIO 0.9-m / $B$ & 6$\times$600s \\
                 & 4-5/10/97 & CTIO 0.9-m / $R$ & 6$\times$300s \\
  ESO-LV 1870510 & 4-5/10/97 & CTIO 0.9-m / $B$ & 6$\times$600s \\
                 & 4-5/10/97 & CTIO 0.9-m / $R$ & 6$\times$300s \\
  ESO-LV 1450250 & 4-5/10/97 & CTIO 0.9-m / $B$ & 6$\times$600s \\
                 & 4-5/10/97 & CTIO 0.9-m / $R$ & 6$\times$300s \\
  \hline  
  \end{tabular}
	\\
\end{table}

\begin{table*}
\begin{minipage}{105mm}
  \caption{Photometric calibration for the optical data }
  \label{tab:optcalib}
  \begin{tabular}{lcccccc}
  \hline
   Tel. & Date & Pass & $Z_0$ & $A$ & $C$ & RMS \\
  \hline
  INT & 25/11/97 & $V$ & 25.54 & 0.12 &  & 0.05 \\
  & & $R$ & 25.48 & 0.07 &  & 0.05 \\
  INT & 19/12/97 & $B$ & 25.47 & 0.24 & +0.025$(B-V)$ & 0.02 \\
  & & $V$ & 25.32 & 0.12 &  & 0.01 \\
  CTIO & 04/10/97 & $B$ & 22.72 & 0.29 & -0.06$(B-R)$ & 0.03 \\
  &  & $R$ & 22.85 & 0.09 &  & 0.025\\
  CTIO & 05/10/97 & $B$ & 22.68 & 0.21 & -0.06$(B-R)$ & 0.015 \\
  &  & $R$ & 22.85 & 0.07 &  & 0.015\\
  JKT & 13/07/98 & $B$ & 22.82 & 0.21 & +0.048$(B-V)$ & 0.018 \\
  &  & $V$ & 22.80 & 0.12 & +0.015$(B-V)$ & 0.015 \\
  &  & $R$ & 22.96 & 0.08 & & 0.016 \\
  \hline
  \end{tabular}
	\\
  The magnitude of an object giving 1 count\,sec$^{-1}$ 
   is $m_{\rm{Pass}} = Z_0 - A \sec z + C$
\end{minipage}
\end{table*}

Around 70 per cent
of the optical photometry preseted in this paper is derived from optical images
kindly provided by Erwin de Blok, Stacy McGaugh,
Karen O'Neil and David Sprayberry.  The remainder of the optical data were
obtained with the Cerro Tololo Interamerican Observatory (CTIO) 0.9-m 
(with a 2048$\times$2048 Tectronix 
CCD and pixel scale of 0.40 arcsec\,pixel$^{-1}$), 
the Jacobus Kapteyn 1.0-m 
(JKT; with a 1024$\times$1024 Tektronix CCD and pixel scale of 
0.33 arcsec\,pixel$^{-1}$), or the Issac Newton 2.5-m telescope 
(INT; as part of its service observing programme; with a 2048$\times$2048 
Loral CCD and pixel scale of 0.37 arcsec\,pixel$^{-1}$).  
A summary of the new optical data
presented in this paper is given in Table \ref{tab:optobs}.

The optical data were overscan corrected, trimmed, and corrected
for any structure in the bias frame by subtracting a 2.5$\sigma$
clipped combination of between 5 and 20 individual bias frames.  
This bias subtraction is typically accurate to $\la\,0.1$ per cent
of the sky level.  
The data were then flat fielded using 2.5$\sigma$ clipped 
combinations of twilight flat frames.  From comparison of 
twilight flats taken at different times, and from inspection of 
the sky level in our galaxy images, we find that the flat
fielding is typically accurate to $\la\,0.4$ per cent.
In addition, the $R$ band frames taken with the INT show low-level
fringing; this fringing has been taken out using a scaled fringe 
frame constructed from all the affected science frames.  The application
of this fringe correction reduces the level of the fringing by a factor
of around four, and allows better determination of the sky level in 
the outermost regions of the image.  Because we average over
large areas, the application of this fringing
correction does not affect the surface brightness profile or 
integrated magnitude to within the uncertainty in 
the sky level. 

If more than one image of the galaxy was taken, the images were 
aligned by shifting the images in $x$ and $y$ so that the 
centroids of bright stars on the image coincided.  The accuracy of
this procedure is typically 
$\sim$ 0.1 arcsec for all of our optical data.  These aligned images
were then averaged (for two or three frames) 
or median combined (for more than three frames) using a 
2.5$\sigma$ clipping algorithm.

The data were calibrated using at least ten standard star fields from Landolt
\shortcite{landolt1992}, with the exception of the INT service data, for
which only one standard field was taken on 25 November 1997 and two
standard fields were taken on 19 December 1997 at similar airmasses 
to the science data.  Mean extinction 
coefficients for La Palma were used for those calibrations, and were checked
in $V$ band by comparison with the results from the Carlsberg Meridian 
Telescope for those nights, whose extinction coefficients were found to 
be identical to within 0.01 mag to La Palma's average values.
The adopted calibrations, along with their RMS scatter, are presented
in Table \ref{tab:optcalib}.

\section{Photometry results} \label{sec:phot}

The calibrated images for a given galaxy were aligned using 
the {\sc iraf} tasks {\sc geomap} and {\sc gregister}.  A minimum of 
five stars were used to transform all of the images of a given 
galaxy to match the $R$ band image.  The typical alignment accuracy
was $\sim$ 0.2 arcsec.  Foreground stars and background objects were 
interactively edited out
using the {\sc iraf} task {\sc imedit}.
For the aperture photometry, these undefined 
areas were smoothly interpolated over using a linear interpolation.  
For the surface photometry, these areas were defined as bad pixels, and
disregarded during the fitting process.  

\subsection{Surface and aperture photometry}

These edited, aligned images were analysed using the 
{\sc stsdas} task {\sc ellipse} and {\sc iraf} task {\sc phot}.  
The central position and ellipse parameters used to fit the galaxy were
defined in $R$ band.  $R$ band was chosen as it was a good comprimise
between signal-to-noise (which is good for the optical $V$ and $R$
images) and the dust and star formation insensitivity of the
redder (and especially near-IR) passbands.
The centre of the galaxy was defined
as the centroid of the brightest portion of the galaxy.
The ellipse parameters were determined by fixing the central 
position of the galaxy, and letting the position angle and ellipticity
vary with radius.  Because of low signal-to-noise in the outer parts of the 
galaxy, the fitting of the ellipse parameters 
was disabled at a suitable radius,
determined iteratively by visually examining the quality of the 
ellipse parameters as a function of radius.
In exceptional cases (usually because the galaxy had a flocculent
disc, with only a few overwhelmingly bright regions, or 
because the galaxy was very irregular and/or lopsided), the ellipse 
parameters were fixed at all radii 
as they were unstable over almost all of the 
galaxy disc.  
In these cases the galaxy ellipticity and position angle were
determined visually.
Galaxies for which the ellipticity and position angle were
fixed are indicated in Table \ref{tab:bulgedisc}.
We also present our `best estimates' of ellipticity and position 
angle, and the radius at which the fitting of the ellipticity and position
angle was disabled in Table \ref{tab:epa}.

The sky level (and its error estimate) 
was estimated using both the outermost regions of the
surface brightness profiles and the mean sky level measured in small
areas of the image which were free of galaxy emission or 
stray starlight.  Owing to the low surface brightness of our sample 
in all passbands, the error in the sky level dominates the
uncertainty in the photometry.  This was also demonstrated using
Monte Carlo simulations, which included the effects of seeing
uncertainty, sky level errors and shot noise.
This sky level, averaged over large areas, is typically accurate to 
a few parts in 10$^5$ of the sky level for the $K'$ images, and better 
than $\sim$ 0.5 per cent of the sky level for the optical and \kd images.

The raw surface photometry profiles in all passbands are presented in Fig.\ 
\ref{fig:surfphot}.  The lower two panels show the ellipticity and 
position angle of the galaxies as a function of radius as determined
from the $R$ band images.  
Also shown are the adopted ellipticities and position angles (dashed lines)
and the largest radius at which they can be measured (arrow).
The upper panel shows the raw surface brightness
profiles in these ellipses in all the available passbands for the sample
galaxies.  
These curves have not been corrected for the effects of seeing, 
galaxy inclination, K-corrections, $(1+z)^4$ surface brightness dimming
or for the effects of foreground galactic extinction.  
The symbols 
illustrate the run of surface brightness with radius in different
passbands, and the dotted lines indicate the effects of adding or subtracting
the estimated sky error from the surface brightness profile.  
The surface
brightness profile is plotted out to the radius where the sky subtraction
error amounts to $\pm$0.2 mag or more in that passband.
The solid line is the best fit bulge/disc profile fit to the surface 
photometry, as described in the next section. 

`Total' aperture magnitudes are derived for the galaxies by extrapolating 
the high signal-to-noise regions of the aperture photometry 
to large radii using the bulge/disc decompositions described in 
the next section.  
The aperture magnitude is determined using the 
{\sc iraf} task {\sc phot} in an annulus large enough to include
as much light as possible, while being small enough to have errors
due to sky level uncertainties smaller than 0.1 mag.  
This aperture magnitude was then extrapolated to infinity using the 
best fit bulge/disc fit to the surface brightness profile.  The quoted
uncertainty in the Table \ref{tab:bulgedisc} primarily reflects the 
uncertainty in the adopted aperture magnitude:  extrapolation errors
are difficult to accurately 
define.  The median extrapolation is $\sim$0.02 mag, and 90 per 
cent of the extrapolations are smaller than $\sim$0.5 mag.  Cases where the
extrapolation has exceed 0.15 mag have been flagged in Table 
\ref{tab:bulgedisc}. 

\subsection{Bulge/disc decompositions}

\begin{table*}
\begin{minipage}{170mm}
\caption{Bulge/disc decompositions}
\label{tab:bulgedisc}
\begin{tabular}{lcccccccc}
\hline
 & & \multicolumn{2}{c}{Disc Parameters} &
\multicolumn{4}{c}{Bulge Parameters} \\
Galaxy & Passband & $\mu_0$ & $h$ &
$\mu_e$ & $r_e$ & B/D & Type & $m_T$ \\
\hline
UGC 128 & $U$ &  23.96$\pm$ 0.15 &  32$\pm$ 6 &
 25.16$\pm$ 0.03 &   8$\pm$ 2 &   0.04$\pm$ 0.01 & e &
 15.0$\pm$  0.2$^b$ \\
  & $B$ &  23.55$\pm$ 0.05 &  24.30$\pm$ 0.6 &
 24.66$\pm$ 0.01 &   6.3$\pm$ 0.3 &   0.05$\pm$ 0.02 & e &
 15.16$\pm$  0.05 \\
  & $V$ &  22.94$\pm$ 0.02 &  24.7$\pm$ 0.5 &
 23.80$\pm$ 0.01 &   7.0$\pm$ 0.1 &   0.07$\pm$ 0.01 & e &
 14.50$\pm$  0.05 \\
  & $R$ &  22.50$\pm$ 0.01 &  22.2$\pm$ 0.1 &
 23.27$\pm$ 0.01 &   6.7$\pm$ 0.1 &   0.08$\pm$ 0.01 & e &
 14.35$\pm$  0.05 \\
  & $I$ &  22.09$\pm$ 0.04 &  21.1$\pm$ 0.3 &
 22.74$\pm$ 0.01 &   6.7$\pm$ 0.2 &   0.11$\pm$ 0.01 & e &
 14.06$\pm$  0.05 \\
  & $K'$ &  20.3$\pm$ 0.2 &  21$\pm$ 4 &
 20.80$\pm$ 0.03 &   7.3$\pm$ 0.6 &   0.14$\pm$ 0.01 & e &
 12.1$\pm$  0.2$^b$ \\
ESO-LV 280140 & $B$ &  23.15$\pm$ 0.02 &  20.8$\pm$ 0.2 & 
--- & --- & --- & --- &  14.94$\pm$  0.10 \\ 
  & $R$ &  21.93$\pm$ 0.01 &  17.6$\pm$ 0.1 & 
--- & --- & --- & --- &  13.95$\pm$  0.07 \\ 
  & $K$ &  20.24$\pm$ 0.03 &  16$\pm$ 1 & 
--- & --- & --- & --- &  12.5$\pm$  0.4$^b$ \\ 
UGC 334$^f$ & $B$ &  23.5$\pm$ 0.1 &  25$\pm$ 10 &
 25.1$\pm$ 0.1 &   5.5$\pm$ 1.5 &   0.02$\pm$ 0.02 & e &
 15.4$\pm$  0.3$^{b,h}$ \\
  & $V$ &  22.6$\pm$ 0.2 &  23$\pm$ 11 &
 24.08$\pm$ 0.2 &   5$\pm$ 1.5 &   0.03$\pm$ 0.03 & e &
 14.7$\pm$  0.3$^{b,h}$ \\
  & $R$ &  22.3$\pm$ 0.2 &  21$\pm$ 7 &
 23.63$\pm$ 0.05 &   5$\pm$ 1 &   0.03$\pm$ 0.03 & e &
 14.4$\pm$  0.3$^{b,h}$ \\
  & $I$ &  21.9$\pm$ 0.1 &  17$\pm$ 3 &
 23.1$\pm$ 0.1 &   5$\pm$ 1 &   0.06$\pm$ 0.04 & e &
 14.4$\pm$  0.2$^{a,h}$ \\
  & $K'$ &  20.0$\pm$ 0.1 &  16$\pm$ 3 &
 21.2$\pm$ 0.1 &   4.5$\pm$ 0.5 &   0.05$\pm$ 0.03 & e &
 12.7$\pm$  0.2$^{b,h}$ \\
0052-0119$^e$ & $B$ &  24.75$\pm$ 0.02 &  36$\pm$ 4 &
 24.06$\pm$0.01 &  12.5$\pm$ 0.1 &   
 0.8$\pm$0.2 & r & 14.7$\pm$  0.1$^b$ \\
          & $V$ &  23.89$\pm$ 0.05 &  30$\pm$ 8 &
 22.86$\pm$ 0.01 &  11.4$\pm$ 0.1 &   
 1.3$\pm$ 0.7 & r & 13.9$\pm$  0.1$^b$ \\
          & $R$ &  23.14$\pm$ 0.03 &  24$\pm$ 4 &
 22.09$\pm$ 0.01 &  9.9$\pm$ 0.1 &   
 1.7$\pm$ 0.7 & r & 13.5$\pm$  0.1$^b$ \\
          & $K'$ & 21.53$\pm$ 0.05 &  20.0$^c$ &
 19.6$\pm$ 0.1 &  13.5$\pm$ 1.5 &  
 10$\pm$ 3 & r & 10.8$\pm$  0.5$^{b}$ \\
UGC 628 & $U$ &  23.0$\pm$ 0.1 &  17$\pm$ 1 & 
 25.18$\pm$ 0.05 &   5.8$\pm$ 0.8 &   0.03$\pm$ 0.02 & e &  15.5$\pm$ 0.1 \\ 
        & $B$ &  23.1$\pm$ 0.1 &  17$\pm$ 2 & 
 24.66$\pm$ 0.05 &   7$\pm$ 1 &   0.07$\pm$ 0.04 & e &  15.6$\pm$ 0.1 \\ 
        & $V$ &  22.55$\pm$ 0.05 &  16.5$\pm$ 0.7 & 
 23.74$\pm$ 0.04 &   7.6$\pm$ 0.4 &   0.13$\pm$ 0.03 & e &  15.1$\pm$ 0.1 \\ 
        & $R$ &  22.14$\pm$ 0.05 &  16.0$\pm$ 0.3 & 
 23.24$\pm$ 0.02 &   7.5$\pm$ 0.2 &   0.15$\pm$ 0.01 & e &  14.7$\pm$ 0.1 \\ 
        & $I$ &  21.65$\pm$ 0.02 &  15.2$\pm$ 0.3 & 
 22.62$\pm$ 0.02 &   8.3$\pm$ 0.1 &   0.23$\pm$ 0.01 & e &  14.20$\pm$ 0.05 \\ 
        & $K'$ &  20.3$\pm$ 0.3 &  16.4$\pm$ 0.6 & 
 20.76$\pm$ 0.05 &  8.2$\pm$ 0.8 &  0.32$\pm$ 0.05 & e &  12.6$\pm$ 0.2$^a$ \\ 
0221+0001$^j$ & $B$ &  23.6$\pm$ 0.1 &   7.7$^c$ &
 23.9$\pm$ 0.4 &   2.10$^c$ &  0.21$\pm$ 0.04 & r & 17.20$\pm$  0.09 \\
          & $V$ &  22.4$\pm$ 0.1 &   7.4$\pm$ 0.5 &
 22.1$\pm$ 0.2 &   1.8$\pm$ 0.3 &  0.27$\pm$ 0.03 & r & 16.12$\pm$  0.09 \\
          & $R$ &  21.92$\pm$ 0.05 &   7.0$\pm$ 0.3 &
 21.98$\pm$ 0.04 &   2.5$\pm$ 0.1 &  0.45$\pm$ 0.03 & r & 15.63$\pm$  0.10 \\
          & $K'$ &  19.5$\pm$ 0.1 &  6.2$\pm$ 0.7 &
 18.76$\pm$ 0.05 &   2.7$\pm$ 0.1 &  1.3$\pm$ 0.4 & r & 12.82$\pm$  0.08 \\
0237-0159$^f$ & $B$ &  23.89$\pm$ 0.08 &  19$\pm$ 3 &
 22.33$\pm$ 0.01 &   3.56$\pm$ 0.05 & 0.29$\pm$ 0.06 & e & 
 15.36$\pm$ 0.10$^a$ \\
          & $V$ &  23.10$\pm$ 0.06 &  17.3$\pm$ 1.4 &
 21.25$\pm$ 0.01 &   3.29$\pm$ 0.03 & 0.38$\pm$ 0.04 & e & 14.68$\pm$  0.10 \\
          & $R$ &  22.45$\pm$ 0.15 &  13$\pm$ 2 &
 20.65$\pm$ 0.01 &   3.13$\pm$ 0.05 &  0.61$\pm$ 0.07 & e & 14.52$\pm$  0.08 \\
          & $K'$ & 19.67$\pm$ 0.05 & 14.60$^c$ &
 17.48$\pm$ 0.01 &  2.45$^c$ & 0.4$\pm$ 0.2 & e & 11.6$\pm$ 0.4$^b$ \\
ESO-LV 2490360 & $B$ &  24.0$\pm$ 0.2 &  37$\pm$ 8 & 
--- & --- & --- & --- &  15.0$\pm$  0.1 \\ 
  & $R$ &  22.9$\pm$ 0.1 &  28$\pm$ 3 & 
--- & --- & --- & --- &  14.3$\pm$  0.1 \\ 
  & $K$ &  20.96$\pm$ 0.07 &  24$\pm$ 3 & 
--- & --- & --- & --- &  12.45$\pm$  0.15$^b$ \\ 
F561-1$^{d,f}$ & $U$ &  23.2$\pm$ 0.2  &  12.50$^c$ &
--- & --- & --- & --- & 16.0$\pm$  0.2 $^b$ \\
       & $B$ &  23.14$\pm$ 0.1 &  11$\pm$ 1 &
--- & --- & --- & --- & 16.18$\pm$  0.09 \\
       & $V$ &  22.51$\pm$ 0.03 &  10.0$\pm$ 0.6 &
--- & --- & --- & --- & 15.66$\pm$  0.09 \\
       & $R$ &  22.15$\pm$ 0.02 &   9.7$\pm$ 0.1 &
--- & --- & --- & --- & 15.43$\pm$  0.09 \\
       & $I$ &  21.63$\pm$ 0.04 &   9$\pm$ 1 &
--- & --- & --- & --- & 15.0$\pm$  0.15 $^a$ \\
       & $K'$ &  20.23$\pm$ 0.04 &  9$\pm$ 1  &
--- & --- & --- & --- & 13.5$\pm$  0.1 $^a$ \\
C1-4 & $B$ &  22.08$\pm$ 0.02 &   8.1$\pm$ 0.15 &
 22.40$\pm$ 0.01 &   2.17$\pm$ 0.04 &   0.10$\pm$ 0.01 & e &
 16.16$\pm$  0.08 \\
  & $V$ &  21.1$\pm$ 0.1 &   7.3$\pm$ 0.5 &
 21.7$\pm$ 0.15 &   1.9$\pm$ 0.4 &   0.07$\pm$ 0.01 & e &
 15.5$\pm$  0.1 \\
  & $R$ &  20.67$\pm$ 0.01 &   8.03$\pm$ 0.02 &
 20.83$\pm$ 0.01 &   2.04$\pm$ 0.01 &   0.11$\pm$ 0.01 & e &
 14.75$\pm$  0.09 \\
  & $I$ &  20.10$\pm$ 0.06 &   7.6$\pm$ 0.3 &
 20.99$\pm$ 0.02 &   2.65$\pm$ 0.15 &   0.10$\pm$ 0.01 & e &
 14.50$\pm$  0.09 \\
  & $K'$ &  18.20$\pm$ 0.02 &   7.6$\pm$ 0.1 &
 17.76$\pm$ 0.01 &   1.65$\pm$ 0.01 &   0.13$\pm$ 0.01 & e &
 12.46$\pm$  0.08 \\
C3-2 & $U$ &  21.90$\pm$ 0.01 &   5.57$\pm$ 0.06 & 
--- & --- & --- & --- &  16.80$\pm$  0.07 \\ 
  & $B$ &  21.74$\pm$ 0.01 &   5.72$\pm$ 0.04 & 
--- & --- & --- & --- &  16.52$\pm$  0.04 \\ 
  & $R$ &  20.08$\pm$ 0.01 &   5.25$\pm$ 0.03 & 
--- & --- & --- & --- &  15.06$\pm$  0.04 \\ 
  & $I$ &  19.43$\pm$ 0.01 &   5.39$\pm$ 0.08 & 
--- & --- & --- & --- &  14.35$\pm$  0.08 \\ 
  & $K'$ &  17.50$\pm$ 0.01 &   4.72$\pm$ 0.03 & 
--- & --- & --- & --- &  12.65$\pm$  0.05 \\ 
F563-V2 & $B$ &  22.16$\pm$ 0.01 &   7.80$^c$ & 
--- & --- & --- & --- &  16.25$\pm$  0.09 \\ 
        & $V$ &  21.64$\pm$ 0.01 &   7.15$\pm$ 0.05 & 
--- & --- & --- & --- &  15.81$\pm$  0.08 \\ 
        & $R$ &  21.24$\pm$ 0.01 &   7.03$\pm$ 0.02 & 
--- & --- & --- & --- &  15.45$\pm$  0.07 \\ 
        & $K'$ &  19.29$\pm$ 0.04 &   6.4$\pm$ 0.5 & 
--- & --- & --- & --- &  13.8$\pm$  0.1$^a$ \\ 
F568-3 & $U$ &  22.5$\pm$ 0.1 &  10$\pm$ 1 &
--- & --- & --- & --- & 16.0$\pm$  0.1 \\
       & $B$ &  22.33$\pm$ 0.03 &   8.9$\pm$ 0.2 &
--- & --- & --- & --- & 16.12$\pm$  0.07 \\
       & $V$ &  21.64$\pm$ 0.01 &   8.3$\pm$ 0.1 &
--- & --- & --- & --- & 15.55$\pm$  0.08 \\
       & $R$ &  21.22$\pm$ 0.01 &   8.0$\pm$ 0.1 &
--- & --- & --- & --- & 15.22$\pm$  0.08 \\
       & $I$ &  20.71$\pm$ 0.01 &   7.8$\pm$ 0.1 &
--- & --- & --- & --- & 14.83$\pm$  0.10 \\
       & $K'$ &  19.16$\pm$ 0.02 &   8.0$\pm$ 0.3 &
--- & --- & --- & --- & 13.11$\pm$  0.08 \\
\hline
\end{tabular}
\end{minipage}
\end{table*}

\begin{table*}
\begin{minipage}{170mm}
\addtocounter{table}{-1}
\caption{Continued.}
\begin{tabular}{lcccccccc}
\hline
 & & \multicolumn{2}{c}{Disc Parameters} &
\multicolumn{4}{c}{Bulge Parameters} \\
Galaxy & Passband & $\mu_0$ & $h$ &
$\mu_e$ & $r_e$ & B/D & Type & $m_T$ \\
\hline
1034+0220 & $B$ &  23.34$\pm$ 0.08 &   10.1$\pm$ 0.4 &
 25.0$\pm$ 0.1 &   8.8$\pm$ 0.7 &  
 0.62$\pm$ 0.08 & r & 16.11$\pm$  0.05 \\
          & $V$ &  22.42$\pm$ 0.08 &   9.5$\pm$ 0.4 &
 23.08$\pm$ 0.08 &   5.0$\pm$ 0.3 &  
 0.55$\pm$ 0.03 & r & 15.35$\pm$  0.05 \\
          & $R$ &  21.84$\pm$ 0.05 &  7.6$\pm$ 0.5 &
 22.48$\pm$ 0.05 &   5.2$\pm$ 0.1 &  
 0.92$\pm$ 0.02 & r & 15.16$\pm$  0.06 \\
          & $K'$ &  20.2$\pm$ 1 &  8$\pm$ 2 &
 19.9$\pm$ 0.2 &   8$\pm$ 2 &  
 4.6$\pm$ 1 & r & 12.20$\pm$  0.2$^a$ \\
N10-2$^d$ & $U$ &  22.8$\pm$ 0.1 &   8$\pm$ 1 &
--- & --- & --- & --- & 17.0$\pm$  0.2$^b$ \\
  & $B$ &  22.20$\pm$ 0.02 &   6.9$\pm$ 0.2 &
--- & --- & --- & --- & 16.61$\pm$  0.08 \\
  & $V$ &  21.16$\pm$ 0.02 &   6.0$\pm$ 0.1 &
--- & --- & --- & --- & 15.85$\pm$  0.09 \\
  & $R$ &  20.62$\pm$ 0.01 &   5.50$\pm$ 0.03 &
--- & --- & --- & --- & 15.50$\pm$  0.08 \\
  & $I$ &  20.12$\pm$ 0.02 &   5.5$\pm$ 0.2 &
--- & --- & --- & --- & 15.15$\pm$  0.1 \\
  & $K'$ &  17.84$\pm$ 0.01 &   4.95$\pm$ 0.05  &
--- & --- & --- & --- & 13.00$\pm$  0.1 \\
1226+0105 & $B$ &  23.7$\pm$ 0.3 &   10.5$^c$ &
 23.8$\pm$ 0.7 &   4.6$\pm$ 2 &  
 0.6$\pm$ 0.3 & r & 16.30$\pm$  0.06 \\
          & $V$ &  22.60$\pm$ 0.06 &   9.8$\pm$ 0.1 &
 22.16$\pm$ 0.07 &   2.9$\pm$ 0.1 &  
 0.48$\pm$ 0.02 & r & 15.50$\pm$  0.09 \\
          & $R$ &  22.4$\pm$ 0.2 & 9.8$\pm$ 0.3 &
 21.9$\pm$ 0.1 &   3.8$\pm$ 0.2 &  
 0.9$\pm$ 0.2 & r & 15.02$\pm$  0.08 \\
          & $K'$ &  19.9$\pm$ 0.5 & 11$\pm$ 5 &
 18.8$\pm$ 0.3 &   3.5$\pm$ 0.8 &  
 0.9$\pm$ 0.2 & r & 12.13$\pm$  0.10 \\
F574-1 & $B$ &  23.00$\pm$  0.04 &  11.5$\pm$  0.3 &
 23.27$\pm$  0.06 &   2.0$\pm$  0.2 &   
 0.04$\pm$  0.01 & e &
 16.67$\pm$  0.09 \\
       & $V$ &  22.40$\pm$  0.03 &  10.9$\pm$  0.15 &
 22.64$\pm$  0.02 &   2.45$\pm$  0.05 &   
 0.08$\pm$  0.01 & e &
 16.18$\pm$  0.09 \\
       & $R$ &  21.95$\pm$  0.06 &  10.3$\pm$  0.4 &
 22.56$\pm$  0.03 &   2.7$\pm$  0.2 &   
 0.07$\pm$  0.01 & e &
 15.9$\pm$  0.1 \\
       & $K'$ &  19.66$\pm$  0.03 &   8.3$\pm$ 0.5 &
 19.32$\pm$  0.01 &   2.19$\pm$  0.03 &   
 0.18$\pm$  0.01 & e &
 13.8$\pm$  0.15 $^a$ \\
F579-V1 & $B$ &  23.1$\pm$ 0.1 &  10.3$\pm$ 0.4 & 
 24.2$\pm$ 0.1 &   1.5$\pm$ 0.1 &  
 0.01$\pm$ 0.01 & e &  16.29$\pm$  0.10 \\ 
        & $V$ &  22.4$\pm$ 0.1 &  10.2$\pm$ 0.4 & 
 23.5$\pm$ 0.4 &   1.8$\pm$ 0.4 &  
 0.02$\pm$ 0.01 & e &  15.59$\pm$  0.09 \\ 
        & $R$ &  21.9$\pm$ 0.1 &   9.8$\pm$ 0.4 & 
 23.1$\pm$ 0.4 &   1.6$\pm$ 0.4 &  
 0.02$\pm$ 0.01 & e &  15.16$\pm$  0.08 \\ 
        & $K'$ &  19.51$\pm$ 0.06 &   9.5$\pm$ 0.6 & 
 20.75$\pm$ 0.06 &   2.3$\pm$ 0.3 &  
 0.03$\pm$ 0.01 & e &  12.68$\pm$  0.09 \\ 
I1-2 & $B$ &  22.9$\pm$ 0.15 &   9.3$\pm$ 0.8 &
 24.09$^c$ &   1.51$^c$ &   0.02$^c$ & e &
 16.5$\pm$  0.1 \\
  & $V$ &  21.93$\pm$ 0.01 &   7.36$\pm$ 0.02 &
 23.06$\pm$ 0.01 &   1.09$^c$ &   0.01$\pm$ 0.01  & e &
 15.75$\pm$  0.1 \\
  & $R$ &  21.48$\pm$ 0.07 &   7.7$\pm$ 0.4 &
 22.1$\pm$ 0.1 &   1.2$\pm$ 0.2 &   0.02$\pm$ 0.01 & e &
 15.45$\pm$  0.08 \\
  & $I$ &  20.78$\pm$ 0.07 &   7.3$\pm$ 0.6 &
 21.6$\pm$ 0.1 &   1.09$^c$ &   0.02$\pm$ 0.01 & e &
 14.7$\pm$  0.1 \\
  & $K'$ &  18.76$\pm$ 0.03 &   6.1$\pm$ 0.2 &
 19.50$\pm$ 0.07 &   1.09$^c$ &   0.03$\pm$ 0.01 & e &
 13.1$\pm$  0.1 \\
F583-1 & $B$ &  23.05$\pm$ 0.04 &  12.40$^c$ &
--- & --- & --- & --- & 16.40$\pm$  0.08 \\
       & $V$ &  22.49$\pm$ 0.09 &  11.4$\pm$ 0.8 &
--- & --- & --- & --- & 16.07$\pm$  0.08 \\
       & $R$ &  22.23$\pm$ 0.03 &  11.40$^c$ &
--- & --- & --- & --- & 15.68$\pm$  0.07 \\
       & $I$ &  21.90$\pm$ 0.08 &  11.00$^c$ &
--- & --- & --- & --- & 15.40$\pm$  0.09 \\
       & $K'$ &  19.8$\pm$ 0.2 &   8$\pm$ 2 &
--- & --- & --- & --- & 13.5$\pm$  0.2$^b$ \\
ESO-LV 1040220 & $B$ &  23.84$\pm$ 0.05 &  40$\pm$ 4 & 
--- & --- & --- & --- &  14.4$\pm$  0.1$^a$ \\ 
  & $R$ &  22.9$\pm$ 0.1 &  37$\pm$ 5 & 
--- & --- & --- & --- &  13.6$\pm$  0.1$^a$ \\ 
  & $K$ &  21.35$\pm$ 0.01 &  39$\pm$ 4 & 
--- & --- & --- & --- &  11.7$\pm$  0.15$^a$ \\ 
ESO-LV 1040440 & $B$ &  23.4$\pm$ 0.1 &  26$\pm$ 2 & 
--- & --- & --- & --- &  14.72$\pm$  0.07 \\ 
  & $R$ &  22.47$\pm$ 0.02 &  25.5$\pm$ 0.7 & 
--- & --- & --- & --- &  13.64$\pm$  0.08 \\ 
  & $K$ &  20.9$\pm$ 0.15 &  23$\pm$ 4 & 
--- & --- & --- & --- &  12.0$\pm$  0.2$^b$ \\ 
ESO-LV 1870510 & $B$ &  23.09$\pm$ 0.05 &  26$\pm$ 2 & 
--- & --- & --- & --- &  14.87$\pm$  0.07 \\ 
  & $R$ &  22.12$\pm$ 0.08 &  26$\pm$ 2 & 
--- & --- & --- & --- &  13.99$\pm$  0.09 \\ 
  & $K$ &  20.55$\pm$ 0.05 &  23$\pm$ 2 & 
--- & --- & --- & --- &  12.6$\pm$  0.2$^b$ \\ 
ESO-LV 1450250 & $B$ &  23.08$\pm$ 0.03 &  32$\pm$ 1 & 
--- & --- & --- & --- &  14.11$\pm$  0.05 \\ 
  & $R$ &  22.05$\pm$ 0.03 &  30$\pm$ 1 & 
--- & --- & --- & --- &  13.31$\pm$  0.05 \\ 
  & $K$ &  20.42$\pm$ 0.03 &  25$\pm$ 1 & 
--- & --- & --- & --- &  11.95$\pm$  0.1$^a$ \\ 
P1-7$^g$ & $B$ &  23.15$\pm$ 0.01 &  15.9$\pm$ 0.1 & 
 24.98$\pm$ 0.03 &   3.2$\pm$ 0.1 &   0.01$\pm$ 0.01 & e & 
 15.36$\pm$  0.05 \\ 
  & $V$ &  22.35$\pm$ 0.02 &  14.9$\pm$ 0.4 & 
 23.86$\pm$ 0.05 &   3.7$\pm$ 0.3 &   0.03$\pm$ 0.01 & e & 
 14.68$\pm$  0.09 \\ 
  & $I$ &  21.22$\pm$ 0.02 &  13.9$\pm$ 0.2 & 
 22.68$\pm$ 0.02 &   4.0$\pm$ 0.2 &   0.04$\pm$ 0.01 & e & 
 13.70$\pm$  0.05 \\ 
  & $K$ &  19.4$\pm$ 0.2 &  12$\pm$ 2 & 
 20.7$\pm$ 0.1 &   4$\pm$ 2 &   0.07$\pm$ 0.03 & e & 
 12.15$\pm$  0.1$^a$ \\ 
2327-0244 & $B$ &  22.44$\pm$ 0.03 &  18.0$^c$ &
 21.08$\pm$ 0.01 &   3.17$\pm$ 0.02 &  
 0.21$\pm$ 0.01 & e & 14.50$\pm$  0.10 \\
          & $V$ &  21.29$\pm$ 0.02 &  16.5$\pm$ 0.4 &
 20.29$\pm$ 0.01 &  3.12$\pm$ 0.01 &  
 0.17$\pm$ 0.01 & e & 13.62$\pm$ 0.10 \\
          & $R$ &  20.58$\pm$ 0.01 &  16.5$^c$ &
 19.62$\pm$ 0.01 &   2.90$\pm$ 0.01 &  
 0.14$\pm$ 0.01 & e & 13.00$\pm$  0.06 \\
          & $K'$ & 17.45$\pm$ 0.03 & 12.6$\pm$ 0.4 &
 16.08$\pm$ 0.01 &   2.23$\pm$ 0.02 &  
 0.21$\pm$ 0.01 & e & 10.32$\pm$  0.07 \\
\hline
\end{tabular}
\\ $\mu_0$ is the bulge/disc decomposition extrapolated central surface 
brightness in mag\,arcsec$^{-2}$, and $h$ is the disc scale length in
arcsec.  $\mu_e$ is the surface brightness at the half-light radius of
the exponential (Type $ = e$) or r$^{1/4}$ law (Type $ = r$) bulge profile 
in mag\,arcsec$^{-2}$, and $r_e$ is the half-light radius in arcsec.  
B/D is the ratio of bulge to disc luminosities, obtained by dividing the total 
luminosities of the fitted bulge by the fitted disc. \\
$^a$ Extrapolation $>0.15$ mag \hspace{0.8cm}
$^b$ Extrapolation $>0.3$ mag \hspace{0.8cm}
$^c$ Held constant \hspace{0.8cm}
$^d$ Disc only fit is chosen for simplicity \\
$^e$ Disc is a very weak component \hspace{0.8cm}
$^f$ The ellipticity and position angle have been constrained \\
$^g$ Fit is carried out using $I$ band images \\
$^h$ The galaxy most likely continues out beyond $\sim$ 50$''$.  We state
	the extrapolated magnitude at 50$''$, along with its uncertainty. \\
$^j$ The choice of bulge profile was essentially arbitrary.\\
\end{minipage}
\end{table*}

\begin{table}
  \caption{Estimated sample ellipticities, position angles and largest radius
	at which they can be measured}
  \label{tab:epa}
  \begin{tabular}{lccc}
  \hline
   Galaxy & $e$ & PA (deg) & $r_{max}$ (arcsec)  \\ 
  \hline
	UGC 128 & 0.43$\pm$ 0.07 & 62$\pm$ 5 & 54 \\
	ESO-LV 280140 & 0.28$\pm$ 0.08 & 125$\pm$ 10 & 160 \\
	UGC 334 & 0.4$\pm$ 0.1 & 45$\pm$ 10 & --- \\
	0052-0119 & 0.25$\pm$ 0.04 & 75$\pm$ 5 & 60 \\
	UGC 628 & 0.45$\pm$ 0.05 & 135$\pm$ 5 & 66 \\ 
	0221+0001 & 0.18$\pm$ 0.05 & 160$^a$ & 29 \\
	0237-0159 & 0.10$\pm$ 0.05 & 115$\pm$ 10 & --- \\
	ESO-LV 2490360 & 0.31$\pm$ 0.08 & 138$\pm$ 7 & 90 \\
	F561-1 & 0.10$\pm$ 0.05  & 60$\pm$ 10 & --- \\
	C1-4 & 0.55$\pm$ 0.05 & 52$\pm$ 2  & 41 \\
	C3-2 & 0.32$\pm$ 0.03 & 29$\pm$ 3 & 30 \\
	F563-V2 & 0.28$\pm$ 0.08 & 150$\pm$ 10 & 37 \\
	F568-3 & 0.22$\pm$ 0.05 & 165$\pm$ 5 & 32 \\
	1034+0220 & 0.25$\pm$ 0.04 & 45$\pm$ 5 & 23 \\
	N10-2 & 0.45$\pm$ 0.05 & 2$\pm$ 3 & 30 \\
	1226+0105 & 0.20$\pm$ 0.05 & 75$\pm$ 25 & 20 \\
	F574-1 & 0.58$\pm$ 0.03 & 85$\pm$ 4 & 46 \\
	F579-V1 & 0.11$\pm$ 0.03 & 123$\pm$ 9 & 33 \\
	I1-2 & 0.18$\pm$ 0.08 & 70$\pm$ 15 & 26 \\
	F583-1 & 0.55$\pm$ 0.06 & 176$\pm$ 4 & 49 \\
	ESO-LV 1040220 & 0.43$\pm$ 0.06 & 99$\pm$ 6 & 70 \\
	ESO-LV 1040440 & 0.14$\pm$ 0.07 & 137$\pm$ 12 & 160 \\
	ESO-LV 1870510 & 0.50$\pm$ 0.08 & 12$\pm$ 5 & 120 \\
	ESO-LV 1450250 & 0.4$\pm$ 0.1 & 150$\pm$ 8 & 120 \\
	P1-7 & 0.31$\pm$ 0.07 & 8$\pm$ 4 & 21 \\
	2327-0244$^b$ & 0.5$\pm$ 0.1 & 115$\pm$ 3 & 66 \\
  \hline  
  \end{tabular}
\hspace{1cm}\\
$^a$ Relatively unconstrained\\
$^b$ These parameters may be affected by the strong bar
\end{table}

\begin{figure*}
\begin{minipage}{11.0cm}
\psfig{figure=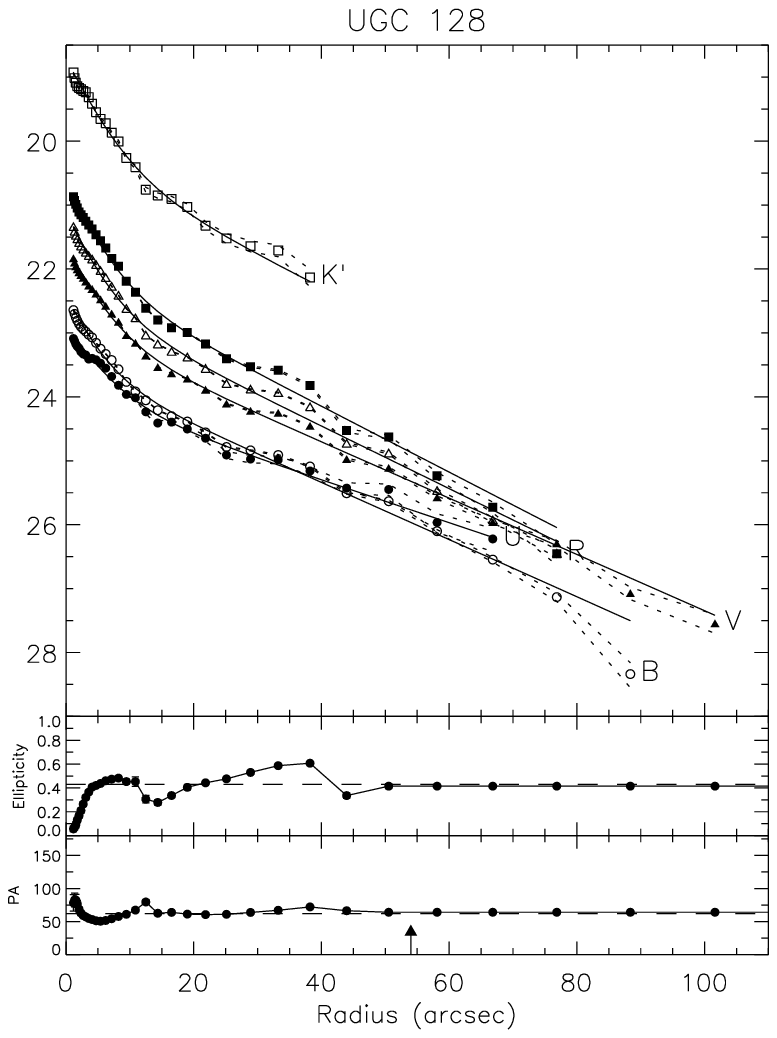}
\end{minipage} 
\begin{minipage}{11.0cm}
\psfig{figure=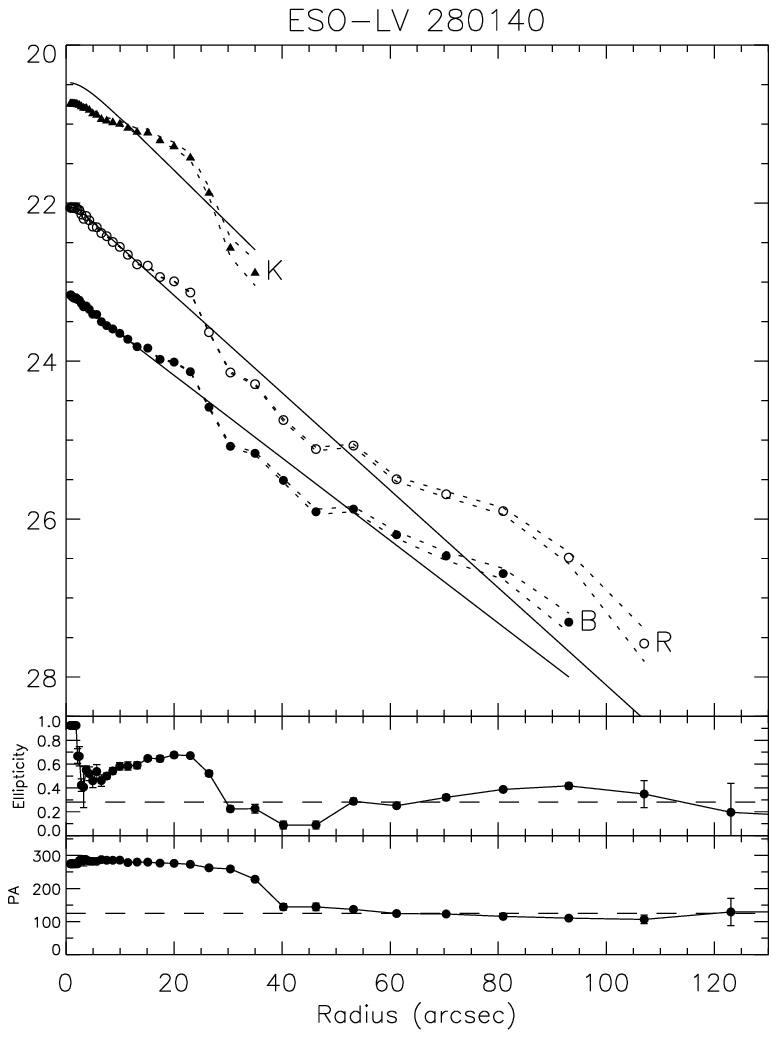}
\end{minipage} 
\begin{minipage}{11.0cm}
\psfig{figure=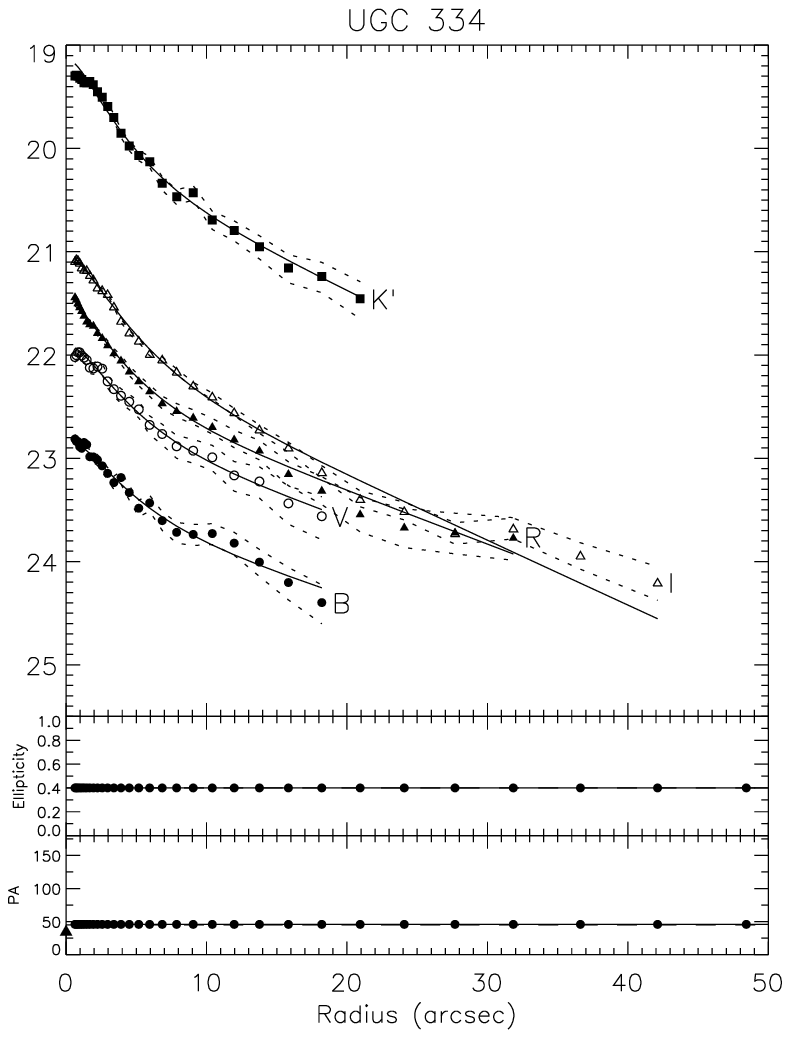}
\end{minipage} 
\begin{minipage}{11.0cm}
\psfig{figure=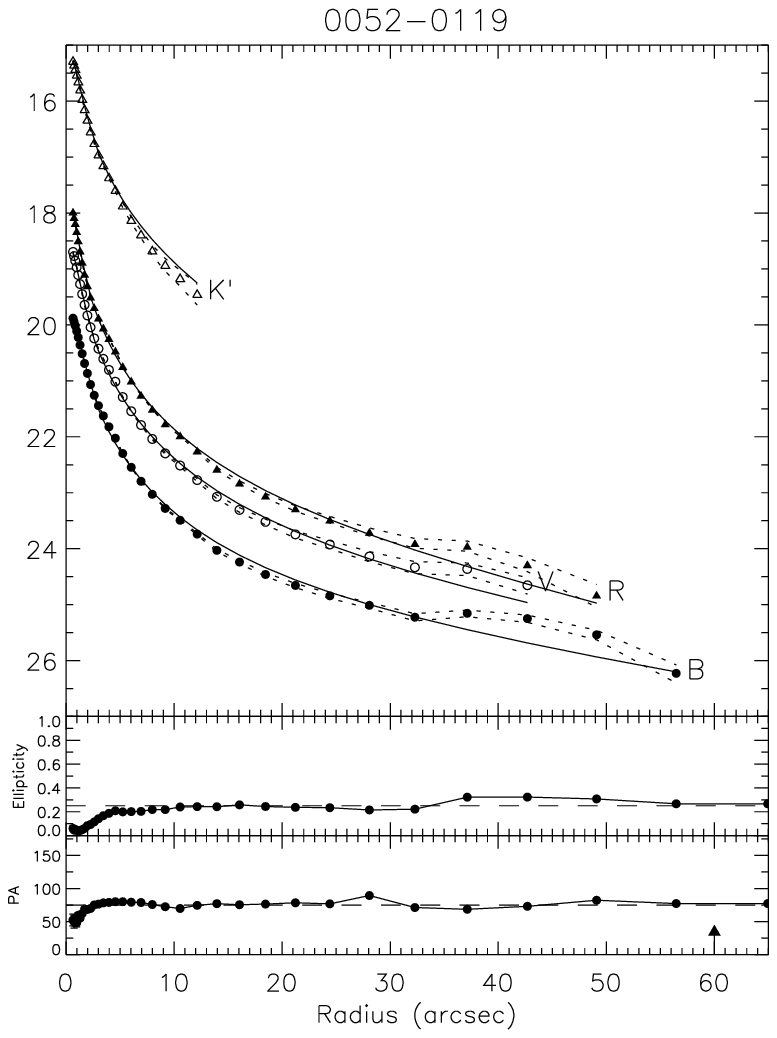}
\end{minipage} 
\caption{Surface photometry in all available passbands, using the 
ellipse parameters determined from the $R$ band image.  The surface
photometry (symbols) is plotted out to the radius where the sky subtraction
error amounts to $\pm$0.2 mag or more.  
Also plotted are the bulge-disc fits to the surface photometry
(solid lines) and the effects of the $\pm$1$\sigma$ 
estimated error in the sky level (dotted lines).  The fitted $R$ band 
ellipticities and position angles are also plotted (with errors,
which are sometimes smaller than the symbol size) as a function of 
radius for each galaxy (out to the maximum radius for which 
they were fit; arrow), along with the estimate for the global
ellipticity and position angle (dashed lines).
The position angle and ellipticity for UGC 334 have been fixed.}
\label{fig:surfphot}
\end{figure*}

\begin{figure*}
\begin{minipage}{11.0cm}
\psfig{figure=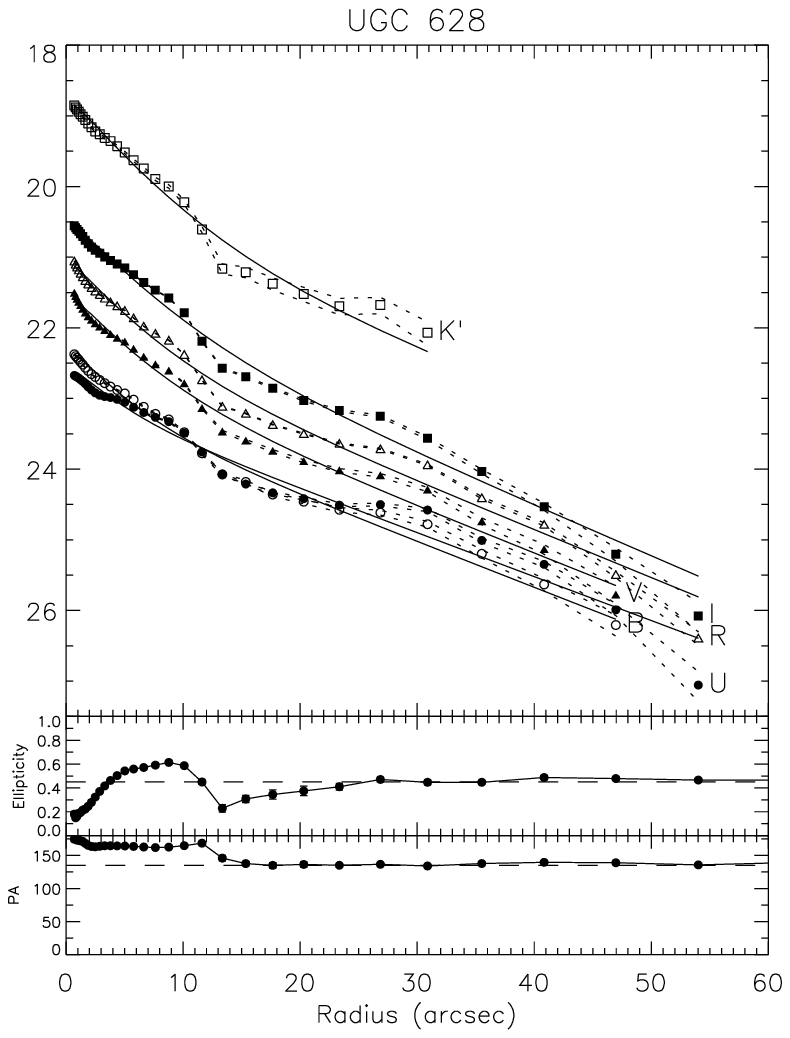}
\end{minipage} 
\begin{minipage}{11.0cm}
\psfig{figure=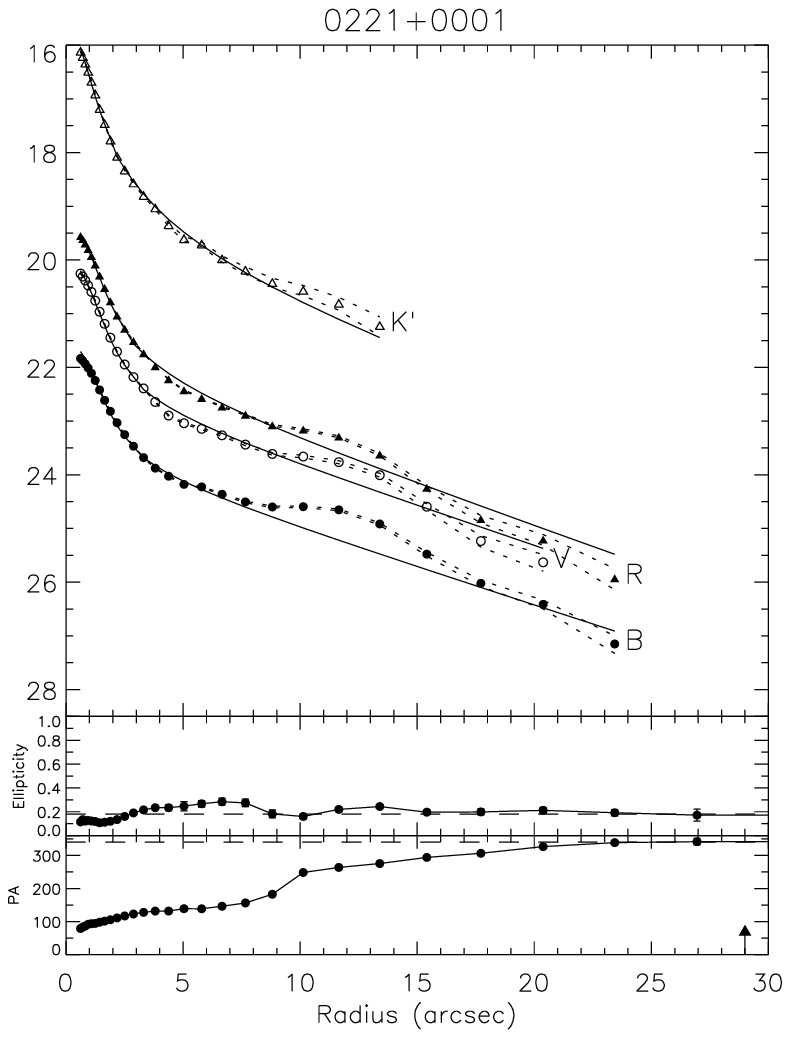}
\end{minipage} 
\begin{minipage}{11.0cm}
\psfig{figure=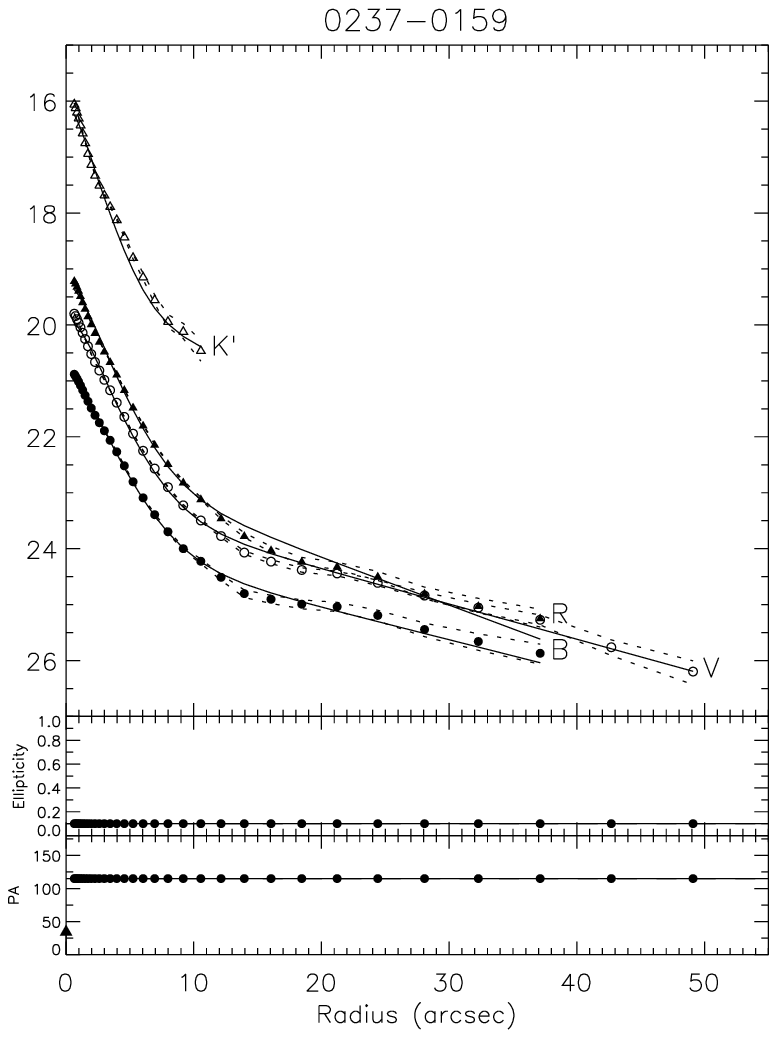}
\end{minipage} 
\begin{minipage}{11.0cm}
\psfig{figure=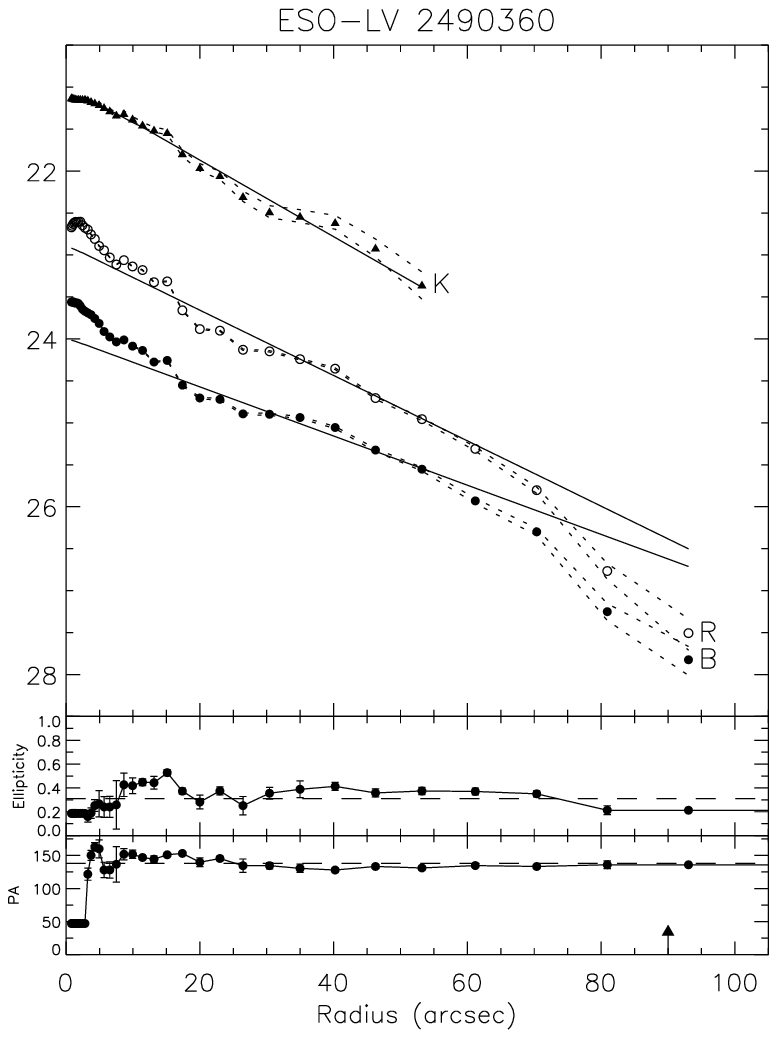}
\end{minipage} 
\addtocounter{figure}{-1}
\caption{Continued. The position angle and ellipticity of 0237-0159 has been
fixed.}
\end{figure*}

\begin{figure*}
\begin{minipage}{11.0cm}
\psfig{figure=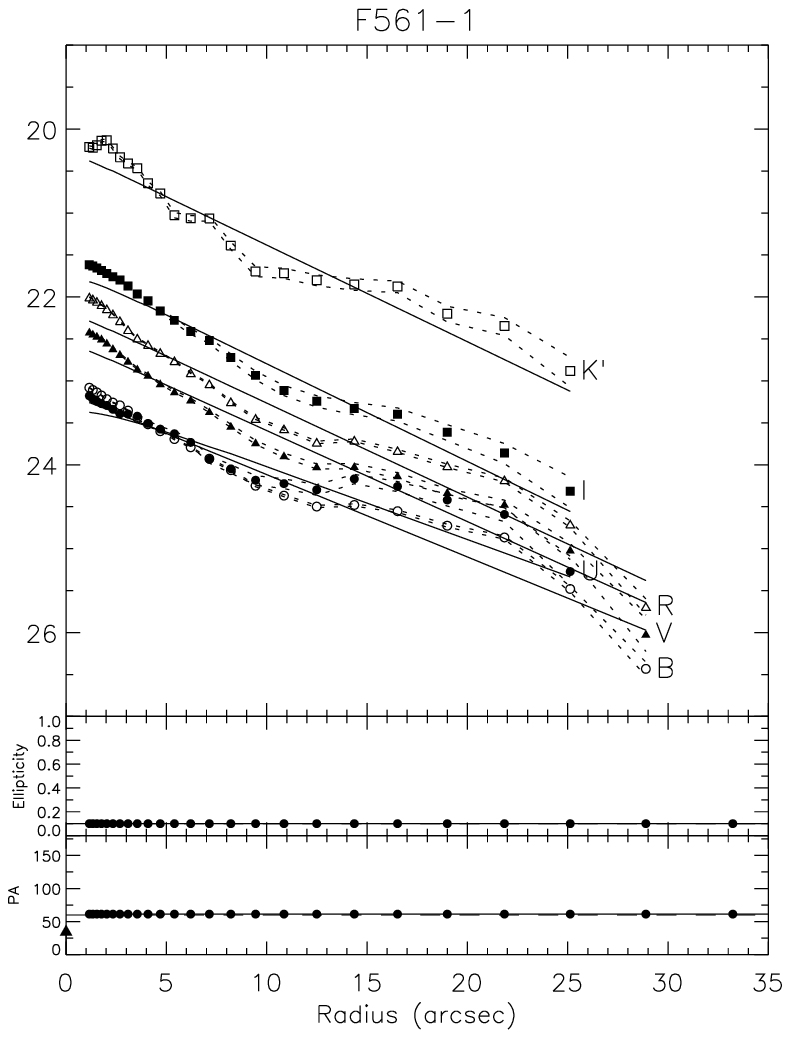}
\end{minipage} 
\begin{minipage}{11.0cm}
\psfig{figure=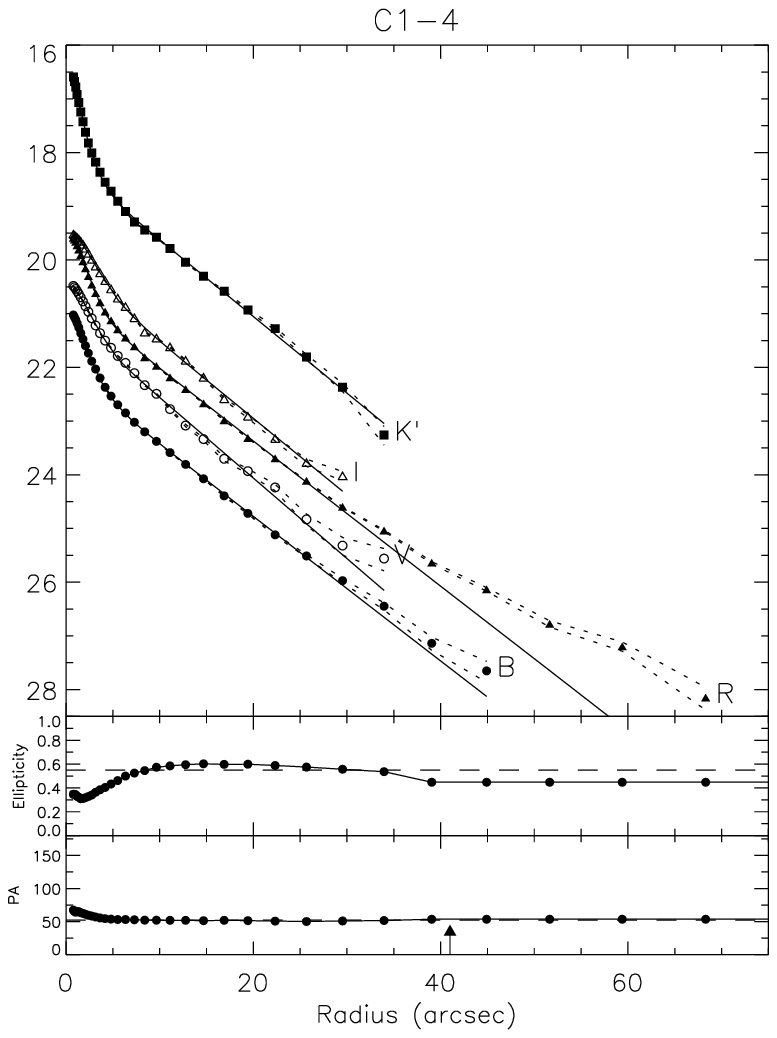}
\end{minipage} 
\begin{minipage}{11.0cm}
\psfig{figure=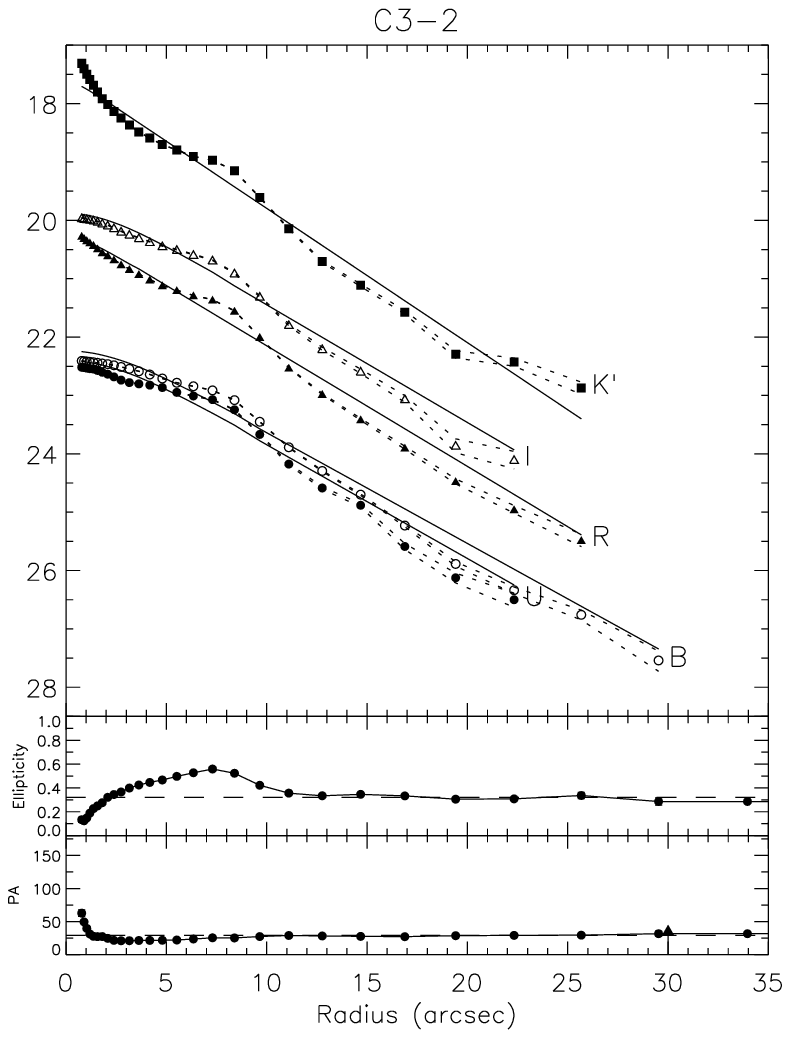}
\end{minipage} 
\begin{minipage}{11.0cm}
\psfig{figure=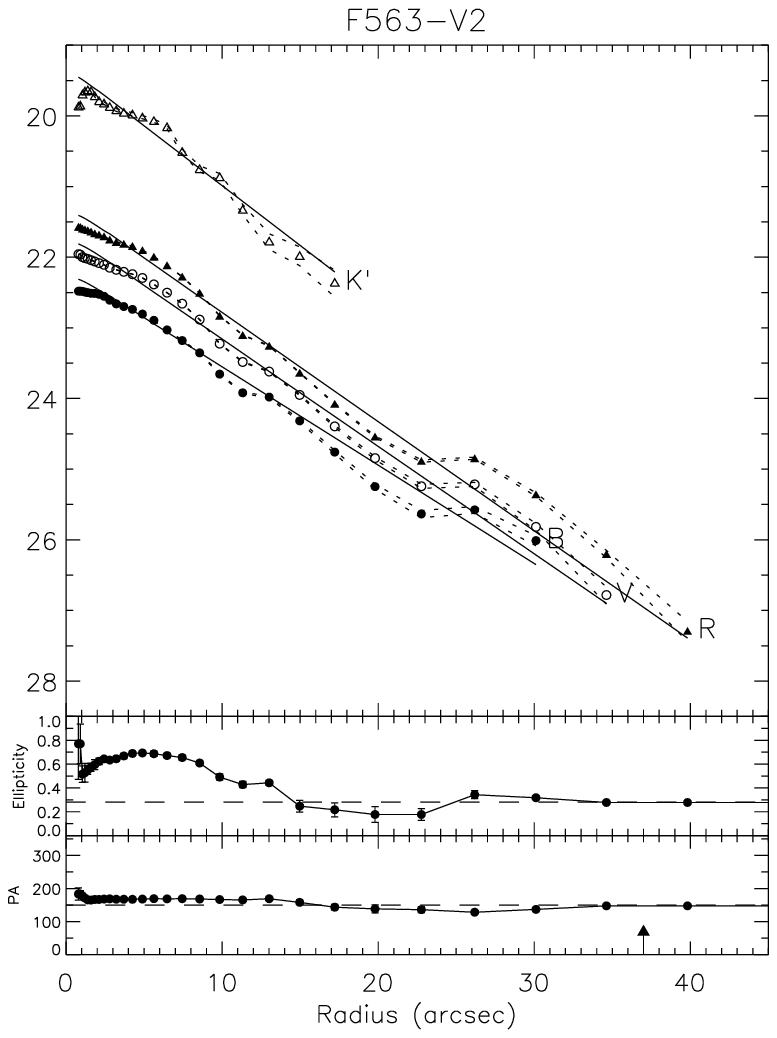}
\end{minipage} 
\addtocounter{figure}{-1}
\caption{Continued. The position angle and ellipticity of F561-1 have
been fixed.  }
\end{figure*}

\begin{figure*}
\begin{minipage}{11.0cm}
\psfig{figure=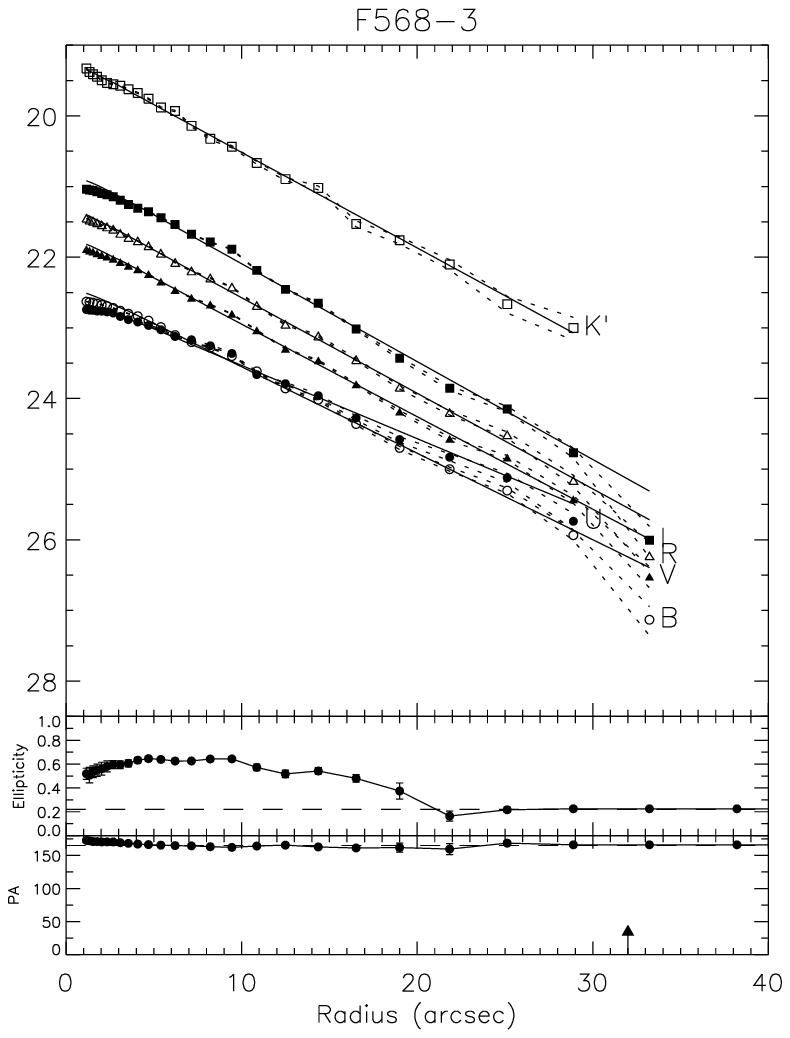}
\end{minipage} 
\begin{minipage}{11.0cm}
\psfig{figure=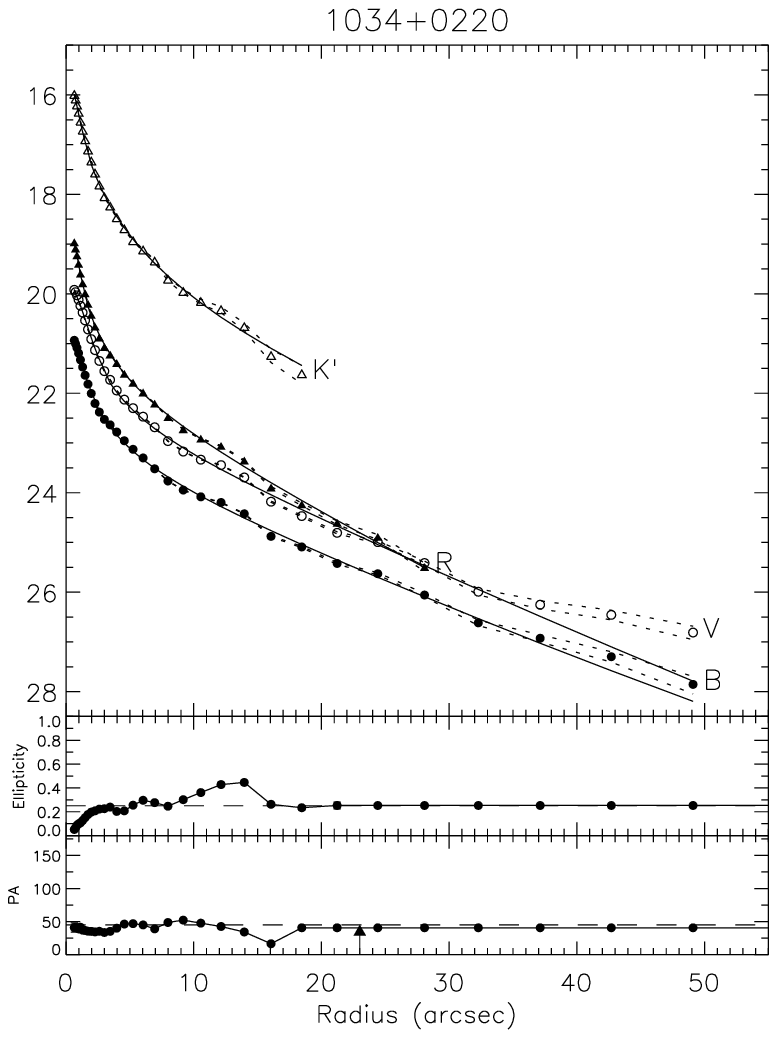}
\end{minipage} 
\begin{minipage}{11.0cm}
\psfig{figure=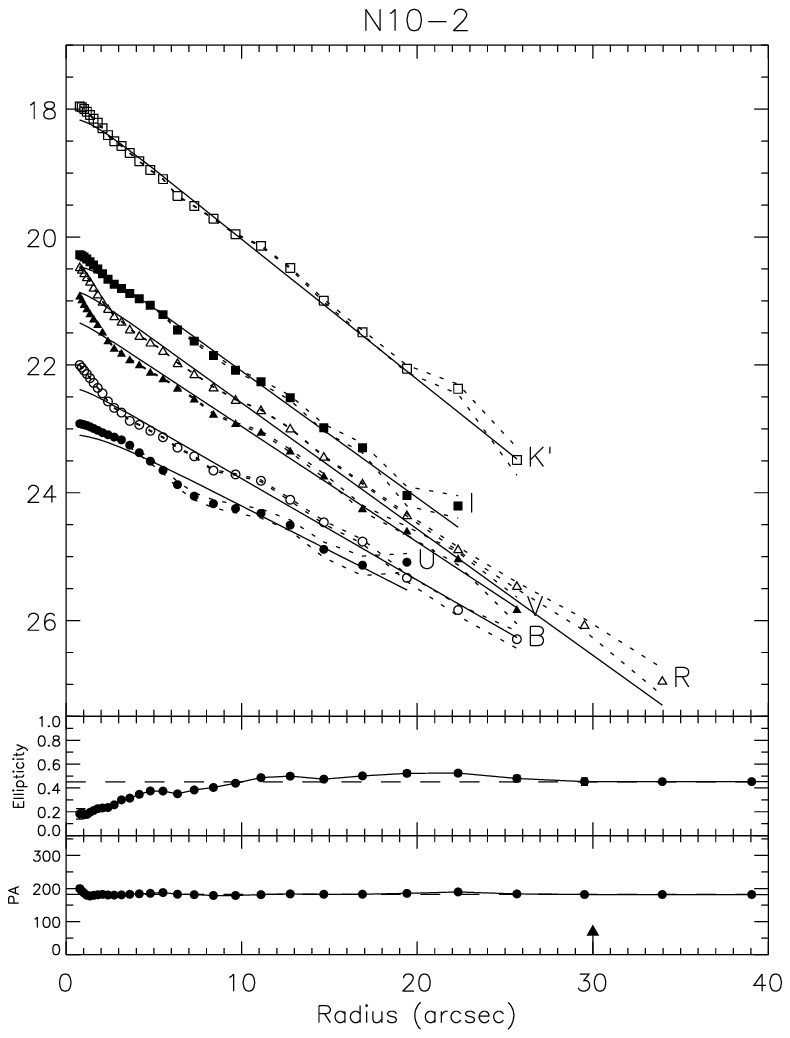}
\end{minipage} 
\begin{minipage}{11.0cm}
\psfig{figure=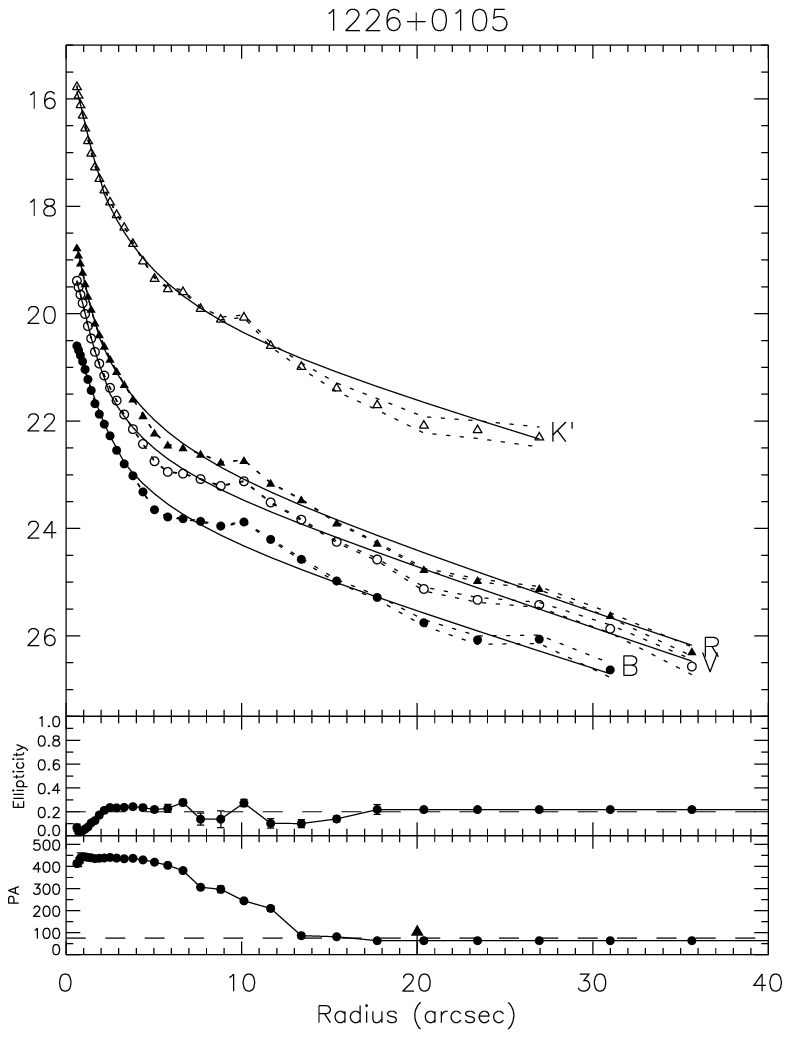}
\end{minipage} 
\addtocounter{figure}{-1}
\caption{Continued.  }
\end{figure*}

\begin{figure*}
\begin{minipage}{11.0cm}
\psfig{figure=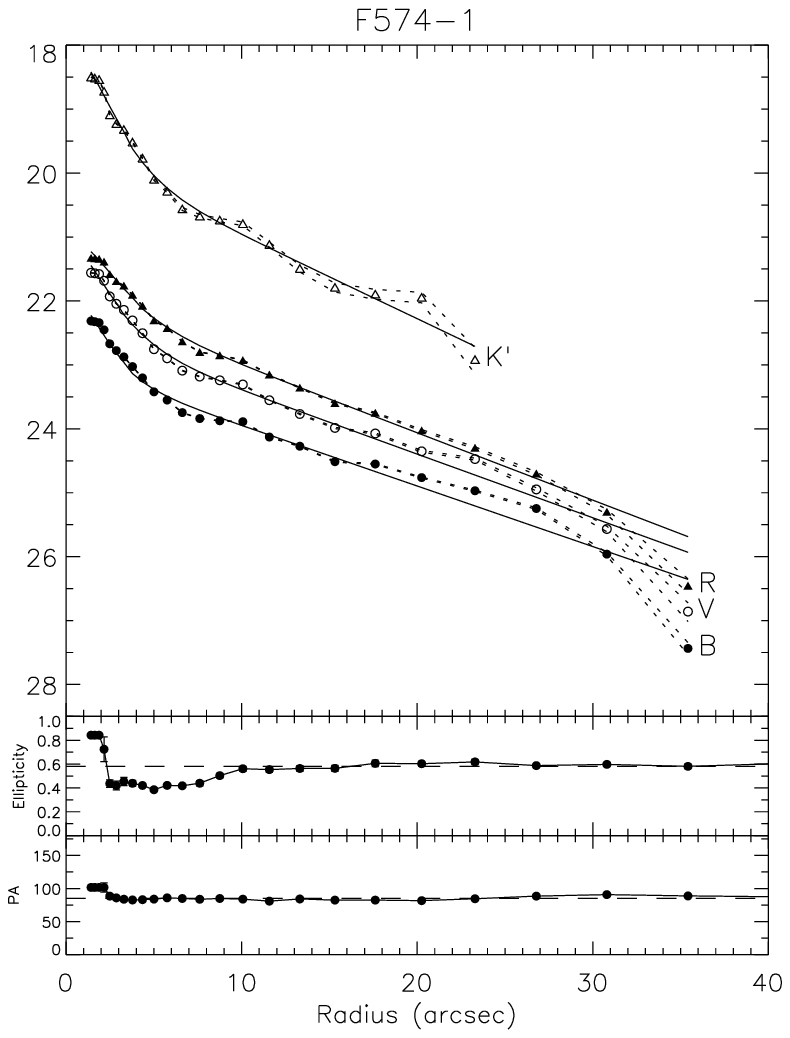}
\end{minipage} 
\begin{minipage}{11.0cm}
\psfig{figure=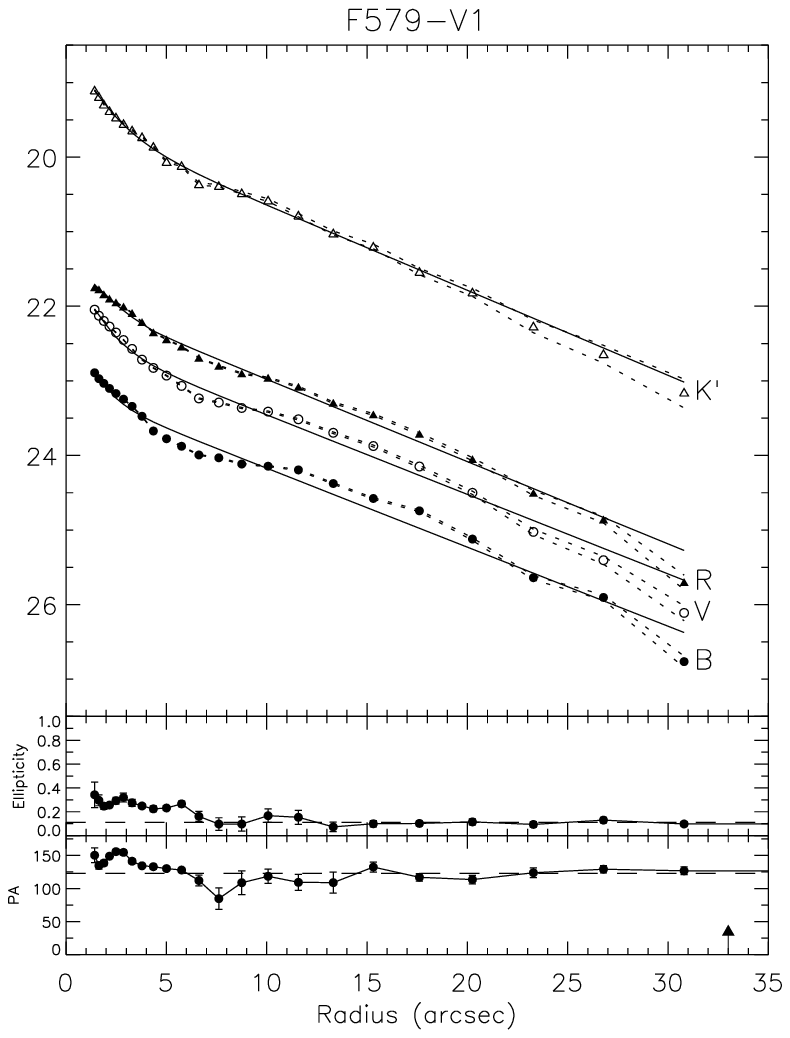}
\end{minipage} 
\begin{minipage}{11.0cm}
\psfig{figure=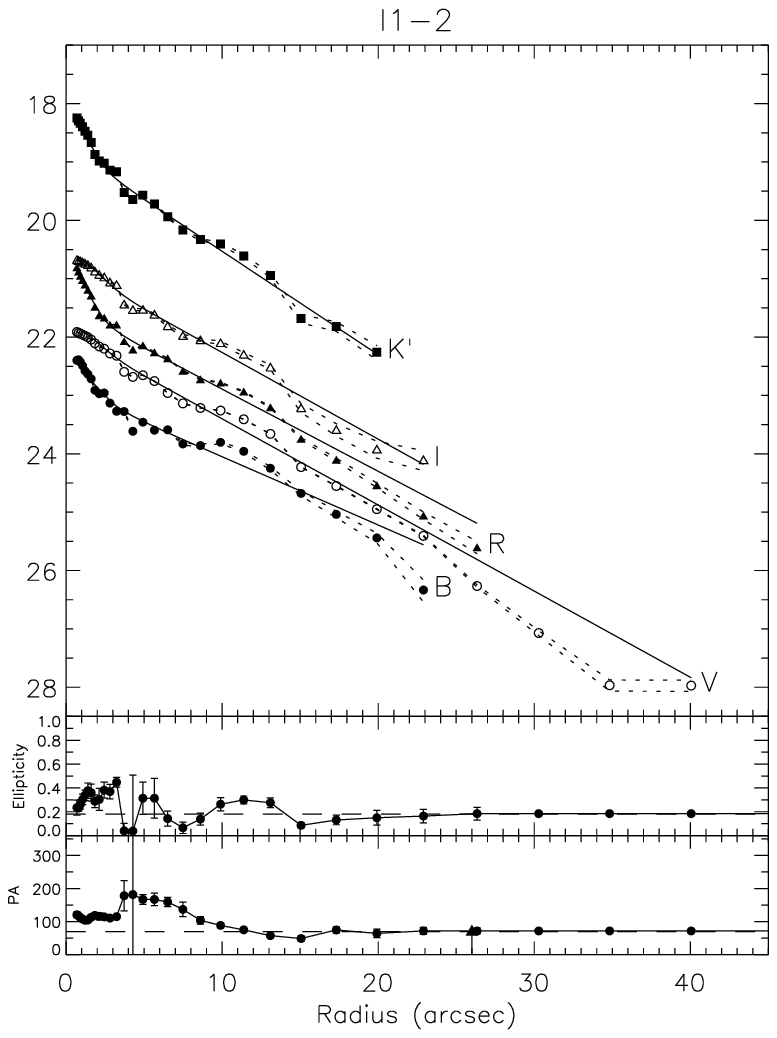}
\end{minipage} 
\begin{minipage}{11.0cm}
\psfig{figure=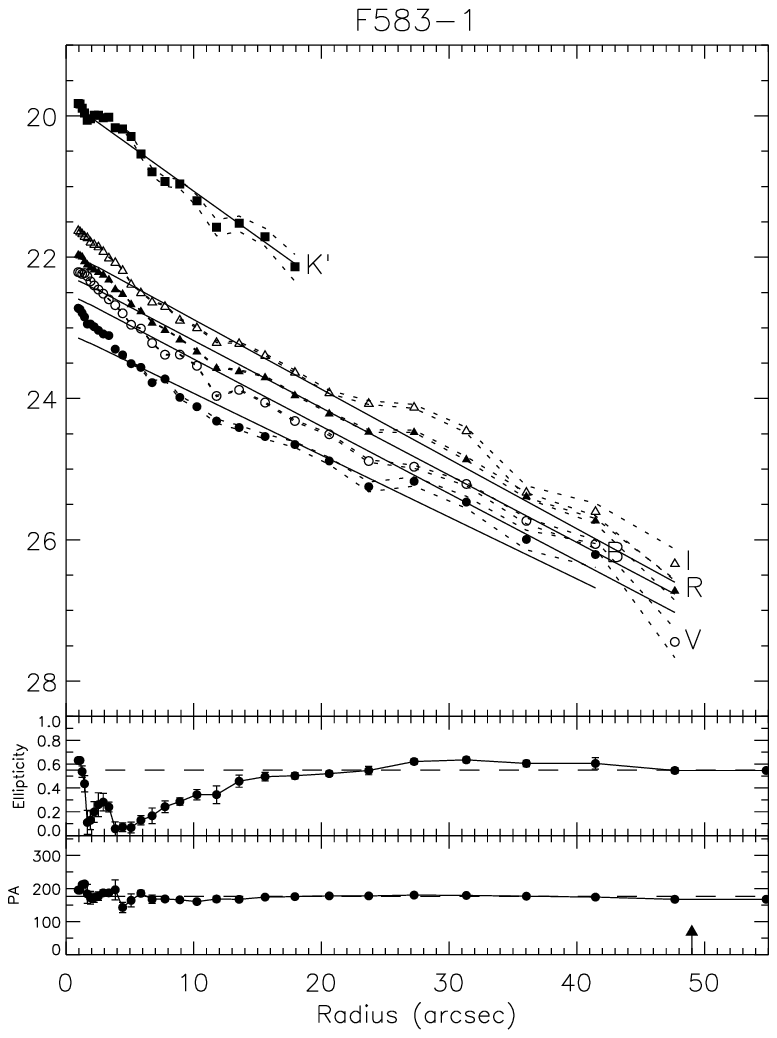}
\end{minipage} 
\addtocounter{figure}{-1}
\caption{Continued. }
\end{figure*}

\begin{figure*}
\begin{minipage}{11.0cm}
\psfig{figure=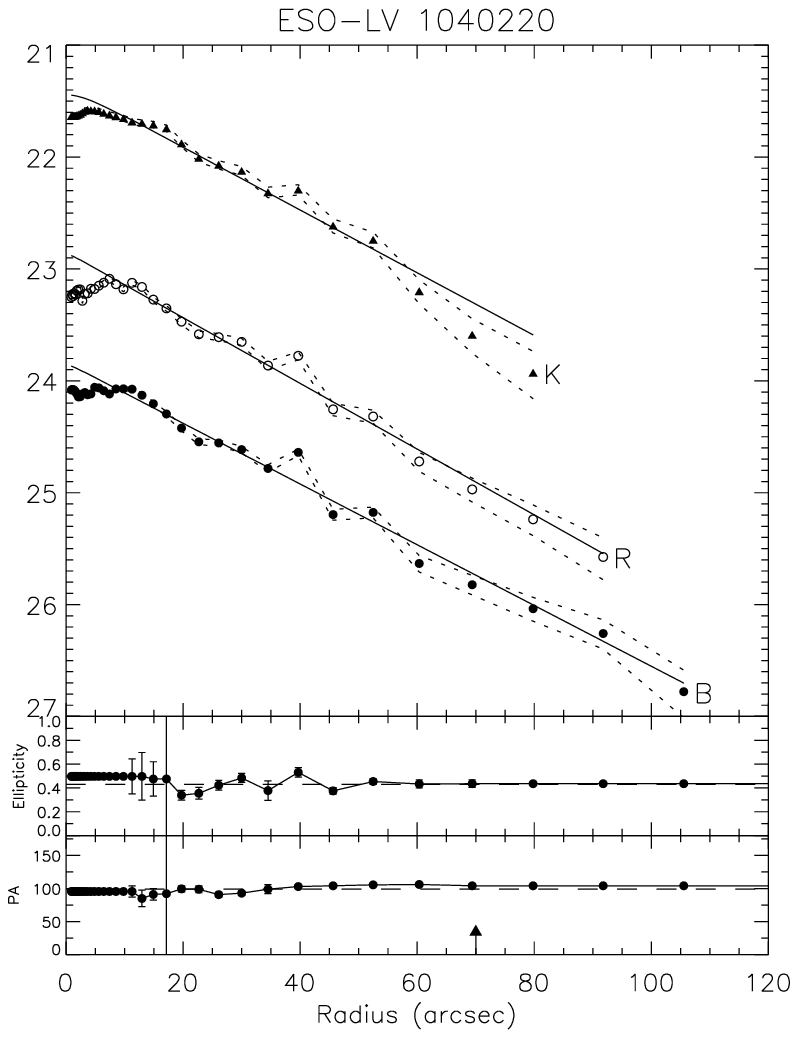}
\end{minipage} 
\begin{minipage}{11.0cm}
\psfig{figure=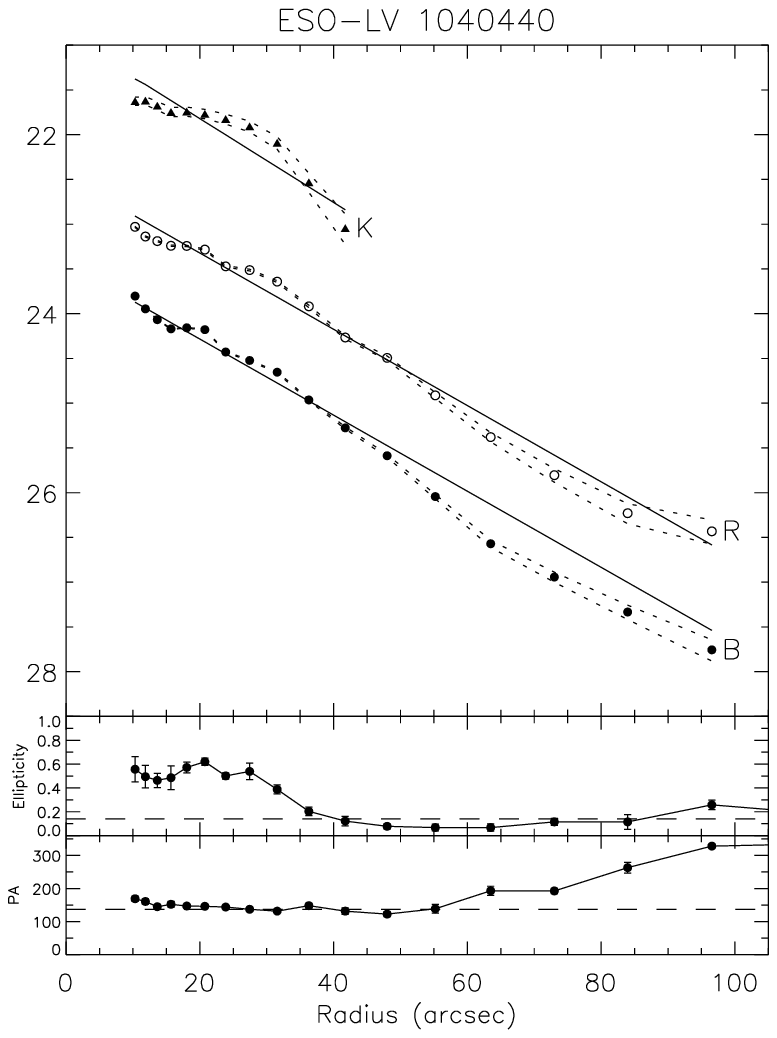}
\end{minipage} 
\begin{minipage}{11.0cm}
\psfig{figure=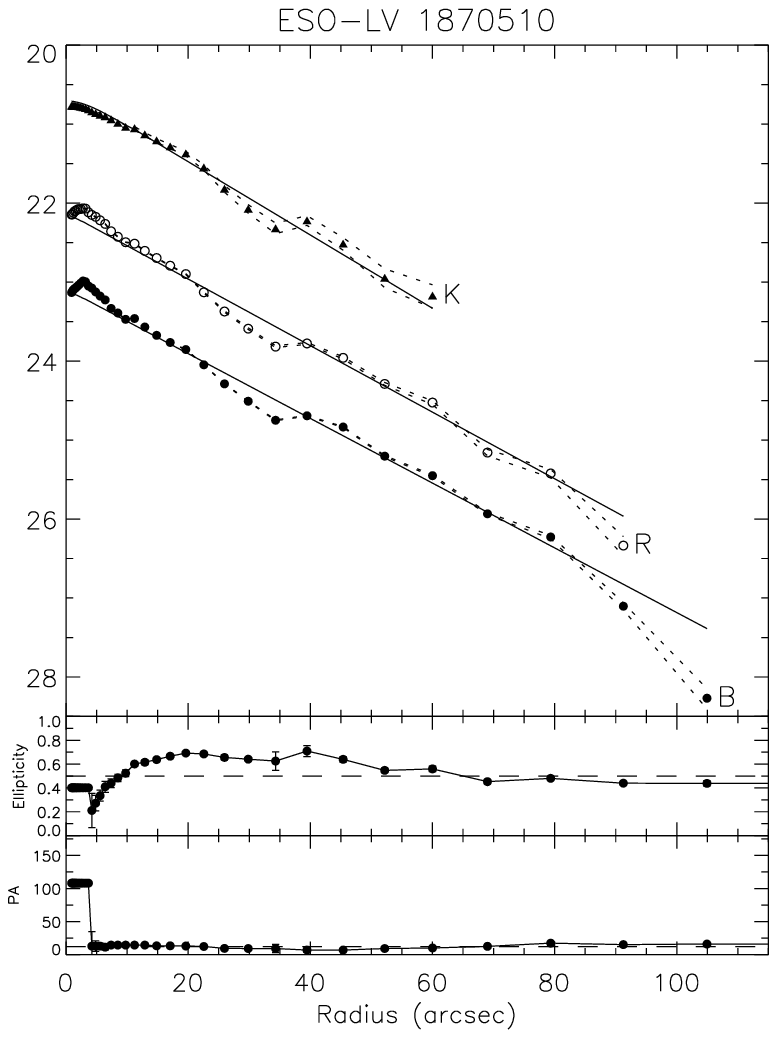}
\end{minipage} 
\begin{minipage}{11.0cm}
\psfig{figure=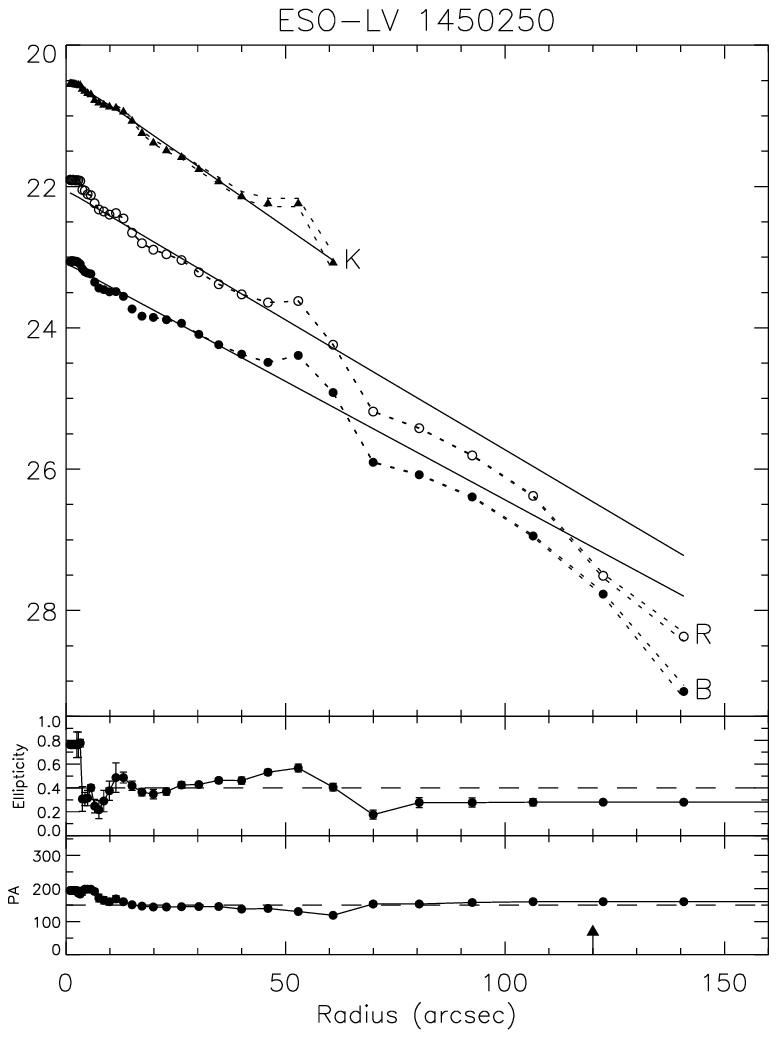}
\end{minipage} 
\addtocounter{figure}{-1}
\caption{Continued. All of the above galaxies have
relatively low surface brightnesses and fairly irregular
structures and therefore have relatively noisy
position angles and/or ellipticities at both small and large radii.}
\end{figure*}

\begin{figure*}
\begin{minipage}{11.0cm}
\psfig{figure=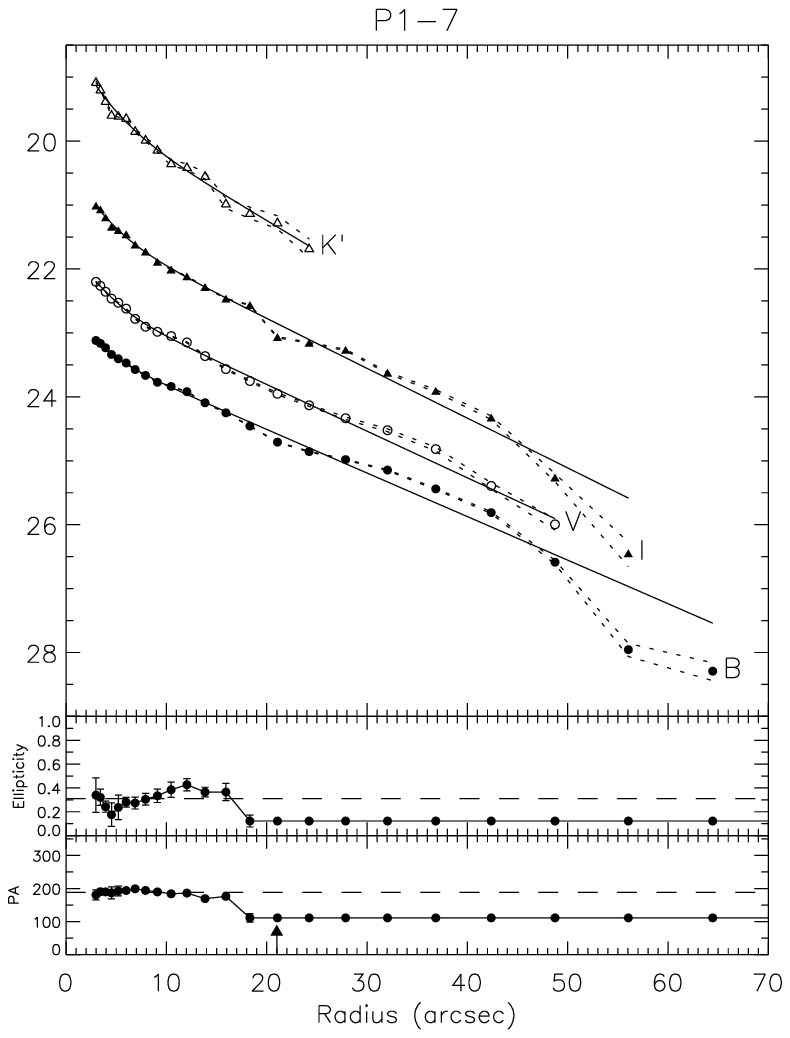}
\end{minipage} 
\begin{minipage}{11.0cm}
\psfig{figure=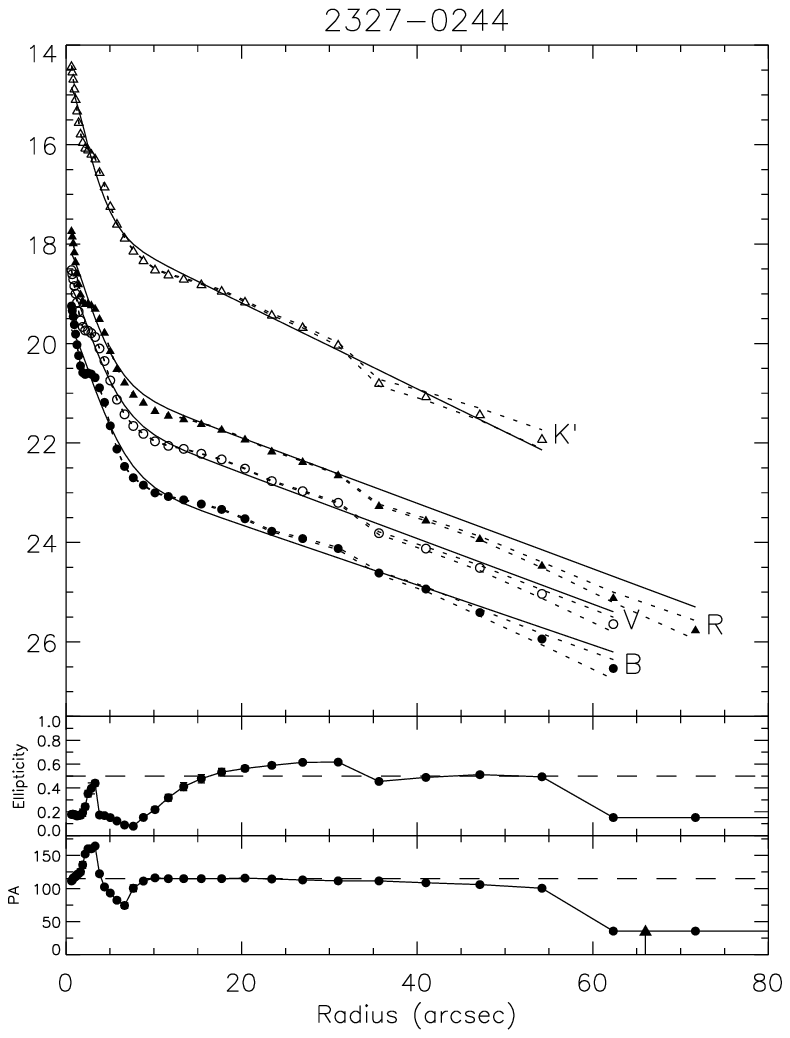}
\end{minipage} 
\addtocounter{figure}{-1}
\caption{Continued. Note that P1-7 has no $R$ band 
data, so $I$ band is used to determine the ellipse parameters.}
\end{figure*}

The surface brightness 
profile $\Sigma(r)$
(in linear flux units)
was fit using one of four possible profiles: 
an exponential disc profile only, 
an exponential bulge plus disc profile, a de Vaucouleurs
$r^{1/4}$ law bulge plus exponential disc profile, and
a de Vaucouleurs $r^{1/4}$ law bulge only.  
The exponential profile has an extrapolated 
central surface brightness $\Sigma_0$ (in linear units), 
and an exponential
disc scale length $h$.  The de Vaucouleurs profile has a 
surface brightness $\Sigma_e$ in linear units at the half-light 
radius $r_e$.  
These exponential or $r^{1/4}$ law model profiles were convolved
with a Gaussian with a FWHM equal to the seeing quality before fitting.
The sky level was not permitted to vary when performing these fits.

Independent fitting for each passband was
carried out using the method of Levenberg and Marquardt 
\cite{numrec}.  The reduced $\chi^2$ statistic was used to determine
the quality of the fit (using the measured RMS in each ellipse
as the error in the flux in that ellipse); 
$\chi^2$ was minimised for each of the four 
types of fit, and the best values of $\chi^2$ for each fitting
function compared.  One fitting function was chosen in preference to 
another if its $\chi^2$ was on average better, over most of 
the passbands.  Most cases 
were clear cut, with reductions in $\chi^2$ by 
a factor of two or more compared
to the other three fitting functions.
However, there were two main types of 
marginal case where the difference between 
minimum $\chi^2$ for two models was small (i.e.\ 
less than $\sim$ 10 per cent), 
and the best model varied between passbands.  
One case is where
there is no evidence for a bulge component in the bluer passbands, 
but tentative evidence for an exponential bulge component in the
red passbands.  In this case, a disc only fit is chosen for simplicity.
The other marginal case arises when a galaxy with a strong bulge is 
fit by an exponential or de Vaucouleurs bulge equally well:  in this 
case, the choice of bulge profile is fairly arbitrary, and the choice is
made to minimise systematic deviations of the fit from the data.  These two
types of case are both flagged in Table \ref{tab:bulgedisc}.

In a number of cases, the disc parameters are relatively
ill-constrained, 
as there is significant structure in the surface brightness profile
in some of the bluer passbands, or in the $K'$ profile.
This structure is typically due to the low signal-to-noise of the
data (especially in $U$ or $K'$ bands), by real structure in 
the surface brightness profile caused by e.g.\ recent star formation 
(affecting the $U$ and $B$ band profiles particularly) or by 
asymmetries in the most irregular LSBGs in our sample.
In these cases, we have chosen to fix some of the fit parameters
to achieve a better convergence.  We usually choose to fix the disc
scale length, as it is possible to use the disc scale lengths in other
passbands, along with the mean variation in scale length with 
passband \cite{dejong1996ii}, to estimate the scale length in this colour.
While clearly not optimal, this 
method allows us to use the same functional form for fitting 
the surface brightness profile
for all the different passbands.  Cases where the fit has been constrained 
have been flagged in Table \ref{tab:bulgedisc}.

\subsection{Comparison with existing data}

\begin{figure*}
\begin{minipage}{17.0cm}
\psfig{figure=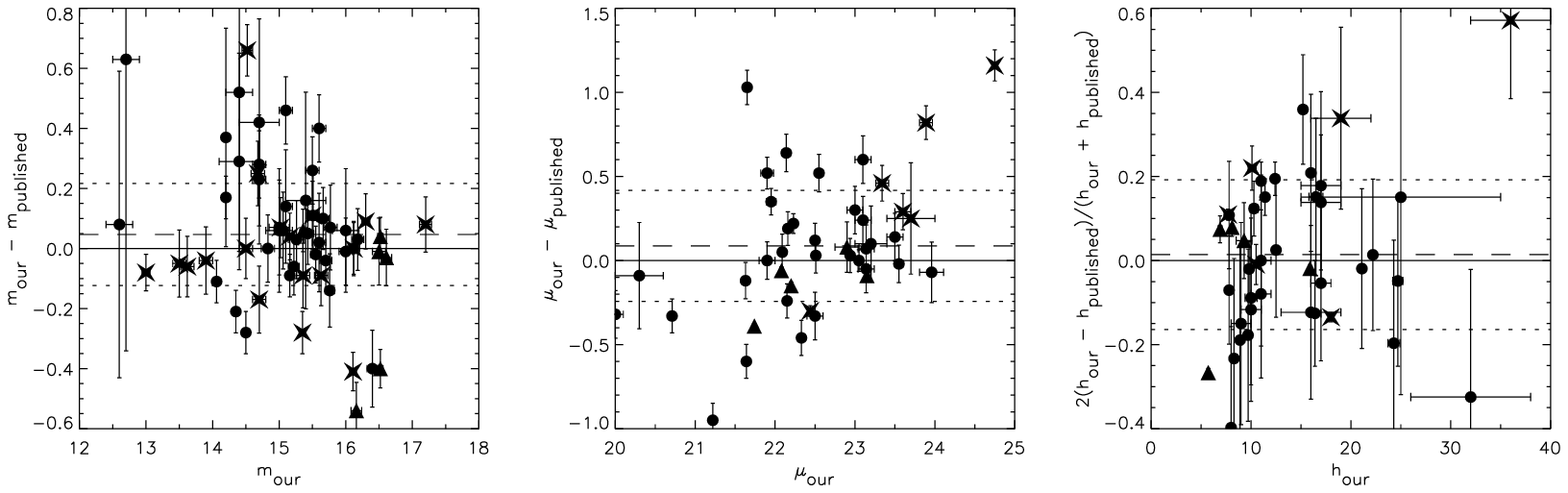}
\end{minipage} 
\caption{Comparison between our measurements of magnitude, central
surface brightness and scale length and those in the literature. 
Solid lines denote the line of equality, dashed lines denote the
mean residual, and dotted lines the 66 per cent interval.
Circles denote the blue selected subsample of LSBGs, triangles the 
red selected subsample, and stars the giant LSBGs from Sprayberry et al.\
\protect\shortcite{spray95}.   }
\label{fig:comp}
\end{figure*}

We show a comparison between our measurements 
of total galaxy magnitude, disc central 
surface brightness and disc scale length and the corresponding
literature measurements in all 
available passbands in Fig.\,\ref{fig:comp}.  
Note that there is a mean offset in magnitude of $-$0.28 mag 
between our magnitudes and those of Sprayberry et al.\ \shortcite{spray95}
(stars in Fig.\,\ref{fig:comp}).
This offset does not significantly affect the 
colours of the galaxies as it is constant over all passbands.
The source of this offset is unknown, however if it is corrected for
their magnitudes agree with ours with a RMS of $\sim$ 0.1 mag.  
Therefore, in Fig.\,\ref{fig:comp} and the subsequent comparison, 
we correct Sprayberry et al.'s magnitudes for this offset.
For the complete sample, 68 per cent of the magnitude differences are
smaller than 0.17 mag (68 per cent interval), and 50 per cent
of the magnitude differences are smaller than 0.09 mag.  For the
central surface brightnesses and scale lengths, the 68 per cent
intervals are 0.33 mag\,arcsec$^{-2}$ and 18 per cent, respectively.
The distributions of all of these residuals are strongly non-Gaussian,
with a relatively narrow `core' of accurate comparisons with a more
extended envelope of less accurate comparisons, with a few
pathological cases.

It is important to understand the origin of the larger magnitude errors.
Many of the less accurate magnitude comparisons 
are from very extended galaxies, which have a large
portion of their flux outside the boundaries of the detector.  Hence their
magnitudes are less accurate due to the more uncertain sky levels and
larger total magnitude extrapolations.  These less accurate comparisons
may be a concern however:  do our quoted uncertainties 
reflect the true uncertainties?  To check this, we compared 
the ratio of the magnitude difference and the combined measurement
errors: the median value of this ratio is $\sim$ 0.8.  Therefore, 
we conclude that our formal magnitude error bars are an accurate
reflection of the true uncertainties.

The central surface brightnesses and scale lengths compare
less favourably, with many of the central surface
brightnesses in particular disagreeing by more than their
combined formal error bars.  However, the formal error bars
only give the estimated errors for a 
given fitting method.  In this case, we have compared our measurements
derived using an automated bulge/disc decomposition with 
measurements derived using
a variety of fitting methods, including `marking the disc' fits
\cite{spray95,deblok1995,deblok1996}, disc only fits 
to the whole surface brightness profile \cite{oneil1997a},
bulge/disc decompositions similar to those presented here 
\cite{mcgaugh1994a}, and sophisticated two-dimensional modelling
of the luminosity distribution \cite{dejong1996ii}.  
Therefore, the formal errors derived using each method are
unlikely to reflect the true uncertainties introduced by 
the use of different fitting methods.
An additional source
of scatter is the presence of a bulge component:  when included, a 
bulge component can take some of the light from the 
disc component, and increase the uncertainty in the disc parameters
accordingly.  
We conclude that the formal errors for the central surface
brightness and scale length are unlikely to represent the true range
of uncertainties introduced by using different fitting methods:  more
representative uncertainties are given by the 68 per cent 
intervals and are $\sim$ 0.3 mag and $\sim$ 20 per cent 
respectively: these uncertainties are comparable to those found 
by e.g.\ de Jong \shortcite{dejong1996ii} and de Blok
et al.\ \shortcite{deblok1995}.

We have also compared our adopted ellipticities and position angles
with those in our sample's source papers.  Our ellipticities compare well with
those in the literature:  most galaxies have ellipticity differences
of $\sim 0.05$ or smaller.   
A few galaxies have ellipticity differences greater than 0.1; 
these differences are typically due to the influence of bars, or due to 
low signal-to-noise in both sets of images.  
Comparison of our adopted position angles with those in the literature 
shows agreement to $\sim$ 7$^{\circ}$.

\subsection{LSBG morphology}

It is interesting to compare and comment on the morphologies 
and surface brightness profiles of LSBGs in the optical and the near-IR.
Bergvall et al.\ \shortcite{bergvall1999} found a tendency towards 
similar morphologies in the optical and near-IR for their sample 
of LSBGs.  We confirm this trend (despite our typically poorer 
signal-to-noise in the near-IR):  this suggests that 
LSBGs lack the dust content and significant amounts of recent star formation
that make morphological classification so passband-dependent for 
galaxies with higher surface brightness \cite{block1994,block1999}.  

Our sample, because of its explicit 
selection to cover as wide a range of LSBG parameters as possible, has
a wide range of morphologies, and as is discussed in the next section, 
SFHs.  The red selected LSBG subsample contains a lenticular
galaxy (C1-4), two early type spirals, and two later type spirals.  
O'Neil et al.\ \shortcite{oneil1997a}, because
of their typically lower spatial resolution, did not find
strong evidence for bulges in most of the red LSBGs.
However, with our higher resolution data 
(usually the INT 2.5-m $R$ band images), we find that four out of the five
red selected LSBGs show evidence for an exponential 
bulge component.  
The LSBG giants all have very strong 
bulges, with bulge to disc ratios of the order of unity in most passbands.
Two LSBG giants are better fit with an exponential bulge component, 
and the rest are better fit (in the $\chi^2$ sense) by a $r^{1/4}$ law
bulge profile.  The discs of LSBG giants usually have pronounced 
spiral structure (see, e.g.\ Sprayberry et al.\ 1995 for images).

The blue selected LSBGs come with a variety of morphologies, from relatively 
well-defined spiral morphologies with weak bulges,
to galaxies with nearly
exponential disc profiles, to galaxies with central `troughs' in 
their luminosity profiles, compared with expectations from an
exponential disc fit.  Two examples of these `trough'
galaxies are ESO-LV 1040220 and 1040440.  
These galaxies have galactic extinction and inclination corrected
central surface brightnesses in $B$ of 
24.1 and 23.4 mag\,arcsec$^{-2}$ respectively.  Thus, in many 
respects, they are similar to the blue LSBGs
studied by Bergvall et al.\ \shortcite{bergvall1999}.  
They found that at surface brightnesses lower than $\sim$ 
23 $B$ mag\,arcsec$^{-2}$, centrally-depressed surface brightness 
profiles are quite common.  On the basis of this limited and
incomplete sample, it is difficult to properly 
confirm their finding, but
our analysis of the surface brightness profiles certainly
tentatively supports their observation.

In Table \ref{tab:bulgedisc}, and to 
a certain extent Fig.\,\ref{fig:surfphot}, it is 
apparent that the disc scale lengths typically decrease and
the bulge to disc ratios increase with increasing wavelength.  These are
clear signatures of colour gradients in our sample of LSBGs.
In the next section, we investigate these colour gradients: we will
argue that these are primarily due to stellar population 
gradients.

\section{Star formation histories} \label{sec:sfh}

\begin{figure*}
\begin{minipage}{17.0cm}
\psfig{figure=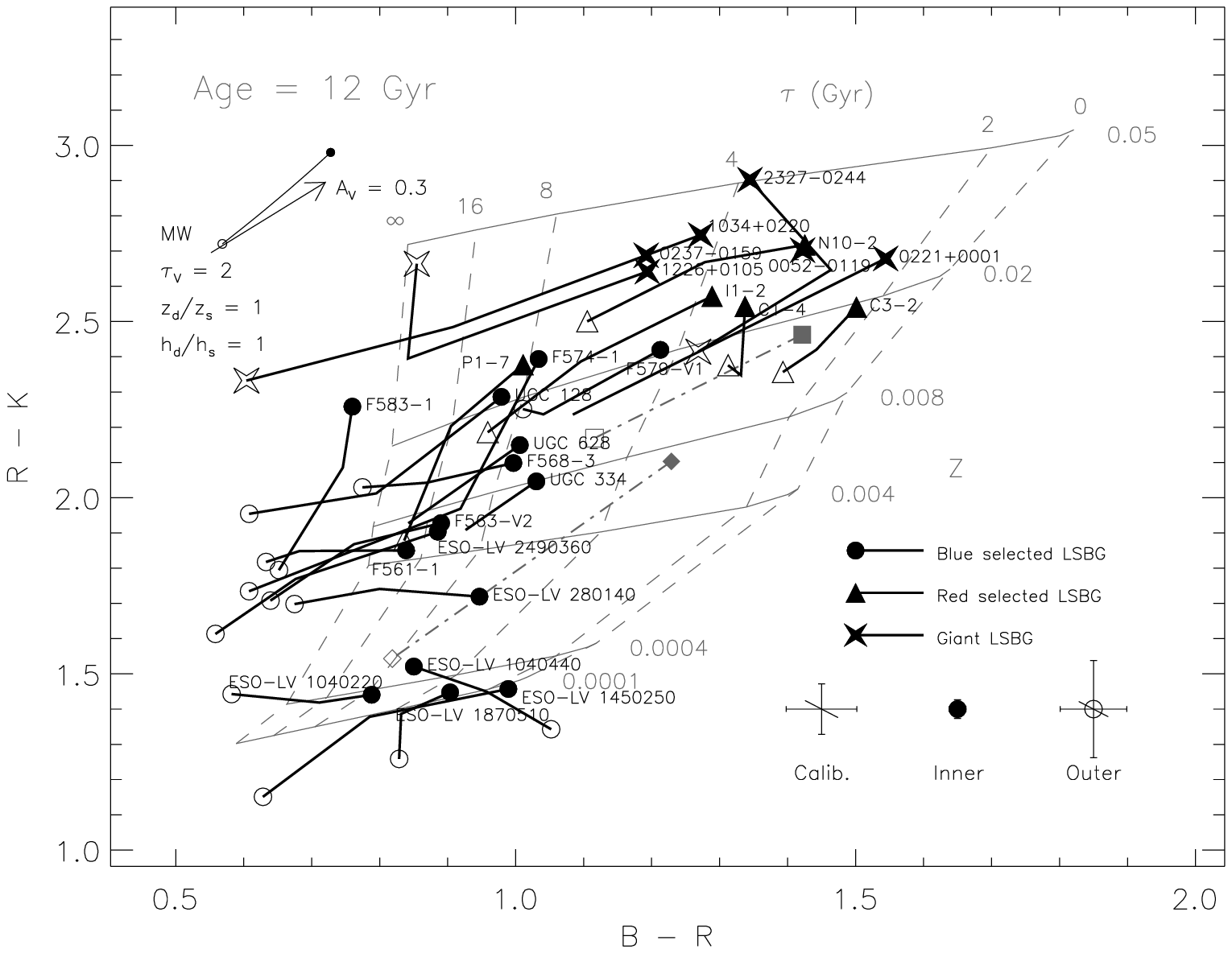}
\end{minipage} 
\caption{Galactic extinction and K-corrected 
$B - R$ and $R - K$ colours for our LSBG sample in 
three radial bins per galaxy: 
$0 < r/h_K < 0.5$ (solid symbols), $0.5 < r/h_K < 1.5$ 
and $1.5 < r/h_K < 2.5$ (open symbols).
Circles denote the blue selected subsample of LSBGs, triangles the 
red selected subsample, and stars the giant LSBGs.
Typical colour uncertainties in the inner and outer bins of 
the sample due to sky level errors are shown in the lower right hand
corner.  Calibration
uncertainties are also shown.  As $R$ band is used to form both colours,
the effects of a typical 1$\sigma$ error in $R$ is also shown (diagonal line).
Over-plotted are Bruzual \& Charlot's (in preparation) {\sc gissel98}
stellar population models, and a foreground screen and Triplex 
model \protect\cite{ddp,evans1994} dust reddening vector (see text for further 
details).  Average Sa--Sc (grey squares) and Sd--Sm (grey diamonds) colours
from de Jong \protect\shortcite{dejong1996iv} at the centre 
(solid symbols) and at
2 disc scale lengths (open symbols) are also shown.}
\label{fig:colcol}
\end{figure*}

The main motivation for this programme of optical and near-IR 
imaging was to study the stellar populations of a diverse sample
of LSBGs.  In this section, we construct accurate colours as a
function of radius for our sample and compare these colours
with model stellar populations.

We first degrade 
all of the edited images of a given galaxy to the same angular 
resolution, and carry out surface photometry in three radial 
bins: $0 < r/h_K < 0.5$, $0.5 < r/h_K < 1.5$ and $1.5 < r/h_K < 2.5$,
where $h_K$ is the $K$ band disc scale length.  Galactic extinction 
corrections were taken from Schlegel et al.\ 
\shortcite[see Table \protect\ref{tab:samplepar}]{sfd}.  K-corrections
for the LSBG giants (as they are at non-negligible redshifts) were computed 
using non-evolving Sbc spectra from King \& Ellis \shortcite{king}, 
and are typically 0.2 mag in $B$, negligible in $R$ and $-0.2$ mag in 
$K$.  If required, $K'$ magnitudes were converted $K$ magnitudes using
the relation in Wainscoat \& Cowie \shortcite{wainscoat1992}, assuming 
a typical $H - K$ colour of $\sim$ 0.3 mag \cite{dejong1996iv}.
The error in this correction is unlikely to exceed 0.05 mag.

We show Galactic extinction and K-corrected $B - R$ and $R - K$ 
colours for our sample
of LSBGs in Fig. \ref{fig:colcol}.  
Measurements in the three radial 
bins per galaxy are connected by solid lines, measurements of the 
central bin are denoted by a solid circle and measurements of the
colours at two $K$ band disc scale lengths are
denoted by open circles.  Note that we only show colours with uncertainties
smaller than 0.3 mag.  Average zero point uncertainties and uncertainties
in the inner and outer points due to sky level errors are also
shown.  Because $R$ band is used in both colour combinations, errors in
$R$ band affect both colours.  Accordingly, the average 1$\sigma$ $R$
band error is also shown (the short diagonal line).
A $B - R$ against $R - K$ colour-colour plot was chosen to maximise 
the number of galaxies on the diagram.  Other colour
combinations (e.g.\ $B - V$ against $R - K$, etc.) limit
the number of galaxies that can be plotted, but  
yield consistent positions on the 
stellar population grids.  Note that P1-7 is placed on the plot assuming 
a $V - R$ colour of $0.40 \pm 0.03$ (derived from the degenerate $B - V$
against $V - R$ colour-colour diagram using a typical galactic
extinction corrected colour for P1-7 of $B - V \sim 0.57$).

Overplotted on the same diagram are stellar population models 
and model dust reddening vectors.  We use the {\sc gissel98} 
implementation of the stellar population 
models of Bruzual \& Charlot (in preparation).  For Fig.\,\ref{fig:colcol}
we adopt a Salpeter \shortcite{sp} initial mass fuction (IMF)
and an exponentially decreasing star formation rate 
characterised by an e-folding timescale $\tau$ and a single, 
fixed stellar metallicity $Z$ (where $Z = 0.02$ is the adopted 
solar metallicity).  For these models, the galaxy age (i.e.\ the time
since star formation first started in the galaxy) is fixed at 12 Gyr.
Note that the shape of the model grid
is almost independent of galaxy age $\ga$ 5 Gyr.
The solid grey lines represent the colours of stellar populations
with a fixed metallicity and a variety of star formation 
timescales.  The dashed grey lines represent the colours of 
stellar populations with different, fixed stellar metallicities and 
a given star formation timescale.  
There is some uncertainty in the shape and placement of the model
grid:  Charlot, Worthey and Bressan \shortcite{charlot1996} discussed
the sources of error in stellar population synthesis models, and 
concluded that the uncertainty in model calibration 
for older stellar populations is 
$\sim$ 0.1 mag in $B - R$ and $\sim$ 0.2 mag in $R - K$, which is 
roughly comparable to the calibration error bars in 
Fig.\,\ref{fig:colcol}.  The colour uncertainties 
are larger for some stellar populations than for others: e.g.\
the optical--near-IR colours for older, near-solar metallicity
stellar populations are relatively secure, whilst the colours
for younger, extreme metallicity stellar populations are much 
more uncertain.  Other sources of systematic error include 
dust reddening and SFH
uncertainties (as our assumption of a smoothly varying
SFR is almost certainly unrealistic).
For these reasons the 
positions of galaxies and colour 
trends between galaxies should be viewed relatively, e.g.\ that 
one class of galaxies is more metal-rich than another class of galaxies.

We also include vectors describing the effect of dust reddening
using either a screen model (arrow) or the more realistic
geometry of a Triplex dust model \cite[curved line]{ddp,evans1994}.  For the 
Triplex model a closed circle denotes the central reddening, and 
an open circle the reddening at 2.5 disc scale lengths.  
For both reddening vectors we use the Milky Way 
extinction law (and for the 
Triplex model, the albedo) from Gordon, Calzetti \& Witt 
\shortcite{gordon1997}.  Our Triplex model assumes equal 
scale length vertically and radially 
exponential dust and stellar distributions, and a central 
optical depth (for viewing a background object) in $V$ band of 2.
These parameters are designed to be a reasonable upper
limit to the effects of reddening in most 
reasonably bright spiral galaxies:  Kuchinski et al.\ \shortcite{k98}
find from their study of 15 highly-inclined spiral galaxies that
they are well-described by the above type of model, with central
$V$ band optical depths of between 0.5 and 2.
For the purposes of calculating the Triplex reddening vector, we use
the optical depth due to absorption only for two reasons.  Firstly,
one might na\"{\i}vely expect that for face-on galaxies at least
as many photons will get scattered into the line of sight as out of it.
Secondly, de Jong \shortcite{dejong1996iv} finds that an absorption-only
Triplex model is a reasonably accurate description of the results of
his realistic Monte-Carlo simulation 
(including the effects of both scattering and 
absorption) of a face-on Triplex geometry spiral galaxy.

\subsection{Results} \label{subsec:sfhres}

In the following interpretation of Fig.\,\ref{fig:colcol}, 
we for the most part {\it explicitly neglect the possible effects of dust 
reddening}.  There are a number of arguments that suggest
that the effects of dust reddening, while present, are less 
important than stellar population differences both within
and between galaxies.  We consider these effects more carefully
in section \ref{subsec:dust}.

\subsubsection{Colour gradients}

The majority of the LSBGs in our sample with colours in 
at least two radial bins have significant 
optical--near-IR colour gradients.  For the most part 
{\it these colour gradients are consistent with the presence of 
a mean stellar age gradient}, where the outer regions of galaxies
are typically younger, on average, than the inner regions of 
galaxies.  This is consistent with the findings
of de Jong \shortcite{dejong1996iv}
who concludes that age gradients are common in spiral galaxies 
of all types.  Note that this is inconsistent with the conclusions
of Bergvall et al.\ \shortcite{bergvall1999}, who find little
evidence for optical--near-IR colour gradients in their sample
of blue LSBGs.  However, their sample
was explicitly selected to lack significant $B - R$ colour gradients,
and so that we disagree with 
their conclusion is not surprising.  The
fact that Bergvall et al.\ managed to 
find galaxies without significant colour gradients is
interesting in itself;  while colour gradients are common 
amongst disc galaxies, {\it they are by no means universally present}.
This is an important point, as by studying the systematic
differences in physical properties between galaxies with and
without colour gradients, it may be possible to identify
the physical mechanism by which an age gradient is generated
in disc galaxies.

There are galaxies which have colour gradients which appear inconsistent
with the presence of an age gradient alone.
The two early-type LSBGs C1-4 and C3-2 have small colour gradients
which are more consistent with small metallicity gradients than with age
gradients.  In addition, a few of the bluer late-type LSBGs  have colour
gradients that are rather steeper than expected on the basis of age gradients
alone, most notably F583-1 and P1-7, both of which appear to have 
colour gradients more consistent with a metallicity gradient.
ESO 1040440 has an `inverse' colour gradient, in that the central 
regions are bluer than the outer regions: this is probably due to a 
combination of very irregular morphology and the effects of foreground
stellar contamination (which in this case 
makes the galaxy colours difficult to accurately estimate).
There are also conspicuous `kinks' in the colour profiles of F574-1 and 
2327-0244, and a possible kink in the colour profile of UGC 128.  
In the case of F574-1, it is likely that the central colours of
the galaxy are heavily affected by dust reddening.  F574-1 is quite
highly inclined ($67^{\circ} \pm 3^{\circ}$, assuming an
intrinsic disc axial ratio $q_0$ of 0.15; Holmberg 1958) and 
shows morphological indications of substantial amounts of 
dust extinction in the INT 2.5-m $B$ and $V$ band images.  
This may also be the case for UGC 128, although the inclination is
smaller in this case and there are no clear-cut morphological indications of 
substantial dust reddening.  The LSBG giant 2327-0244 has a red nucleus in the 
near-IR, but a relatively blue nucleus in the optical.  However, 2327-0244 is 
is a Seyfert 1 and has a central starburst \cite{terlevich1991}, 
making our interpretation of the central colours in 
terms of exponentially decreasing SFR models invalid.

\subsubsection{Colour differences between galaxies}

In considering colour trends between LSBGs, we will 
first consider the star formation histories of different classes
of LSBG, and then look at the star formation histories in a more global 
context.  

\paragraph{Blue selected LSBGs}
The majority of blue selected LSBGs, 
e.g.\ F561-1, UGC 334 and ESO-LV 2490360, are
also blue in the near-IR.
Their colours indicate 
that most blue selected LSBGs are younger than HSB
late-types, with rather similar metallicities
(compare the positions of the main body of blue selected LSBGs to
the average Sd--Sm galaxy from de Jong's sample).
This relative youth, compared to the
typically higher surface brightness sample 
of de Jong \shortcite{dejong1996iv}
suggests that the age of a galaxy
may be more closely related to its surface brightness than the metallicity is. 
We will come back to this point later in section
\ref{subsec:correl}.

Four out of the fifteen blue selected 
LSBGs fail to fit this trend: ESO-LV 1040220, 
ESO-LV 1040440, ESO-LV 1450250 and ESO-LV 1870510 all fall substantially
($\sim 0.3$ mag) bluewards of the main body of blue LSBGs in $R - K$ 
colour, indicating a lower average metallicity.  While it should 
be noted that the South Pole subsample all have larger zero point
uncertainties than the northern hemisphere sample, we feel that it 
is unlikely that the zero point could be underestimated
so substantially in such a large number of cases.
While the models are tremendously uncertain at such young ages and 
low metallicities, these optical--near-IR colours 
suggest metallicities $\la 1/10$ solar.
This raises an interesting point:  most low metallicity galaxies 
in the literature are relatively high density blue compact dwarf
galaxies \cite{thuan1999,izotov1997,hunter1995}.
Thus, these galaxies and e.g.\ the blue LSBGs from 
Bergvall et al.\ \shortcite{bergvall1999} 
offer a rare opportunity to study 
galaxy evolution at both low metallicities {\it and} densities, perhaps 
giving us quite a different view of how star formation 
and galaxy evolution work at low metallicity.

\paragraph{Red selected LSBGs}
In Bell et al.\ \shortcite{b99}, we found that two red selected 
LSBGs (C1-4 and C3-2) were old and metal rich, indicating that they are more 
evolved than blue selected LSBGs.  In this larger sample of five
red selected LSBGs, we find that this is not always the case.
The galaxies N10-2, C3-2, 
and C1-4 (with $B$ band central surface brightnesses $\sim$
22.5 mag\,arcsec$^{-2}$) 
{\it are} comparitively old and metal rich, with central colours
similar to old, 
near-solar metallicity stellar populations.  However, I1-2
and P1-7 (with lower $B$ band central surface brightnesses $\sim$
23.1 mag\,arcsec$^{-2}$) both have reasonably blue 
galactic extinction corrected optical--near-IR colours.
I1-2 has a stellar population similar to de Blok et al.'s 
\shortcite{deblok1996} F579-V1, which is part of our blue selected
subsample: these galaxies lie in the overlap between the two samples.
P1-7 is relatively similar to other, brighter blue LSBGs
such as UGC 128 and F574-1: it was included in the red selected 
subsample only by virtue of 
its relatively high foreground extinction $A_B = 0.45$ mag.  

Our limited data suggests that the red-selected LSBGs catalogued
by O'Neil et al.\ \shortcite{oneil1997a,oneil1997b} are a 
very heterogeneous group.  Unlike the blue or giant LSBGs, 
the red-selected LSBGs seem to have relatively few common traits.
The five red-selected LSBGs in this study seem to be a mix of two 
types of galaxy: i) early-type spirals or lenticulars which 
are genuinely red but have surface brightnesses 
$\mu_{B,0} \sim$ 22.5 mag\,arcsec$^{-2}$ at the upper range of 
the LSB class, and ii) objects with low surface brightnesses
but colours that are not genuinely red.  Objects in class ii)
appear in the red-selected subsample due to 
large galactic foreground reddening and photometric errors.

Recently, O'Neil et al. \shortcite{oneil1999} reported the detection
of a small number of red, gas-rich LSBGs.  These galaxies would clearly
contradict our above interpretation of the red LSBGs, suggesting that
at least some of the red LSBGs are a distinct (though potentially
quite rare) class of galaxy.  It is interesting to note that 
Gerritsen \& de Blok 
\shortcite{gerritsen1999} predict that around 20 per cent
of LSBGs should have $B - V \sim 1$ and relatively low 
surface brightnesses $\mu_{B,0} \ga 24.0$ mag\,arcsec$^{-2}$: 
these galaxies, which represent the fraction of the 
LSBG population that lack recent star formation, 
would appear to be both red and gas rich (blue LSBGs would 
be galaxies with exactly the same past SFH, but more recent star 
formation).
However, note that O'Neil et al.'s result is subject to 
significant observational uncertainties: for example, the $B - V$ colour
of P1-7 adopted by O'Neil et al.\ \shortcite{oneil1999} is 0.9, whereas
the foreground galactic extinction-corrected colour of P1-7 in this
study is found to be 0.57$\pm$0.1.  P1-7 is on the verge of being 
classified as a red, gas rich LSBG with a $B - V$ colour of 0.9, however
it lies well within the envelope of blue, gas rich LSBGs with 
a galactic extinction corrected $B - V$ colour of 0.57.  
Proper observational 
characterisation of these red, gas rich LSBG candidates may 
prove quite crucial in testing LSBG formation and evolution models.

\paragraph{Giant LSBGs}
LSBG giants have 
galactic extinction and K-corrected 
optical--near-IR colours which are similar to 
the redder LSBGs from O'Neil et al.\ \shortcite{oneil1997a,oneil1997b},
indicating central stellar populations that are reasonably old 
with roughly solar metallicity.
However these central colours
are not consistent with an old, single burst stellar 
population, although this conclusion is somewhat dependent
on the choice of K-correction and model uncertainties 
(K-correction uncertainties 
are typically $\la$ 0.05 mag in each axis at 20000 km\,s$^{-1}$ 
for a change in galaxy type from Sbc to Sab, or Sbc to Scd).
The outer regions of LSBG giants are much younger, on average, 
whilst still retaining near-solar metallicities.  
Comparing the $B - V$ and $R - K$ colours
for the LSBG giants with  
some of the bluer HSB Sa--Sc galaxies in de Jong's \shortcite{dejong1996iv}
sample shows that both sets of galaxies 
have similar stellar populations.   Thus, LSBG giants do not have
unique SFHs (implying that there need be no difference in e.g.\
star formation mechanisms between the HSB Sa--Sc galaxies and the 
LSBG giants), however it is interesting that
LSBG giants can have both a substantial young stellar population 
and a high metallicity, whilst possessing such low stellar surface
densities (assuming reasonable stellar mass to light ratios).  

\paragraph{Summary}
Overall, there is a 
clear age and metallicity sequence, with red 
LSBGs and LSBG giants in the high stellar metallicity and older average 
age corner of the plot, 
progressing to the lower stellar metallicity and younger average
age region of the plot for the blue LSBGs.
{\it This suggests that LSBGs, just like HSBGs, come in a variety of
morphological types and SFHs; moreover, 
their morphologies and SFHs are linked}.

\subsection{Are the stellar population differences real?} 
	\label{subsec:discreal}

\subsubsection{Are IMF uncertainties important?}

We have constructed Fig.\,\ref{fig:colcol} 
using a Salpeter \shortcite{sp} IMF.  
However, the IMF is still reasonably uncertain, especially at the 
very low mass and high mass ends (see e.g.\ Elmegreen 1999, Scalo 1998).
Variation of the low mass end of the IMF only significantly
changes the (relatively ill-constrained) stellar mass to light ratio.
Variation of the high mass end of the IMF, as illustrated by
using e.g.\ the Scalo \shortcite{sc} or Miller \& Scalo \shortcite{ms}
IMF, only significantly changes the high metallicity, young
stellar populations corner of the colour-colour plane:  the colours in 
these regions are subject to considerable modelling uncertainties 
at any rate.  We conclude that our results are robust to reasonable
IMF uncertainties. 

\subsubsection{Dust reddening}
	\label{subsec:dust}

We have interpreted Fig.\,\ref{fig:colcol} in terms of stellar 
population differences, however we have so far neglected the 
effects of dust reddening on the colours of our stellar 
populations.  In Fig.\,\ref{fig:colcol} we show 
both a screen $A_V = 0.3$ and absorption-only Triplex model 
$\tau_V = 2$ dust reddening vector, both assuming Milky 
Way extinction and albedo curves.  SMC extinction and albedo
curves make little difference to the direction of the
reddening vectors, but increases the length of the 
Triplex model reddening vector slightly because of the 
lower albedo at all wavelengths, compared to the Milky Way dust properties.
The screen model vector is shown for
illustrative purposes only, as the dust will be 
distributed roughly similarly to the starlight in 
realistic spiral galaxies, as assumed in the Triplex model.

The colour gradients in our sample of LSBGs are unlikely to be 
due to the effects of dust reddening for four reasons.  
\begin{enumerate}
\item The optical
depth that we choose for the Triplex model is on the
high end of the plausible dust optical depths derived
by Kuchinski et al.\ \shortcite{k98}.  
\item The Triplex model assumes that all of the dust is distributed 
smoothly throughout the galaxy.  In real galaxies
typically 1/3 of the dust is gathered into optically
thick clumps, which, when viewed face-on, simply tend to 
`drill holes' in the light distribution but not
produce any significant colour changes \cite{dejong1996iv,k98}.  
This dust would be 
more easily found in more edge-on galaxies, as in the edge-on orientation
the probability of a clump along the line of sight is much 
larger than in a face-on disc.  Therefore, Kuchinski et al.'s
dust optical depths are likely to account for both the clumped
and unclumped dust.
\item Most LSBGs have lower metallicity than their higher
surface brightness counterparts.  This implies that the dust to gas
ratio should be lower in LSBGs than in HSBGs.  
\item Only one galaxy in our sample, the starbursting
Seyfert 1 2327-0244, has been detected 
by IRAS.  This is in stark contrast to the diameter
limited sample of de Jong \& van der Kruit \shortcite{dejong1994},
for which 78 per cent of the sample was detected by IRAS.
This suggests that dust is much less important in LSBGs than in 
their higher surface brightness counterparts: indeed, many of the 
non-detections in de Jong \& van der Kruit's sample were LSBGs. 
\end{enumerate}

The above arguments suggest that the Triplex dust 
reddening vector in Fig.\,\ref{fig:colcol} is an upper
limit to the real effects of dust reddening.  Comparison of
this upper limit on the dust reddening effects with the colour
gradients in our sample suggests that most galaxies must have
stellar population gradients to produce such pronounced colour
gradients.  This conclusion is supported by the direct modelling
of Kuchinski et al.\ \shortcite{k98}, who use their own best-fit
galaxy models to estimate the colour gradients in face-on spiral
galaxies:  they conclude that colour gradients as large as the ones
we observe in LSBGs, or observed by de Jong \shortcite{dejong1996iv},
are too large to arise solely from reddening by realistic amounts of dust.
The same argument applies to relating different populations
of LSBG:  the amount of reddening required to make an intrinsically
blue LSBG appear red is too large to be compatible with 
the observational evidence cited above.

\section{Discussion} \label{sec:disc}

\subsection{Quantifying the colour-colour plane}

In section \ref{subsec:sfhres}, we saw a pronounced age/metallicity
sequence.  LSBGs can have a diverse range of stellar populations: 
red selected LSBGs can be quite old and metal rich, whilst
most blue selected LSBGs are much younger, and more metal poor.
However, from Fig.\,\ref{fig:colcol} alone, it is difficult to 
see {\it why} the star formation histories of low surface 
brightness galaxies are as diverse as they are.  For this reason, 
we have chosen to quantify the positions of galaxies on the colour-colour
plane by assigning average ages and metallicities from the {\sc gissel98} 
stellar population models of Bruzual \& Charlot (in preparation).

A finely interpolated grid of stellar population models was generated,
covering a range of star formation timescales and metallicities.
The metallicity ranges 
from $\langle Z \rangle = \log_{10}(Z/Z_{\sun}) = -2.0$ (1/100 solar) 
to $\langle Z \rangle = 0.4$ (2.5 times solar).
In Fig.\,\ref{fig:colcol}, we can see that some galaxies have 
optical--near-IR colours too blue to be described adequately by 
a constant star formation rate:  for this reason, we must include 
exponentially {\it increasing} SFHs in our model grid. 
We include star formation timescales $\tau$ ranging from
0 (an average age of 12 Gyr) to $\infty$ (constant star formation 
with an average age of 6 Gyr) to $-1$ Gyr (an average age of 1 Gyr).
We can see from the range of models included above that 
$\tau$ is an inappropriate parameter for describing the SFHs:  
we parameterise the SFH using the average age of the stellar population.

For each galaxy, the galactic extinction and K-corrected 
$B - R$ and $R - K$ colours between $0.5 < r/h_K < 1.5$ (when 
available; 0052-0119 and 0237-0159 have only central colours
and are omitted from further consideration)
are compared with the colours of the finely interpolated
stellar populations grid. The quantity 
$\chi^2 = \sum_{i=(B-V),(R-K)} (M_i - O_i)^2/\sigma_i^2$ is minimised,
where $M_i$ is the model colour, $O_i$ is the observed colour and
$\sigma_i$ is the error in the observed colour.  Error bars are
obtained for the average ages and metallicities by fitting the 
colours plus or minus their 1$\sigma$ sky level and zero 
point errors added in quadrature. 

This procedure has a number of limitations.  
Firstly, we neglect the effects of dust reddening.
We considered the possible effects of 
dust earlier in section \ref{subsec:dust}, and concluded
that the effects of dust are fairly minimal for this sample.
Our use of the colours between 
$0.5 < r/h_K < 1.5$ should act to further lessen the importance
of dust reddening on our results.
Secondly, because we use simple SFHs assuming a single metallicity, 
and because of stellar population model uncertainties, the ages and 
metallicities we determine are unlikely to be accurate in absolute
terms.  However, the idea of this simple analysis
is to identify relative trends in SFH
and metallicity, and their possible causes.  The ages and metallicities
we derive here will serve this purpose well:  the optical--near-IR colours
provide a robust way to order galaxies in terms of the relative importance
of $\ga$ 5 Gyr old stars compared to younger $\la$ 2 Gyr old stars, 
which we parameterise using the average age of the stellar population.
To better understand the relative uncertainties, we have also 
carried out the analysis using the stellar population models of 
Kodama \& Arimoto \shortcite{ka97}:  the results obtained with
their models are indistinguishable (in a relative sense) to the 
results presented in Figs. \ref{fig:correl} and 
\ref{fig:correl2}.  An extended version of this analysis using 
all the available colour 
combinations for a much larger sample of galaxies will be presented
in Bell \& de Jong \shortcite{papii}.

\subsection{The correlations} \label{subsec:correl}

\begin{figure*}
\begin{minipage}{17.0cm}
\psfig{figure=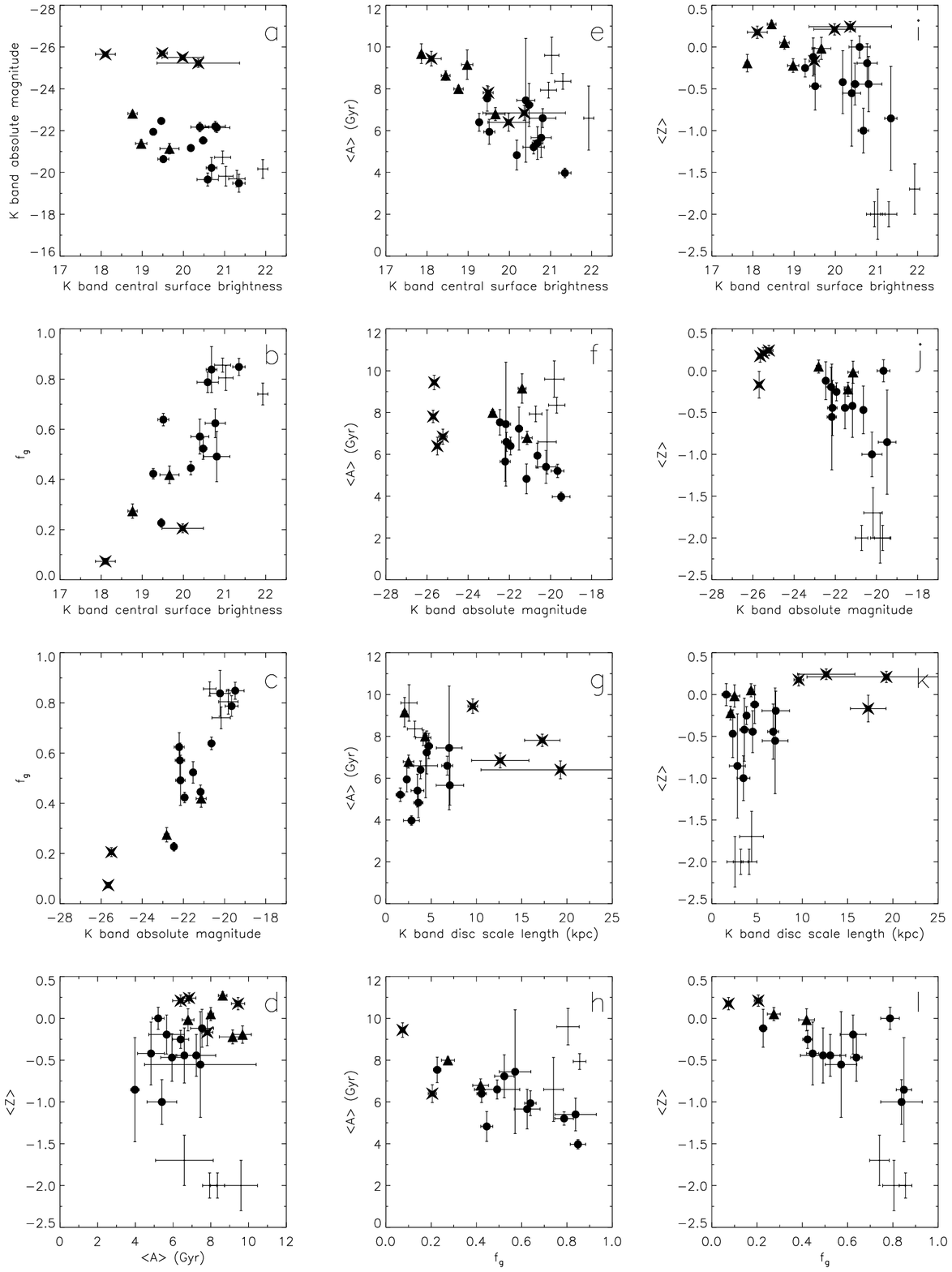}
\end{minipage} 
\caption{Correlations between best-fit average ages, metallicities,
$K$ central surface brightnesses, $K$ absolute magnitudes, $K$
disc scale lengths and gas fractions.  
Circles denote the blue selected subsample of LSBGs, triangles the 
red selected subsample, and stars the giant LSBGs.
Note that the ages for the 
four ESO-LV galaxies near the low metallicity edge of the grid are 
highly uncertain: these points are denoted by naked error bars in all plots.}
\label{fig:correl}
\end{figure*}

In order to understand what physical mechanisms might be driving
the SFHs of LSBGs, we plot the best-fit
average ages and metallicities against physical 
parameters, such as 
$K$ central surface brightnesses, $K$ absolute magnitudes, $K$
disc scale lengths and gas fractions.  These correlations are
presented in Fig.\,\ref{fig:correl}.  Note that we do not attempt
to provide a best fit or attach any kind of statistical 
significance to these relationships.  We feel that this is
over-interpreting this relatively limited dataset at this stage.

Galactic extinction and K-corrected 
$K$ absolute magnitudes are calculated using distances
from Table \ref{tab:samplepar}.
These absolute magnitudes are combined with 
the \hi masses (multiplied by 1.33 to account for 
the helium mass fraction) from Table \ref{tab:samplepar} to
derive the gas fractions, using a $K$ band $M/L$ of 0.6 $M_{\sun}/L_{\sun}$
(c.f.\ Verheijen 1998; Chapter 6) and
assuming a solar $K$ band absolute magnitude of 3.41 \cite{allen}.
Note that these gas fractions do not account for any 
molecular or hot gas component. 
The same distances are used to derive the 
$K$ band disc scale lengths in kpc.
The $K$ band surface brightnesses have been corrected
for galactic extinction, $K$ corrections, $(1+z)^4$ surface
brightness dimming and inclination.  Inclination corrections $C^i$ have
been determined using an intrinsic edge-on disc axial ratio $q_0 = 0.15$ 
assuming that the disc is transparent;
these corrections are given by:
\begin{equation}
C^i = -2.5 \log_{10} (\frac{(1-e)^2-q_0^2}{1-q_0^2})^{1/2},
\end{equation} 
where $e$ is the adopted disc ellipticity from Table \ref{tab:epa}
\cite{holmberg1958}.

Before discussing Fig.\,\ref{fig:correl}, it is useful to note that
the average ages of ESO-LV 1040220, 1040440, 1450250 and 1870510 are
all potentially quite uncertain.  This is due to their low metallicities:
they have optical--near-IR colours suggesting metallicities $\la$ 1/100
solar metallicity (which falls off the model grid). 
At these metallicities, the
model predictions (especially for young ages) are all rather uncertain.  
Therefore, we feel that their ages are less 
constrained (even in a relative sense) than their error bars otherwise 
suggest.  These points are denoted by a naked set of error bars 
in Figs.\,\ref{fig:correl} and \ref{fig:correl2}.
In the discussion that follows, we typically ignore the 
average ages for these four data points.

\subsubsection{Limitations of our dataset}

In Fig.\,\ref{fig:correl}, panel a, we show the relationship between 
$K$ band central surface brightness and $K$ band absolute magnitude
in our dataset. 
In panels b and c, we show the relationship between gas fraction 
and $K$ band surface brightness and magnitude respectively. 
The main purpose of these panels is to simply show that
in our sample of LSBGs, surface brightness, magnitude and gas fraction 
are quite strongly
inter-related, and it is expected that correlations of
any given quantity with 
surface brightness will be reflected by correlations of that same
quantity with magnitude and gas fraction.  
Thus, on the basis of this dataset alone, 
it will be difficult to 
unambiguously differentiate  
between trends in the stellar populations of LSBGs driven by magnitude,
trends driven by surface brightness, or trends driven
by the gas fraction.  Note that because of selection effects
(primarily that we select only galaxies with 
low $B$ band surface brightness),
panels a, b and c are unlikely
to be representative of any universal correlation 
between surface brightness, magnitude and gas fraction.  

\subsubsection{Ages and metallicities}

Fig.\,\ref{fig:correl}, panel d shows the correlation between the best-fit
average age and metallicity.  Ignoring the uncertain ages of the 
ultra-low metallicity LSBGs, there is a correlation, albeit a noisy
one, between age and metallicity, in the sense that older galaxies 
tend to be more metal rich.  This correlation quantifies the 
global trend between age and metallicity 
observed earlier in Fig.\,\ref{fig:colcol}.  This correlation is
expected:  galaxies appear old through a lack of recent star formation, 
implying a lack of gas (see also panel h, the age--gas fraction
correlation).  Most of our sample LSBGs are reasonably massive, 
meaning that their gas is unlikely to have been ejected by 
e.g.\ supernova driven winds.  Therefore, their low gas fraction
is likely to be due to consumption by
star formation:  this results in 
reasonably high stellar metallicities (see e.g.\ Pagel 1998).  
Note that some 
low surface brightness dwarf spheroidals 
may run counter to this argument, having low metallicities 
and reasonably old average ages (e.g.\ Mateo 1998):  
to explain this one would have to invoke 
removal of the gas from these low mass galaxies by e.g.\
supernova-driven winds \cite{dekel} or, for cluster 
dwarf spheroidals, ram-pressure stripping
\cite{abadi}.

\subsubsection{Trends in LSBG age}

In panels e, f, g and h, we show the trends in age with 
$K$ central surface brightness, $K$ absolute magnitude, 
$K$ disc scale length and gas fraction respectively.
Age does not appear to correlate with disc
scale length.  Neglecting the uncertain ages
of the four low metallicity ESO-LV galaxies, there appear 
to be trends in age with surface brightness, magnitude
and gas fraction, in the sense that lower $K$ band surface brightness, 
lower $K$ luminosity and higher gas fraction LSBGs all have 
younger average ages.  Furthermore, the scatter between age and 
surface brightness appears smaller than the scatter in the
age--magnitude and age--gas fraction correlations.  This suggests, 
tentatively, that the age--surface brightness correlation is the
main correlation in this dataset, and the other correlations are
the result of the magnitude--surface brightness and gas 
fraction--surface brightness correlations.  
We return to this issue later in section \ref{subsubsec:mod}.

\subsubsection{Trends in LSBG metallicity}

In panels i, j, k and l of Fig.\,\ref{fig:correl}, 
we show the trends in LSBG metallicity
with $K$ band surface brightness, magnitude, disc scale length
and gas fraction.  Note that the metallicities of the four
low metallicity ESO-LV galaxies are reasonably well-constrained, 
therefore it is fair to include them in the discussion of these trends.
Metallicity, unlike age, seems to be correlated with $K$ band disc scale 
length.  This seems primarily because of the LSBG giants: they 
are bright and have large scale lengths, so their high metallicities will
produce correlations between metallicity and magnitude and metallicity
and scale length.  It is likely that a metallicity--magnitude 
relation is more fundamental, however it is impossible to 
rule out scale length dependence in the metallicity on the 
basis of this dataset.

Metallicity, like age, seems to correlate well
with $K$ band surface brightness, magnitude and gas fraction
in the sense that LSBGs with lower metallicities have
lower surface brightnesses, lower luminosities and higher gas
fractions than LSBGs with higher metallicities.  
Both age and metallicity are also linked (panel d), so the similarity
in trends with physical parameters between age and metallicity
is not surprising.  Again, as was the case for the age, it is
more or less impossible to tell, on the basis of this dataset alone, 
which of the magnitude--metallicity or surface brightness--metallicity
correlations are more fundamental.  

\subsection{SFH as a function of total mass and density}
 \label{subsubsec:mod}

\begin{figure*}
\begin{minipage}{17.0cm}
\psfig{figure=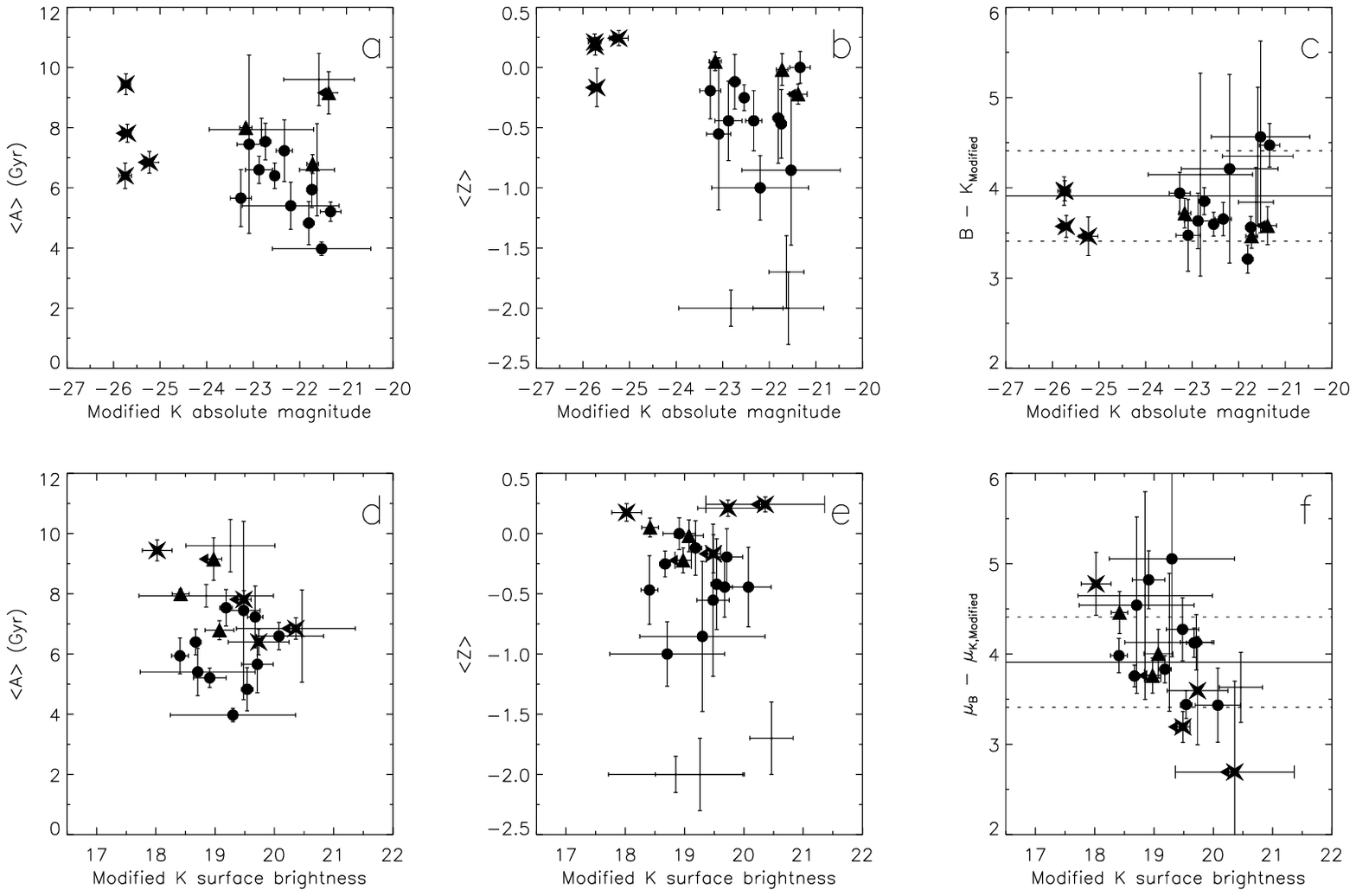}
\end{minipage} 
\caption{Correlations between best-fit average ages and metallicities
and $K$ band surface brightnesses and magnitudes, modified
by turning all of the available gas mass into stars with a $K$ band
mass to light ratio of 0.6 $M_{\sun}/L_{\sun}$.  
Circles denote the blue selected subsample of LSBGs, triangles the 
red selected subsample, and stars the giant LSBGs:
naked error bars denote the four
low metallicity galaxies with potentially large age errors.
Red selected or giant LSBGs at known distances 
for which there are no gas masses
are shown as lower limits by error bars, with a solid triangle
pointing towards the direction in which the galaxies move with
increasing gas fraction:  the vertex of the solid triangle denotes
the modified surface brightness or magnitude assuming a gas fraction 
of 0.2.  Panels c and f show the difference between the $B$ band 
absolute magnitude and the modified $K$ absolute magnitude and 
the difference between the $B$ band central surface brightness and
modified $K$ band central surface brightness, respectively.
Solid lines denote an average `colour' $B - K_{\rm Modified} = 3.9$,
and the dashed line a scatter of 0.5 mag around this average.}
\label{fig:correl2}
\end{figure*}

In order to investigate how the {\it baryonic} mass and surface 
density affect the SFHs of LSBGs, we have modified
the $K$ surface brightnesses and absolute magnitudes by turning 
the neutral gas fraction of the galaxy into stars using the 
correction to the magnitude $+ 2.5 \log_{10} (1 - f_g)$.  
This correction assumes that the gas will turn into 
stars with the same $K$ band mass to light ratio as was
assumed for the stellar population, which in this case is 
0.6 $M_{\sun}/L_{\sun}$.  This correction makes a number of assumptions.
\begin{enumerate}
\item The $K$ band mass-to-light ratio is expected to be relatively 
robust to the presence of young stellar populations, however
our assumption of a constant $K$ band mass to light ratio is
still a crude assumption.  Note however that the 
relative trends in Fig.\,\ref{fig:correl2} are quite robust
to changes in stellar mass-to-light ratio: as the stellar mass-to-light ratio
increases, the modified magnitudes creep closer to their unmodified
values asymptotically. 
\item The presence of molecular gas is 
not accounted for in this correction.
Assuming a typical galactic CO to H$_2$ conversion ratio, the 
non-detections of LSBGs by Schombert et al.\ \shortcite{schombert1990}
and de Blok \& van der Hulst \shortcite{deblok1998co} imply 
H$_2$ to \hi ratios smaller than 0.25, and exceptionally as low as 0.04.
This result is complicated by the expected 
metallicity dependence in the 
CO to H$_2$ conversion ratio \cite{mihos1999}, however, 
molecular hydrogen is unlikely to dominate for all plausible
values of the conversion factor.
\item The correction
to the surface brightness 
implicitly assumes that
the gas will turn into stars with the same spatial distribution 
as the present-day stellar component.  This is a poor
assumption as the gas distribution is usually much more
extended than the stellar distribution; most LSBG gas
distributions are centrally peaked however, and so the 
correction is unlikely to be completely wrong.  
\end{enumerate}
All of the above arguments suggest that the modified magnitudes and 
surface brightnesses are unlikely to give an accurate measure of the true
masses and densities of LSBGs.  These quantities do however better
reflect the total mass and density of the galaxy than the $K$ band
magnitudes and surface brightnesses alone, and so are useful
as an indication of the kind of trends we might see if
we could work out the baryonic masses and densities of LSBGs.
The median
correction is -0.9 mag, and 68 per cent of the corrections lie
within 0.7 mag of this value.

The trends in age and metallicity with modified $K$ band
central surface brightness and absolute magnitude are shown in 
Fig.\,\ref{fig:correl2}.  
Three red and giant LSBGs with known redshifts but unknown (or 
unmeasurable) \hi fluxes have been placed on Fig.\,\ref{fig:correl2}
also:  error bars denote their (unmodified) magnitudes or surface
brightnesses, and solid triangles denote the direction that
these galaxies move in with increasing gas fraction, with the vertex
of the triangle indicating the offset that is produced by converting a 
gas fraction of 0.2 into stars.
For reference, in panels c and f,
we also show the relationship
between $B$ band magnitude and surface brightness and 
the modified $K$ band magnitude and surface brightness.
Surprisingly, we find that $B$ band magnitude and surface brightness
are reasonable reflections of the modified magnitudes and 
surface brightnesses 
of LSBGs (remembering the uncertain assumptions that
went into constructing the modified magnitudes and surface brightnesses).
Both the $B$ band magnitudes and surface brightnesses 
have a mean offset from the modified $K$ band values 
consistent with a +3.9 mag offset,
with an RMS of around 0.5 mag (solid and dotted lines in panels c and f).
This relationship may be caused by 
the sensitivity of the optical mass to light ratios
to recent star formation:  $B$ band mass to light ratios decrease
for younger populations in a such a way as to make the increase in
brightness for younger populations offset almost exactly (to within a 
factor of 60 per cent or so, from the 0.5 mag scatter) by their larger
gas fractions, making the $B$ band a reasonable total mass indicator
in LSBGs (assuming that the $K$ band stellar mass to light 
ratio is fairly constant).

In panels a and b of Fig.\,\ref{fig:correl2}, we see that the 
LSBG ages and metallicities correlate with their modified
$K$ band magnitudes (baryonic masses).  In stark contrast, 
in panels d and e, we see that the ages and metallicities
correlate poorly with the modified $K$ band surface brightnesses
(baryonic surface densities).  This result is puzzling, especially in 
the light of the tight correlation between average age and 
surface brightness in panel e of Fig.\,\ref{fig:correl}.  There are
two possible interpretations of this huge scatter in the 
age--modified surface brightness plane.

Is it possible that there is little or no correlation 
between the SFH and surface density of a galaxy, as panels d and e suggest?
In this case, we must explain the strong correlations between 
age and $K$ band surface brightness and metallicity and 
$K$ band surface brightness in panels e and i of Fig.\,\ref{fig:correl}.
This may be possible to explain via selection effects:
because we selected our galaxies to have a relatively narrow
range of $B$ band surface brightness, and older, more metal rich
galaxies are redder, we would naturally expect to see an artificial
correlation between $K$ band surface brightness (which would be higher
because of the redder colours) and age (which correlates with
these red colours).  Note, however, that this interpretation has real
difficulty in explaining the tightness 
and dynamic range of the age--$K$ band surface brightness
correlation.

Conversely, 
is it possible that there is a correlation between baryonic 
surface density and age, and that for some reason it is masked in 
this dataset?  
There is quite a narrow range (only $\sim$ 2 mag\,arcsec$^{-2}$)
of modified surface brightness:  the close
correspondence between $B$ band and 
modified $K$ band surface brightnesses (panel f of Fig.\,\ref{fig:correl2})
means that because we selected our galaxies to have a relatively narrow
range of $B$ band surface brightness, we have implicitly selected
for a narrow range in baryonic central surface density.  
Also, our modifications to the surface brightnesses were large
for gas-rich galaxies with low $K$ band surface brightnesses and much 
smaller for gas-poor galaxies with higher $K$ band surface brightnesses, 
steepening the relationship between SFH and modified surface brightness
considerably.  Thus, we may
simply lack an adequate surface density range to be able to distinguish
a steep trend in age with surface density, especially in the presence of 
significant uncertainties in the ages and surface densities. 
Note that this is the interpretation that we prefer:
the correlation between age and $K$ band surface
brightness is quite tight, which is quite suggestive of 
some kind of correlation between density and SFH.  
Note that the age--mass relation would survive in this case relatively 
unscathed:  the dynamic range for the age--mass relation is in excess 
of 5 magnitudes, which makes it quite robust to changes $\sim$ 1 mag
in the relative positioning of the data points along the trend.

\subsection{A unifying view}

All the above correlations are consistent with the proposition 
that {\it the age of a LSBG stellar population is primarily
correlated with its surface density and that its metallicity is 
correlated with both surface density and mass}, albeit with
considerable scatter.  Panels e, f, i and 
j of Fig.\,\ref{fig:correl} all strongly support this 
scheme:  in particular, LSBG giants have a range of 
$K$ band surface brightnesses (and ages; panel f) 
but have bright $K$ band absolute
magnitudes (and high metallicities; panel j).  The main obstacle
for such a scenario is the lack of correlation between
modified $K$ band surface brightnesses and age and metallicity, which 
{\it may} be due to our explicit selection of galaxies with 
only a relatively narrow range of $B$ band central surface brightness.

An important question to ask is if this scenario is 
physically plausible.
A relationship between surface brightness/density and average age would
be easy to understand:  either a density dependent (e.g.\ Schmidt 1959)
star formation law or a star formation timescale which is 
proportional to the local dynamical timescale 
would result in such a correlation (note that
a magnitude--age correlation might suggest a global dynamical
timescale dependence in the star formation law; e.g.\ Kennicutt 1998).  
Such a scenario would also very naturally account for the existence of
age gradients in LSBGs:  the outer regions of LSBGs are less dense, 
and would form stars less quickly than their inner regions, resulting in 
an age gradient.

Such a scheme is qualitatively consistent with those proposed
by e.g.\ de Blok et al.\ \shortcite{deblok1996}, 
Jimenez et al.\ \shortcite{jimenez1998} or
Gerritsen \& de Blok \shortcite{gerritsen1999}.  In these schemes, 
it is primarily the low surface density that slows down the 
evolution of LSBGs (either explicitly through a density-dependent
star formation law or implicitly, through an inability to build up
a high gas metallicity).  Metallicity's mass dependence is 
likely to stem from the relative importance of feedback:  
the efficiency with which a galaxy ejects its metals in 
a supernova-driven wind is primarily driven by its mass 
\cite{maclow1999}.

While the above scenario is consistent with the data, it is unlikely to be 
unique:  because of the correlation between surface brightness and 
magnitude, a purely mass-dependent age and metallicity may be plausible.
Note, however that a mass-dependent age has some disadvantages:
the ages of LSBG giants support a surface brightness dependent age,
and age gradients would have to be explained using a different 
mechanism.  

\section{Conclusions} \label{sec:conc}

We have performed deep imaging of 
a sample of 26 LSBGs
in the optical and near-IR in order to study their 
stellar populations.  By comparing their optical--near-IR 
colours with the latest stellar population models, 
it is possible to constrain the young to old
star ratio (parameterised by the average age of the galaxy) and
the galaxy metallicity.  
We have found the following.
\begin{itemize}
\item Optical--near-IR colour gradients are common in LSBGs.  Most
colour gradients are consistent with a mean stellar age gradient,
with the outer regions of galaxies appearing younger than the inner
region of galaxies.  We argue against the effects of dust reddening
as the only cause of LSBG colour gradients.  As common as colour
gradients are, they are not 
present in all LSBGs \cite{bergvall1999}; this is important, because
it may provide a chance to directly observe what drives the age
gradient in LSBGs.
\item We find that LSBGs have a wide range in morphologies and 
stellar populations, ranging from old, near solar metallicity populations
for the very reddest LSBGs, to younger, high metallicity populations
in the LSBG giants, to young and metal poor populations in the blue
gas-rich LSBGs.   
\item By comparing the observed optical--near-IR colours between
0.5 and 1.5 disc scale lengths with stellar population models, 
we have determined best-fit average ages and metallicities that are
robust in a relative sense, so that trends in age and metallicity
will be quite secure.
\item When the highly uncertain ages of the lowest metallicity 
galaxies are excluded, there are strong trends between
both age and metallicity and $K$ band surface brightness, absolute 
magnitude and gas fraction.
LSBGs with low $K$ band surface brightnesses, low $K$ luminosities 
and high gas fractions are all fairly unevolved, young, low metallicity
systems.  In contrast, LSBGs with higher $K$ band surface brightnesses,
higher $K$ band luminosities and smaller gas fractions appear much 
more evolved, with older average ages and higher (near solar) metallicities.
\item We have constructed crude estimates of the total 
mass and total densities of the atomic hydrogen, helium 
and stars in LSBGs by
turning the gas into stars with a constant $K$ band 
mass to light ratio of 0.6 $M_{\sun}/L_{\sun}$.  
Surprisingly, we find that the $B$ band 
absolute magnitudes and surface brightnesses
are reasonable predictors of the mass and density estimators, 
to within a factor of $\sim$ 60 per cent or so.
We find that the ages and metallicities of LSBGs correlate
well with our mass estimator, but correlate poorly with our
surface density estimator.
We argue that the poor correlation between our 
surface density estimator and SFH is the result of the 
narrow dynamic range in surface density probed by our sample, compounded
by errors in SFH and density determination.
\item Our results are consistent with a scenario in which 
the age of a LSBG is correlated primarily with its surface
density, and the metallicity of a LSBG is correlated
with both its surface density and mass (albeit with much scatter).
This kind of correlation would be observed if 
the star formation law depended either explicitly on gas
surface density or on the local dynamical timescale, and 
if the efficiency with which a galaxy retained its newly-synthesised
metal content was a function of its mass.
\end{itemize}

\section*{Acknowledgements}

We would like to thank Erwin de Blok, Karen O'Neil,
Stacy McGaugh and David Sprayberry for providing
surface photometry and images of galaxies in their sample, and for
helpful discussions.  In particular, we would like to thank Karen
O'Neil for providing information about her LSBG sample before their
publication, and Enzo Branchini for determining the peculiar motions
of the galaxy sample.
We would also like to thank the referee for useful comments on the manuscript.
EFB would like to thank the Isle of Man Education Department for their
generous support.
Support for RSdJ was provided by NASA through Hubble Fellowship
grant \#HF-01106.01-98A from the Space Telescope Science Institute,
which is operated by the Association of Universities for Research in
Astronomy, Inc., under NASA contract NAS5-26555.
The northern hemisphere near-IR observations were obtained
using the Apache Point Observatory 3.5-m telescope, which is owned
and operated by the Astrophysical Research Consortium.
The United States National Science Foundation supported 
the near-IR observations made at the South Pole through a 
cooperative agreement with the Center for Astrophysical Research 
in Antarctica, Grant No. NSF OPP-8920223.
Some of the observations described in this paper were made during
service time at the Isaac Newton Telescope and at the United Kingdom
Infrared Telescope.  This project made use of STARLINK 
computing facilities in Durham.
This research has made use of the NASA/IPAC Extragalactic Database (NED)
which is operated by the Jet Propulsion Laboratory, California Institute of
Technology, under contract with the National Aeronautics and Space
Administration.


\begin{thebibliography}{}

\bibitem[\protect\citename{Abadi, Moore \& Bower }1999]{abadi}
  Abadi M. G., Moore B., Bower R. G., 1999, MNRAS in press, astro-ph/9903436

\bibitem[\protect\citename{Allen }1973]{allen}
  Allen  C. W., 1973, ``Astrophysical Quantities'', University of 
  London, The Athlone Press

\bibitem[\protect\citename{Barnaby et al.\ }1999]{barn}
  Barnaby D., Harper D. A., Loewenstein, R. F., Mrozek F., Thoma M.,
   Lloyd J. P., Rauscher B. J., Hereld M., Severson S. A., 1999, PASP, 
   submitted 

\bibitem[\protect\citename{Bell et al.\ }1999]{b99}
  Bell E. F., Bower R. G., de Jong R. S., Hereld M., Rauscher B. J.,
  1999, MNRAS, 302, L55

\bibitem[\protect\citename{Bell \& de Jong }1999]{papii}
  Bell E. F., de Jong R. S., 1999, MNRAS, submitted

\bibitem[\protect\citename{Bergvall et al.\ }1999]{bergvall1999}
  Bergvall  N., R\"{o}nnback  J., Masegosa  J., \"{O}stlin  G., 
  1999, A\&A, 341, 697

\bibitem[\protect\citename{Block et al.\ }1994]{block1994}
  Block D. L., Bertin G., Stockton A., Grosb{\o}l P., Moorwood A. F. M., 
	Peletier R. F., 1994, A\&A, 288, 365
 
\bibitem[\protect\citename{Block \& Iv\^{a}nio }1999]{block1999}
  Block D. L., Iv\^{a}nio P., 1999, A\&A, 342, 627
 
\bibitem[\protect\citename{Branchini et al.\ }1999]{enzo}
  Branchini E. F., Teodoro L., Frenk C. S., Schmoldt I., Efstathiou G., 
    White S. D. M., Saunders W., Rowan-Robinson M., Keeble O., Tadros H.,
    Maddox S., Oliver S., Sutherland W., 1999, MNRAS, in press

 
\bibitem[\protect\citename{Casali \& Hawarden }1992]{casali1992}
  Casali M. M., Hawarden T. G., 1992, JCMT-UKIRT Newsletter, 3, 33
 
\bibitem[\protect\citename{Charlot, Worthey \& Bressan }1996]{charlot1996}
  Charlot S., Worthey G., Bressan A., 1996, ApJ, 457, 625
 
\bibitem[\protect\citename{de Blok \& van der Hulst }1998a]{deblok1998spec}
  de Blok W. J. G., van der Hulst J. M., 1998a, A\&A, 335, 421

\bibitem[\protect\citename{de Blok \& van der Hulst }1998b]{deblok1998co}
  de Blok W. J. G., van der Hulst J. M., 1998b, A\&A, 336, 49

\bibitem[\protect\citename{de Blok, McGaugh \& van der Hulst }1996]{deblok1996}
  de Blok W. J. G., McGaugh S. S., van der Hulst J. M., 1996, MNRAS,
  283, 18

\bibitem[\protect\citename{de Blok, van der Hulst \& Bothun }1995]{deblok1995}
  de Blok W. J. G., van der Hulst J. M., Bothun G. D., 1995, MNRAS,
  274, 235


\bibitem[\protect\citename{de Jong }1996a]{dejong1996iv}
  de Jong R. S., 1996a, A\&A, 313, 377

\bibitem[\protect\citename{de Jong }1996b]{dejong1996ii}
  de Jong R. S., 1996b, A\&AS, 118, 557

\bibitem[\protect\citename{de Jong \& van der Kruit }1994]{dejong1994}
  de Jong  R. S., van der Kruit  P. C., 1994, A\&AS, 106, 451

\bibitem[\protect\citename{Dekel \& Silk }1986]{dekel}
  Dekel A., Silk J., 1986, ApJ, 303, 39


\bibitem[\protect\citename{Disney, Davies \& Phillipps }1989]{ddp}
  Disney M. J., Davies J. I., Phillipps S., 1989, MNRAS, 239, 939 

\bibitem[\protect\citename{Elias et al.\ }1982]{elias}
  Elias  J. H., Frogel  J. A., Matthews  K., Neugebauer  G., 1982, 
  AJ, 87, 1029 

\bibitem[\protect\citename{Elmegreen }1999]{elmegreen1999}
  Elmegreen B. G., 1999, in 
	``The Evolution of Galaxies on Cosmological Timescales'',  
	eds. Beckman J. E. \& Mahoney T. J., ASP Conference Series

\bibitem[\protect\citename{Evans }1994]{evans1994}
  Evans R., 1994, PhD Thesis, University of Cardiff 

\bibitem[\protect\citename{Gerritsen \& de Blok }1999]{gerritsen1999}
  Gerritsen J. P. E., de Blok W. J. G., 1999, A\&A, 342, 655

\bibitem[\protect\citename{Gordon et al.\ }1997]{gordon1997}
  Gordon K. D., Calzetti D., Witt A. N., 1997, ApJ, 487, 625

\bibitem[\protect\citename{Holmberg }1958]{holmberg1958} 
  Holmberg E., 1958, Medd.\ Lunds Astron.\ Obs.\ Ser., 2, No.\ 136

\bibitem[\protect\citename{Huchtmeier \& Richter }1989]{huchtmeier1989}
  Huchtmeier W. K., Richter O.-G., 1989, ``A General Catalog of
  \hi Observations of Galaxies'', New York, Springer-Verlag

\bibitem[\protect\citename{Hunt et al.\ }1998]{Hunt1998}
  Hunt L. K., Mannucci F., Testi L., Migliorini S., Stanga R. M.,
  Baffa C., Lisi F., Vanzi L., 1998, AJ, 115, 2594

\bibitem[\protect\citename{Hunter \& Thronson }1995]{hunter1995}
  Hunter D. A., Thronson H. A., 1995, ApJ, 452, 238


\bibitem[\protect\citename{Izotov et al.\ }1997]{izotov1997}
  Izotov Y. I., Lipovetsky V. A., Chaffee F. H., Foltz C. B., 
     Guzeva N. G., Kniazev A. Y., 1997, ApJ, 476, 698

\bibitem[\protect\citename{Jimenez et al.\ }1998]{jimenez1998}
  Jimenez R., Padoan P., Matteucci F., Heavens A. F., 1998, MNRAS, 299, 123 

\bibitem[\protect\citename{Kennicutt }1998]{kennicutt1998}
  Kennicutt R. C., Jr., 1998, ApJ, 498, 181 

\bibitem[\protect\citename{King \& Ellis }1985]{king}
  King C. R., Ellis R. S., 1985, ApJ, 288, 456

\bibitem[\protect\citename{Kodama \& Arimoto }1997]{ka97} 
  Kodama T. \& Arimoto N., 1997, A\&A, 320, 41 (KA97)

\bibitem[\protect\citename{Kuchinski et al.\ }1998]{k98}
  Kuchinski L. E., Terndrup D. M., Gordon K. D., Witt A. N., 1998, AJ,
  115, 1438 

\bibitem[\protect\citename{Landolt }1992]{landolt1992}
  Landolt A. U., 1992, AJ, 104, 372

\bibitem[\protect\citename{Lauberts \& Valentijn }1989]{esolv}
  Lauberts A., Valentijn E. A., 1989, ``The surface photometry
  catalogue of the ESO-Uppsula galaxies.''

\bibitem[\protect\citename{MacLow \& Ferrera }1999]{maclow1999}
  MacLow M.-M., Ferrera A., 1999, ApJ, 513, 142

\bibitem[\protect\citename{Mateo }1998]{mateo}
  Mateo M., 1998, ARA\&A, 36, 435

\bibitem[\protect\citename{McGaugh }1994]{mcgaugh1994b}
  McGaugh S. S., 1994, ApJ, 426, 135

\bibitem[\protect\citename{McGaugh \& Bothun }1994]{mcgaugh1994a}
  McGaugh S. S., Bothun G. D., 1994, AJ, 107, 530

\bibitem[\protect\citename{McGaugh \& de Blok }1997]{mcgaugh1997}
  McGaugh S. S., de Blok W. J. G., 1997, ApJ, 481, 689


\bibitem[\protect\citename{McGaugh, Schombert \& Bothun }1995]{mcgaugh1995mor}
  McGaugh S. S., Schombert J. M., Bothun G. D., 1995, AJ, 109, 2019

\bibitem[\protect\citename{Mihos, Spaans \& McGaugh }1999]{mihos1999}
  Mihos J. C., Spaans M., McGaugh S. S., 1999, ApJ, 515, 89

\bibitem[\protect\citename{Miller \& Scalo }1979]{ms}
  Miller G. E., Scalo J. M., 1979, ApJS, 41, 513

\bibitem[\protect\citename{Moshir et al.\ }1990]{iras}
Moshir M., Kopan G., Conrow T., McCallon H., Hacking P., 
Gregorich D., Rohrbach G., Melnyk M., Rice W., Fullmer L., et al., 1990,
`Infrared Astronomical Satellite Catalogs---The Faint Source Catalog', 
Version 2.0.

\bibitem[\protect\citename{Nguyen et al.\ }1996]{nguyen1996}
  Nguyen H. T., Rauscher B. J., Severson S. A., Hereld M., et al.\ ,
  1996, PASP, 108, 718

\bibitem[\protect\citename{O'Neil et al.\ }1997a]{oneil1997a}
  O'Neil K., Bothun G. D., Cornell M. E., 1997a, AJ, 113, 1212

\bibitem[\protect\citename{O'Neil et al.\ }1997b]{oneil1997b}
  O'Neil K., Bothun G. D., Schombert J. M., Cornell M. E., Impey C. D.,
  1997b, AJ, 114, 2448

\bibitem[\protect\citename{O'Neil et al.\ }1999]{oneil1999}
  O'Neil K., Bothun G. D., Schombert J. M., 1999, AJ, accepted, 
  astro-ph/9909129

\bibitem[\protect\citename{Padoan, Jimenez \& Antonuccio-Delogu }1997]
  {padoan1997}
  Padoan P., Jimenez R., Antonuccio-Delogu V., 1997, ApJ, 481, L27

\bibitem[\protect\citename{Pagel }1998]{pagel}
  Pagel B. E. J., 1998, ``Nucleosynthesis and Chemical Evolution
     of Galaxies'' (Cambridge University Press, Cambridge)
 


\bibitem[\protect\citename{Persson et al.\ }1998]{nicmos}
  Persson S. E., Murphy D. C., Krzeminski W., Roth M., Rieke M. J.,
  1998, AJ, 116, 2475

\bibitem[\protect\citename{Press et al.\ }1986]{numrec}
  Press W. H., Flannery B. P., Teukolsky S. A., Vetterling W. T.,
  1986, Numerical Recipes---The Art of Scientific Computing
  (Cambridge University Press, Cambridge).

\bibitem[\protect\citename{Quillen \& Pickering }1997]{quillen1997}
  Quillen A. C., Pickering T. E., 1997, astro-ph/9705115

\bibitem[\protect\citename{Rauscher et al.\ }1998]{rauscher1998}
  Rauscher B. J., Lloyd J. P., et al., 1998, ApJ, 506, 116

\bibitem[\protect\citename{Rauscher et al.\ }1999]{rauscher1999}
  Rauscher B. J., Lloyd J. P., et al., 1999, MNRAS, submitted


\bibitem[\protect\citename{Richter, Tammann \& Huchtmeier }1987]{richter}
  Richter O.-G., Tammann G. A., Huchtmeier W. K., 1987, A\&A, 171, 33

\bibitem[\protect\citename{R\"{o}nnback \& Bergvall }1995]{ronnback1995}
  R\"{o}nnback J., Bergvall N., 1995, A\&A, 302, 353

\bibitem[\protect\citename{Salpeter }1955]{sp}
  Salpeter E. E., 1955, ApJ, 121, 61

\bibitem[\protect\citename{Scalo }1986]{sc}
  Scalo J. M., 1986, Fundam. Cosmic Phys., 11, 1

\bibitem[\protect\citename{Scalo }1998]{sc98}
  Scalo J. M., 1998, in ``The Stellar Initial Mass Function'', eds.
	Gilmore G. \& Howell D.,  ASP Conference Series,
                   142, 201

\bibitem[\protect\citename{Schlegel, Finkbeiner \& Davis }1998]{sfd}
  Schlegel D. J., Finkbeiner D. P., Davis M., 1998, ApJ, 500, 525 

\bibitem[\protect\citename{Schmidt }1959]{schmidt1959}
  Schmidt M., 1959, ApJ, 129, 243

\bibitem[\protect\citename{Schneider et al.\ }1992]{schneider1992}
  Schneider S. E., Thuan T. X., Magnum J. G., Miller J., 
  1992, ApJS, 81, 5


\bibitem[\protect\citename{Schombert et al.\ }1990]{schombert1990}
  Schombert J. M., Bothun G. D., Impey C. D., Mundy L. G., 1990, AJ,
  100, 1523

\bibitem[\protect\citename{Sprayberry et al.\ }1995]{spray95}
Sprayberry D., Impey C. D., Bothun G. D., Irwin M. J., 1995, AJ, 
109, 558

\bibitem[\protect\citename{Terlevich et al.\ }1991]{terlevich1991}
Terlevich R., Melnick J., Masegosa J., Moles M., Copetti M. V. F., 
1991, A\&AS, 91, 285

\bibitem[\protect\citename{Theureau et al.\ }1998]{theureau1998}
Theureau G., et al., 1998, A\&AS, 130, 333

\bibitem[\protect\citename{Thuan, Izotov \& Foltz }1999]{thuan1999}
  Thuan T. X., Izotov Y. I., Foltz C. B., 1999, astro-ph/9905345

\bibitem[\protect\citename{van der Hulst et al.\ }1993]{vdh93}
  van der Hulst J. M., Skillman E. D., Smith T. R., Bothun G. D.,
  McGaugh S. S., de Blok W. J. G., 1993, AJ, 106, 548

\bibitem[\protect\citename{van Zee, Haynes \& Salzer }1997]{vanzeesfr}
  van Zee L., Haynes M. P., Salzer J. J., 1997, AJ, 114, 2479

\bibitem[\protect\citename{Verheijen }1998]{verheijen1998}
  Verheijen M. A. W., 1998, PhD Thesis, University of Groningen

\bibitem[\protect\citename{Wainscoat \& Cowie }1992]{wainscoat1992}
  Wainscoat R. J., Cowie L. L., 1992, AJ, 103, 332

\bibitem[\protect\citename{Wegner, Haynes \& Giovanelli }1993]{wegner1993}
  Wegner G., Haynes M. P., Giovanelli R., 1993, AJ, 105, 1251


\end{thebibliography}
\end{document}